\documentclass[journal]{IEEEtran}
\usepackage{ifpdf}

\usepackage{amsmath}
\usepackage{algorithm}
\usepackage[noend]{algpseudocode}

\makeatletter
\def\BState{\State\hskip-\ALG@thistlm}
\makeatother

\usepackage{cite}

\usepackage{lipsum}

\usepackage{amsmath}
\usepackage{amssymb}
\usepackage{flexisym}
\usepackage{array}
\newcolumntype{C}[1]{>{\centering\let\newline\\\arraybackslash\hspace{0pt}}m{#1}}

\usepackage{color}
\definecolor{dblue}{rgb}{0,0,0.8}

\usepackage{xcolor}

\usepackage{nomencl}
\makenomenclature
\usepackage{etoolbox}

\usepackage{subfig}
\usepackage{graphicx}
\usepackage{fixltx2e}

\usepackage{stfloats}
\usepackage{nomencl}
\makenomenclature

\usepackage[justification=centering]{caption}

\usepackage{nomencl}
\makenomenclature

\usepackage{setspace}

\usepackage[utf8]{inputenc}

\usepackage{amssymb}

\usepackage{nomencl}
\makenomenclature
\usepackage{etoolbox}
\usepackage{hyperref}

\usepackage{url}

\usepackage{mhchem}

\usepackage{nomencl}
\makenomenclature

\usepackage[justification=centering]{caption}

\usepackage{nomencl}
\makenomenclature

\usepackage{algorithm}
\usepackage{algpseudocode}
\usepackage{amsmath}
\usepackage{algorithm}
\usepackage{algpseudocode}
\usepackage{amsmath}

\usepackage[table]{xcolor} 
\usepackage{caption}       

\usepackage{hyperref}
\makenomenclature
 \usepackage{algorithm}
\usepackage{algpseudocode}
\usepackage{multirow}
\usepackage{graphicx}
\usepackage{cite}
\usepackage{amsmath,amssymb,amsfonts}
\usepackage{graphicx}
\usepackage{textcomp}
\usepackage{booktabs}
\usepackage{tabularx}
\usepackage{ifpdf}

\usepackage{amsmath}
\usepackage{algorithm}
\usepackage[noend]{algpseudocode}

\makeatletter
\def\BState{\State\hskip-\ALG@thistlm}
\makeatother

\usepackage{cite}

\usepackage{amsmath} 
\usepackage{amssymb} 
\usepackage{longtable} 
\usepackage{array} 
\usepackage{booktabs} 
\usepackage{nomencl} 
\makenomenclature

\ifCLASSINFOpdf
   \DeclareGraphicsExtensions{.pdf,.jpeg,.png}
\else
\fi
\usepackage{lipsum}

\usepackage{amsmath}
\usepackage{amssymb}
\usepackage{flexisym}
\usepackage{array}
\newcolumntype{C}[1]{>{\centering\let\newline\\\arraybackslash\hspace{0pt}}m{#1}}

\usepackage{color}
\definecolor{dblue}{rgb}{0,0,0.8}
\usepackage{xcolor}

\usepackage{nomencl}
\makenomenclature
\usepackage{etoolbox}

\usepackage{subfig}
\usepackage{graphicx}
\usepackage{fixltx2e}

\usepackage{stfloats}
\usepackage{nomencl}
\makenomenclature

\usepackage[justification=centering]{caption}

\usepackage{nomencl}
\makenomenclature
\usepackage{enumitem} 
\usepackage{setspace}

\usepackage[utf8]{inputenc}

\usepackage{amssymb}

\usepackage{nomencl}
\makenomenclature
\usepackage{etoolbox}
\usepackage{hyperref}

\usepackage{url}

\usepackage{mhchem}

\usepackage{nomencl}
\makenomenclature

\usepackage[justification=centering]{caption}

\usepackage{algpseudocode}
\usepackage{amsmath}

\usepackage{array}

\usepackage[table]{xcolor} 
\usepackage{caption}       

\usepackage{hyperref}
\makenomenclature
 \usepackage{algorithm}
\usepackage{algpseudocode}

\begin{document}
%


\title{Resilient Design and Optimal Operation of Battery Energy Storage Systems for Behind-the-Meter Data Center Microgrids}


%

\author{Hamed Haggi,~\IEEEmembership{Member, IEEE}, Chinmay Morankar,~\IEEEmembership{Senior Member, IEEE}\\ \vspace{0.1cm}
\IEEEauthorblockA{WSP USA Inc., Houston, TX, USA}


\thanks{Hamed Haggi and Chinmay Morankar are with the WSP USA Inc, Houston, TX, USA. Emails: hamed.haggi@wsp.com, chinmay.morankar@wsp.com.}
}

\maketitle

\begin{abstract} Increasing deployment of AI-driven hyperscale data centers, combined with escalating power requirements, interconnection bottlenecks, and high reliability standards, has accelerated the adoption of behind-the-meter (BTM) microgrids utilizing dedicated on-site generation assets. However, the highly dynamic and stochastic nature of modern data center loads creates operational and planning challenges that are not adequately captured by conventional load representations. To this end, this paper proposes a high-fidelity hyperscale data center load modeling framework that disaggregates facility demand into detailed IT and non-IT components and represents load behavior at second- and minute-level resolutions. Building upon this foundation, a comprehensive multi-phase optimization framework is developed for the planning and resilient design of battery energy storage systems (BESS) for BTM data center microgrids. The framework deploys BESS on the load side to act as a dynamic buffer between highly stochastic demand and generation assets. By smoothing the load spikes and excursions before they are observed by gas turbines, the load-side BESS reduces operational stress and limits conditions that can contribute to sub-synchronous torsional interactions (SSTI), shaft fatigue, system trips, equipment damage, and forced outages. Simultaneously, generation-side BESS resources are optimally sized to provide spinning reserves and enhance system resilience. The proposed methodology co-optimizes BESS with solar, wind, natural gas fuel cells, simple-cycle gas turbines, and combined-cycle gas turbines while incorporating practical operational, reliability, resiliency, and technology-specific constraints. Simulation results demonstrate the proposed methodology's ability to identify cost-effective, resilient, and operationally feasible designs that satisfy the demanding reliability requirements and energy market challenges of next-generation hyperscale data centers while providing valuable planning insights for investors, developers, and system operators. \end{abstract}

\begin{IEEEkeywords}
Battery Energy Storage System, Carbon Free Energy, Data Center Load Model, Fuel Cell, Gas Turbines, Load Smoothing, Locational Marginal Pricing, Microgrid, Operation, Optimization, Renewable Energy, Sub Synchronous Torsional Interactions.
\end{IEEEkeywords}

\IEEEpeerreviewmaketitle

\renewcommand\nomgroup[1]{%
  \item[\bfseries
  \ifstrequal{#1}{N}{{\textbf{\textit{Sets and Indices}}}}{%
    \ifstrequal{#1}{P}{{\textbf{\textit{Parameters and Variables}}}}{}%
  }]
}
\mbox{}

\nomenclature[N]{{$\mathcal{I}$}}{{Set of gas turbines, indexed by $i$.}}
\nomenclature[N]{{$\mathcal{T}$}}{{Set of time intervals, indexed by $t$.}}

\nomenclature[P]{{$\Delta t$}}{{Time interval (hr, min, or s).}}
\nomenclature[P]{{$P_{[t]}^{Load}$}}{{Data center load (MW).}}
\nomenclature[P]{{$P_{[t]}^{Shed}$}}{{Load shedding amount (MW).}}
\nomenclature[P]{{$P_{[t]}^{Net}$}}{{Smoothed load variable (MW).}}
\nomenclature[P]{{$P_{[t]}^{Imp}$}}{{Imported power from the grid (MW).}}
\nomenclature[P]{{$P_{[t]}^{Exp}$}}{{Exported power to the grid (MW).}}
\nomenclature[P]{{$R^{X,Dn}$}}{{Technology $X$ ramp-down limit (MW/hr).}}
\nomenclature[P]{{$R^{X,Up}$}}{{Technology $X$ ramp-up limit (MW/hr).}}
\nomenclature[P]{{$\overline{X}$}}{{Technology $X$ maximum limit.}}
\nomenclature[P]{{$\underline{X}$}}{{Technology $X$ minimum limit.}}
\nomenclature[P]{{${\delta}_{[t]}$}}{{Binary variable for avoiding simultaneous grid power import and export.}}
\nomenclature[P]{{$VOLL$}}{{Value of lost load (\$/MWh).}}
\nomenclature[P]{{$RRP$}}{{Ramp-rate violation penalty term (\$/MWh).}}
\nomenclature[P]{{$LMP$}}{{Locational marginal price (\$/MWh).}}
\nomenclature[P]{{$PPA$}}{{Power purchase agreement price (\$/MWh).}}
\nomenclature[P]{{$\alpha_L$}}{{Data center load error coefficient.}}

\nomenclature[P]{{$P_{[t]}^{GT}$}}{{Gas turbine active power generation (MW).}}
\nomenclature[P]{{$P_{[t]}^{GT,Res}$}}{{Gas turbine reserve power (MW).}}
\nomenclature[P]{{$R_{[t]}^{Slack}$}}{{Slack reserve variable for feasibility (MW).}}
\nomenclature[P]{{${U}_{[t]}$}}{{Binary variable for the on/off status of a turbine.}}
\nomenclature[P]{{${SU}_{[t]}$}}{{Binary variable for the startup status of a turbine.}}
\nomenclature[P]{{${SD}_{[t]}$}}{{Binary variable for the shutdown status of a turbine.}}
\nomenclature[P]{{$R^{GT,SU}$}}{{Gas turbine startup ramp limit (MW/hr).}}
\nomenclature[P]{{$R^{GT,SD}$}}{{Gas turbine shutdown ramp limit (MW/hr).}}
\nomenclature[P]{{$UT_i$}}{{Minimum uptime for gas turbine $i$ (hr).}}
\nomenclature[P]{{$DT_i$}}{{Minimum downtime for gas turbine $i$ (hr).}}
\nomenclature[P]{{$R_{[t]}^{Req}$}}{{Minimum system reserve requirement (MW).}}
\nomenclature[P]{{$C_{[i]}^{GT,SRMC}$}}{{Short-run marginal cost of gas turbine $i$ (\$/MWh).}}
\nomenclature[P]{{$C_{[i]}^{GT,NL}$}}{{No-load cost of gas turbine $i$ (\$/hr).}}
\nomenclature[P]{{$\gamma^{Crtl}$}}{{Penalty cost for maximizing gas turbine usage (\$/MWh).}}
\nomenclature[P]{{$\mu^{Crtl}$}}{{Gas turbine reserve feasibility penalty term (\$/MWh).}}
\nomenclature[P]{{$GT^{HR}$}}{{Gas turbine heat rate (MMBtu/MWh).}}
\nomenclature[P]{{$V^{O\&M}$}}{{Variable operation and maintenance cost of gas turbines (\$/MWh).}}
\nomenclature[P]{{$FP$}}{{Fuel price for gas turbines (\$/MMBtu).}}

\nomenclature[P]{{$P_{[t]}^{PV}$}}{{Dispatched PV power (MW).}}
\nomenclature[P]{{$P_{[t]}^{PV_{Crtl}}$}}{{Curtailed PV power (MW).}}
\nomenclature[P]{{$C_{[i]}^{PV}$}}{{Annualized fixed operation and maintenance cost of PV.}}
\nomenclature[P]{{$\rho^{Crtl}$}}{{PV curtailment penalty cost(\$/MWh).}}

\nomenclature[P]{{$P_{[t]}^{Wind}$}}{{Dispatched wind power (MW).}}
\nomenclature[P]{{$P_{[t]}^{Wind_{Crtl}}$}}{{Curtailed wind power (MW).}}
\nomenclature[P]{{$C_{[i]}^{Wind}$}}{{Annualized fixed operation and maintenance cost of wind generation.}}
\nomenclature[P]{{$\zeta^{Crtl}$}}{{Wind curtailment penalty cost(\$/MWh).}}
\nomenclature[P]{{$\alpha_P/\alpha_W$}}{{PV and wind forecast error coefficients.}}

\nomenclature[P]{{$P_{[t]}^{ch}$}}{{BESS charging power (MW).}}
\nomenclature[P]{{$P_{[t]}^{dis}$}}{{BESS discharging power (MW).}}
\nomenclature[P]{{${SOC}_{[t]}$}}{{BESS state of charge (\%).}}
\nomenclature[P]{{$P_{[t]}^{BESS,Res}$}}{{BESS reserve power (MW).}}
\nomenclature[P]{{${E}_{[t]}$}}{{BESS energy level (MWh).}}
\nomenclature[P]{{$\eta^{ch}/\eta^{dis}$}}{{BESS charging/discharging efficiency (\%).}}
\nomenclature[P]{{$f^{EOL}$}}{{BESS end-of-life factor (degradation related).}}
\nomenclature[P]{{$f^{PCS}$}}{{BESS power conversion system loss factor.}}
\nomenclature[P]{{$f^{Temp}$}}{{BESS temperature derating factor.}}
\nomenclature[P]{{$f^{Avail}$}}{{BESS availability factor.}}
\nomenclature[P]{{$f^{Res}$}}{{BESS design reserve margin factor.}}
\nomenclature[P]{{$E^{R}/P^{R}$}}{{BESS rated energy and power (MWh/MW).}}
\nomenclature[P]{{${SOC}^{init}$}}{{BESS initial state of charge at $t=0$ (\%).}}
\nomenclature[P]{{${SOC}^{end}$}}{{BESS final state of charge at $t=T$ (\%).}}
\nomenclature[P]{{${\phi}_{[t]}$}}{{Binary variable for avoiding simultaneous BESS charging and discharging.}}
\nomenclature[P]{{$IC_{Energy}$}}{{Annualized and horizon-adjusted BESS energy investment cost coefficient (\$/MWh).}}
\nomenclature[P]{{$IC_{Power}$}}{{Annualized and horizon-adjusted BESS power investment cost coefficient (\$/MW).}}
\nomenclature[P]{{${E}_{[t]}^{Brdg}$}}{{Required BESS reserve energy for bridging (MWh).}}
\nomenclature[P]{{${E}_{[t]}^{Resi}$}}{{Required BESS reserve energy for resilience (MWh).}}
\nomenclature[P]{{${E}_{[t]}^{BS}$}}{{Required BESS reserve energy for black start (MWh).}}
\nomenclature[P]{{${E}_{[t]}^{Res}$}}{{Required BESS reserve energy for dispatch (MWh).}}
\nomenclature[P]{{$\tau$}}{{Minimum duration for BESS reserve calculations (min).}}
\nomenclature[P]{{$C_{en}$}}{{BESS energy-related capital cost (\$/MWh).}}
\nomenclature[P]{{$C_{pow}$}}{{BESS power-related capital cost (\$/MW).}}
\nomenclature[P]{{$r$}}{{Annual discount rate (\%).}}
\nomenclature[P]{{$h$}}{{BESS lifetime (years).}}
\nomenclature[P]{{$N_{year}$}}{{Number of days in a year.}}

\nomenclature[P]{{$P_{[t]}^{FC}$}}{{Fuel cell power generation (MW).}}
\nomenclature[P]{{$FC^{HR}$}}{{Fuel cell heat rate (MMBtu/MWh).}}
\nomenclature[P]{{$C_{[t]}^{NG}$}}{{Natural gas price (\$/MMBtu).}}
\nomenclature[P]{{$P_{[t]}^{FC,Res}$}}{{Fuel cell reserve power (MW).}}
\nomenclature[P]{{$\theta^{X}$}}{{Availability of asset $X$.}}


\printnomenclature

\section{Introduction}

\IEEEPARstart{T}{he} continued expansion of hyperscale data centers is creating unprecedented growth in electricity demand across power systems worldwide. Driven by cloud computing, artificial intelligence, machine learning, and high-performance computing workloads, these facilities have evolved into mission-critical infrastructure that supports digital transformation, economic activity, and emerging AI applications. The recent emergence of generative AI and large language models has accelerated this trend dramatically, requiring large fleets of power-intensive processors operating continuously at high utilization levels. As a result, data center electricity consumption is expected to grow substantially over the coming decade, making data centers one of the most significant new sources of load growth in electric power systems \cite{meier2026reliability}.

This transformation is particularly evident in Texas, where abundant energy resources, extensive natural gas infrastructure, land availability, favorable market conditions, and a competitive electricity market have positioned the Electric Reliability Council of Texas (ERCOT) \cite{ercot2026about}, as a premier destination for hyperscale AI and cloud computing investments. As a result, proposed data center campuses are rapidly increasing in scale, with many projects requiring several hundred megawatts of capacity and some approaching gigawatt-level demand. This unprecedented growth presents significant challenges for ERCOT, including generation adequacy, transmission expansion, interconnection backlogs, operational flexibility, and long-term system reliability. Compounding these challenges, the speed of data center development often outpace the deployment of new generation and transmission infrastructure, raising growing concerns over resource adequacy, transmission congestion, interconnection readiness, and the ability of the grid to reliably support future load growth \cite{mural2026ai}.

Unlike conventional commercial facilities, hyperscale data centers require extremely high levels of reliability, resiliency, and power quality. Even brief interruptions can result in operational disruptions, computational failures, and significant economic consequences. Historically, these requirements have been addressed through utility grid interconnections supplemented by backup generators. However, the combination of increasing electricity demand, interconnection constraints, market volatility, and corporate decarbonization commitments has exposed the limitations of traditional supply architectures. As a result, both operators and utilities are increasingly exploring alternative approaches capable of providing greater operational flexibility, energy security, and environmental performance. In this context, behind-the-meter (BTM) microgrids have emerged as a promising solution for future hyperscale data centers. By integrating solar photovoltaic (PV) systems, wind generation, battery energy storage systems (BESS), fuel cells (FC), and dispatchable thermal resources such as simple-cycle gas turbines (SCGTs) and combined-cycle gas turbines (CCGTs), data center microgrids can provide firm power, renewable integration, resiliency enhancement, and cost optimization within a unified energy framework. Such systems have the potential to reduce dependence on external grid infrastructure while supporting both reliability, and long term decarbonization objectives \cite{ERCOT_LTLF_2025}. Despite these advantages, the planning, operation, and optimization of data center microgrids remain highly challenging. Renewable resources introduce variability and uncertainty, energy storage systems are constrained by state-of-charge and degradation considerations, fuel cells and gas turbines exhibit operational limitations, and electricity market prices fluctuate continuously over time. Simultaneously, AI-driven data center loads are becoming increasingly dynamic because of workload variability, advanced cooling technologies, rapidly evolving computational requirements, and stochastic high power bursts which increase the risk of failures. Therefore, optimal operation requires the coordinated management of diverse resources while balancing techno-economic-environmental, and reliability objectives \cite{Springer2024}.

Although extensive research has been conducted on data center models, microgrid optimization, and renewable integration, important gaps remain in the literature. Existing studies often focus on individual technologies or limited resource portfolios, while comparatively few investigate the coordinated optimization of PV, wind, BESS, fuel cells, SCGTs, and CCGTs within a comprehensive BTM architecture specifically designed for hyperscale AI-driven facilities operating under ERCOT market conditions. Given the scale of projected load growth in Texas and the strategic role of data centers in future power system planning, a systematic evaluation of integrated microgrid solutions is increasingly important. This paper addresses this need through a comprehensive review of existing research and the development of an optimization framework for the resilient planning and operation of BESS for hyperscale data center microgrids in ERCOT region.

Several studies have highlighted the emerging impact of AI-driven data centers on power systems and grid infrastructure. Authors of \cite{mural2026ai} analyzed the unprecedented growth of AI data centers in major U.S. regions (Texas and Virginia) and emphasized the resulting challenges related to grid reliability, regulatory frameworks, and equitable cost allocation. Similarly, industry reports have stressed the need for standardized requirements for large-load interconnection and operation. The white paper in \cite{analysis2026white} reviewed global efforts toward grid readiness for data centers and advocated for unified IEEE standards addressing interconnection, modeling, and performance requirements. Likewise, the report in \cite{zahedi2026best} examined best practices for integrating large electrical loads into the North American grid, highlighting concerns related to power quality, ride-through capabilities, operational visibility, and post-disturbance recovery.

Focusing on data center load models, Li et al in \cite{li2025ai} investigated the interactions between large-scale AI workloads and data center power electronics, identified key power conversion constraints associated with rapid GPU-driven load dynamics, and highlighted advanced power delivery solutions for future hyperscale AI infrastructures. Chen et al. in \cite{chen2025electricity} reviewed the evolving relationship between AI data centers and power systems, characterized AI-driven electricity demand across multiple operational timescales, and identified key challenges and solution pathways for achieving reliable, efficient, and sustainable grid integration. This reference \cite{saxena2023sustainable} proposed a sustainable and secure cloud data center load management framework that reduced energy consumption and carbon emissions while improving resource utilization. A modular data center load model was developed in \cite{rahmani2018complete} that captures component-level power consumption, environmental impacts, and hourly energy profiles for flexible analysis and planning of diverse data center designs. Du et al. in \cite{du2026tokens} proposed a token-based load modeling framework that links LLM inference workloads and model quantization with data center power consumption for grid-responsive energy management. The interaction between computing workloads and energy systems has also attracted increasing attention. Liu et al. \cite{liu2024optimal} addressed renewable energy integration challenges by developing a scheduling framework that exploits the flexibility of delay-tolerant workloads. Guan et al. \cite{guan2026unlocking} proposed a two-stage optimization strategy that jointly schedules workloads and uninterruptible power supply (UPS) systems for participation in energy and reserve markets under uncertainty. Furthermore, Liu et al. \cite{liu2022optimal} developed an integrated planning and operational framework based on a physics-informed data-driven model to reduce energy costs while improving data center reliability. These studies demonstrate the growing importance of workload-aware energy management approaches for future sustainable computing infrastructures.

Given the increasing grid impacts of large-scale computing facilities, significant research efforts have focused on energy storage systems as a means of improving operational flexibility and reliability. Hybrid energy storage systems have been proposed to mitigate rapid power fluctuations associated with data center operations. Ko et al. \cite{ko2025mitigation} developed a framework combining batteries and supercapacitors to reduce demand fluctuations, while Wang et al. \cite{wang2026mitigating} introduced a hybrid energy storage planning methodology for renewable-powered computing centers that captures multi-timescale fluctuations associated with large language model (LLM) training workloads. BESS have also been investigated as a key solution for managing the highly dynamic behavior of AI data center loads \cite{azizi2026strengthening}. Furthermore, Ju et al. \cite{ju2026integrating} demonstrated how BESS can help hyperscale facilities overcome grid constraints, accelerate power access, and generate additional revenue streams through participation in electricity markets. Lu et al. \cite{lu2026battery} proposed a battery-assisted operational framework for hyperscale AI data centers in which on-site BESS buffers fast internal load dynamics against time-varying grid constraints. Their results demonstrated that BESS can simultaneously enhance day-ahead workload commitment and real-time delivery reliability while transitioning from reliability-oriented assets to economic flexibility resources as grid constraints relax. In addition, Chen et al. \cite{chen2026multiphysics} developed an aging-aware optimization framework that incorporates detailed battery degradation dynamics into data center microgrid scheduling, showing that dynamic aging feedback can significantly reduce battery degradation costs and extend asset lifetime.

Parallel research efforts have focused on the planning and operation of data center microgrids. Jones et al. \cite{jones2026megawatts} proposed a techno-economic framework for BTM AI data center microgrids and demonstrated that hybrid configurations combining renewable resources and battery storage can reduce costs while improving reliability. Guo et al. \cite{guo2026integrated} developed an integrated planning framework that co-optimizes wind generation, solar PV systems, battery storage, and grid interactions to achieve highly reliable clean energy delivery. For islanded data center operations, Lian et al. \cite{lian2023robust} proposed a robust multi-objective optimization model minimizing operational costs, renewable curtailment, and computational resource over-provisioning under renewable and workload uncertainties. The broader microgrid literature provides additional methodologies applicable to data center energy systems. Ding et al. \cite{ding2015stochastic} introduced stochastic scheduling approaches that simultaneously consider economic performance, reliability, and greenhouse gas emissions under renewable uncertainty. Khodaei \cite{khodaei2014resiliency} proposed a resilience-oriented microgrid scheduling framework designed to minimize load curtailment during grid disruptions through robust optimization techniques. Song et al. \cite{song2026co} further extended microgrid coordination by proposing a co-optimized scheduling framework enabling shared energy storage leasing among multiple microgrids while preserving participant privacy using distributed optimization techniques. Beyond steady-state operational planning, researchers have increasingly investigated dynamic stability and grid support services provided by data centers. Ross et al. \cite{ross2026using} developed a high-fidelity electromagnetic transient model of a hyperscale AI data center, including cooling systems, UPS infrastructure, co-generation units, and grid-forming energy storage systems. Their results demonstrated that properly controlled grid-forming storage can significantly enhance power system stability by mitigating frequency excursions and improving fault recovery performance during rapid AI-driven load fluctuations.

In addition to energy management and power system integration studies, several review papers have summarized the state of the art in data center optimization and reliability. Ahmed et al. \cite{ahmed2021review} surveyed models for energy consumption and reliability assessment in data centers, identifying research gaps related to internal power conditioning and cooling system reliability while emphasizing the importance of practical and accessible modeling approaches. Similarly, Srikandabala et al. \cite{srikandabala2025optimization} reviewed recent developments in data center optimization, highlighting the growing role of artificial intelligence and machine learning while identifying the need for integrated, scalable, explainable, and sustainable end-to-end frameworks. Zhang et al. \cite{zhang2026bi} proposed a bi-level scheduling framework for data center clusters with shared energy storage that combines game-theoretic scheduling and revenue allocation mechanisms. Their results demonstrated improved energy efficiency, reduced electricity costs, and enhanced incentives for cooperative participation.


Although substantial progress has been made in data center energy management, existing studies generally focus on individual aspects such as battery operation, workload scheduling, renewable integration, microgrid planning, or grid interaction. A comprehensive framework that simultaneously considers behind-the-meter generation resources, battery storage systems, natural gas generation technologies, workload dynamics, reliability requirements, economic optimization, and grid-connected operation remains limited. This gap motivates the development of integrated co-optimization frameworks capable of supporting the reliable, resilient, and economically efficient operation of future AI-driven data center microgrids.

To this end, the main contributions of this paper are summarized as follows:

\begin{itemize}
    \item A realistic hyperscale data center load modeling framework is developed and implemented, featuring a detailed decomposition of both IT and non-IT loads with second and minute based resolutions. IT load consists of critical computing, interactive applications, AI inference, AI training, batch processing, storage, network infrastructure, and GPU transient workloads. The non-IT load includes cooling systems, auxiliary equipment, and electrical losses associated with UPS systems, transformers, power distribution units (PDUs), substations, and supporting infrastructure. In addition, stochastic burst events and sharp load spikes are incorporated to realistically emulate the highly dynamic operating behavior of modern hyperscale data centers that could create SSTI issues for gas turbines by contributing as negative damping.
    
    \item A comprehensive multi-phase optimization framework is proposed for the optimal planning and resilient design of BESS as part of BTM data center microgrids operating in both load side (for load smoothing application and potential SSTI mitigation) and generation side (for spinning reserve and resilience purposes). The framework integrates BESS, solar, wind generation, natural gas powered FC, and gas turbines (combined-cycle and simple-cycle), while explicitly incorporating industry-level actual BESS operational and technical constraints encountered in real-world projects. 

    \item A comprehensive techno-economic assessment is conducted for the ERCOT market, incorporating solar and wind resource data, locational marginal prices (LMPs), and other key technical and economic inputs. Monte Carlo–based scenario generation is utilized using statistical data, while sample average approximation (SAA) method is then applied to reduce the dataset to 12 representative days that capture monthly and seasonal variations, thereby improving computational efficiency without compromising solution fidelity.
    
\end{itemize}

The rest of the paper is organized as follows: Section II will discuss the proposed BTM hyperscale data center microgrid optimization framework. Section III focuses on high fidelity data center load model, and then Section IV will focus on data center load smoothing optimization and potentail SSTI mitigation for generation side. Section V will present the problem formulation for resilient design and optimal operational planning of BESS within BTM data center framework. Section VI will present simulation results and numerical analysis while Section VII concludes the paper and presents future directions.

\section{Proposed Multi-Phase BTM Hyperscale Data Center Microgrid Framework}

\begin{figure}
\centering
\footnotesize
\captionsetup{justification=raggedright, singlelinecheck=false, font={footnotesize}}
\includegraphics[width=3.5in]{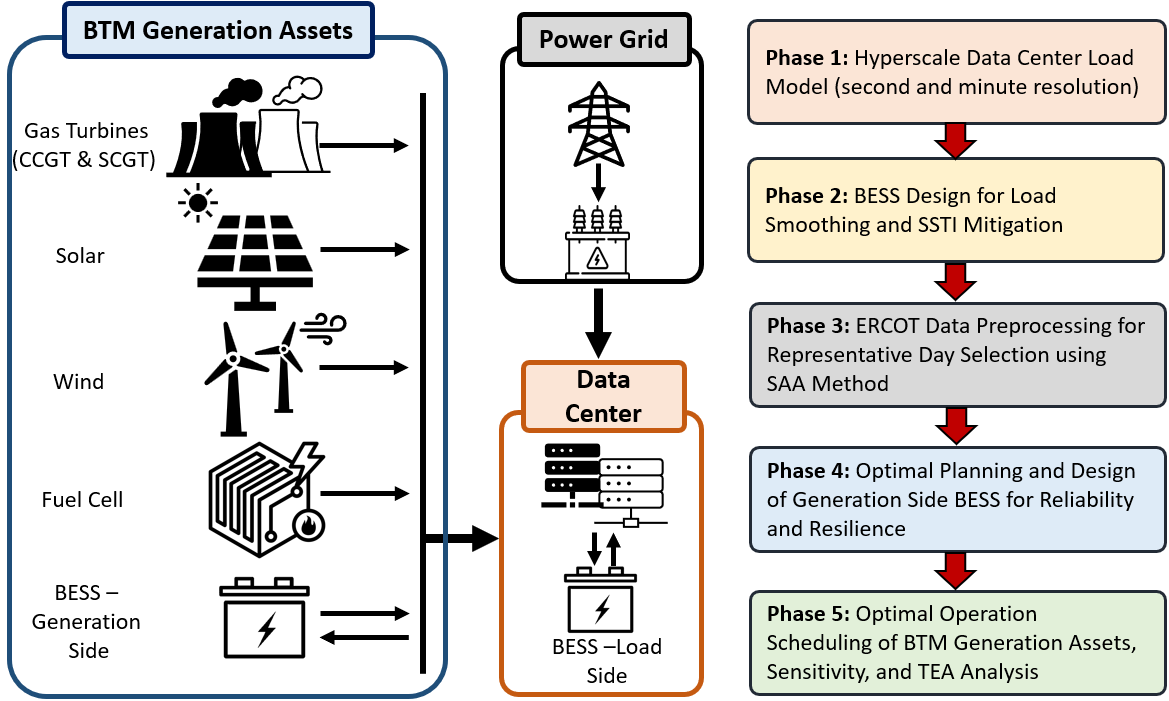}
\caption{Behind-the-meter Data Center framework with potential grid connection, solar, wind, natural gas power plants (CCGT and SCGT), modular BESS and Fuel Cells.}
\label{MG_schematic}
\end{figure}

The proposed multi-stage framework presented in Figure (\ref{MG_schematic}) consists of five phases: Phase 1 focuses on the development of a high-fidelity synthetic hyperscale data center load model at both second-level and minute-level resolutions. The facility demand is decomposed into IT and non-IT components to accurately represent the contributions of different subsystems. The IT load consists of critical computing, interactive applications, AI inference, AI training, batch processing, storage, network infrastructure, and GPU transient workloads. The non-IT load includes cooling systems, auxiliary equipment, and electrical losses associated with UPS systems, transformers, power distribution units (PDUs), substations, and supporting infrastructure. In addition, stochastic burst events and rapid load spikes are incorporated to realistically emulate the highly dynamic operating behavior of modern hyperscale data centers.

Phase 2 introduces a BESS directly coupled to the data center load to mitigate high-frequency demand fluctuations and smooth the net load observed by generation resources. First, second-level load variability and ramp-rate analyses are performed to identify sharp demand spikes, and the load-side BESS is optimally sized to reliably absorb these transient shocks throughout its operational lifetime. Subsequently, an optimization-based load-smoothing strategy is implemented in which rapid ramp-rate variations are penalized to reduce generator cycling and improve operational stability. By acting as a dynamic buffer between highly variable AI-driven loads and generation assets, the BESS prevents sharp load excursions from being transferred to gas turbines, thereby reducing the risk of SSTI, negative damping effects, shaft fatigue, equipment damage, and potential system trip.

Phase 3 focuses on data collection and scenario generation. Historical ERCOT West Hub data, including solar generation, wind generation, and locational marginal prices (LMPs), are first cleaned, validated, and preprocessed. Key statistical characteristics, including mean, variance, covariance, and temporal correlations, are then extracted and used to generate 10,000 stochastic scenarios through Monte Carlo simulation. To improve computational tractability while preserving the underlying statistical properties and seasonal variability of the dataset, the SAA method is subsequently employed to identify twelve representative operating days, one for each month with hourly resolution, resulting in a final optimization horizon of 288 hours. This approach effectively captures uncertainty and long-term variability while significantly reducing computational burden.

Phase 4 focuses on the optimal planning and resilient design of the generation side BESS for the BTM generation portfolio, which consists of several SCGTs, CCGT, solar PV, wind generation, natural gas FC. A planning optimization and energy reserve dispatch operation optimization based on security-constrained unit commitment (SCUC) formulation implemented is used to determine the optimal sizing and configuration of BESS unit while simultaneously satisfying techno-economic, reliability, resilience, and reserve requirements.

Finally, Phase 5 focuses on the optimal operational scheduling of the designed microgrid. The framework coordinates generation dispatch, renewable utilization, storage operation, reserve management, carbon free energy (CFE) calculation, N-1 reliability analysis, and grid interactions. Collectively, the proposed framework provides an integrated and computationally efficient platform that combines high-resolution load modeling, BESS-enabled load smoothing, uncertainty-aware scenario generation, resilient resource planning, and optimal operational scheduling for next-generation hyperscale data center microgrids.

\section{Hyperscale AI Data Center Load Modeling Framework}
\label{DC_load_model_section}

This section presents the overall methodology used to construct the high fidelity hyperscale AI data center load profile employed throughout this work. Rather than directly generating a single aggregate facility demand profile, the proposed framework builds the total campus load from the bottom up by explicitly representing the underlying computational workloads, electrical infrastructure, thermal-management systems, and auxiliary facility services that collectively determine power consumption. This hierarchical modeling strategy preserves the physical relationships between AI workloads and supporting infrastructure while producing a realistic time-varying electrical demand profile suitable for power-system planning, reliability assessment, microgrid design, energy-storage sizing, and operational studies.

\subsection{Campus Architecture}

The modeled campus consists of four representative buildings supplied through two utility substations. Each building represents a distinct mix of hyperscale computing services. The first building primarily hosts large-scale AI training clusters, the second building hosts AI inference clusters responsible for serving models to users, the third building represents cloud-computing services, and the fourth building primarily contains storage and networking infrastructure. Buildings dedicated to AI accelerators employ liquid-cooling technologies, whereas buildings with lower power densities employ conventional air-cooling systems. From an electrical perspective, the campus is represented as a collection of interconnected computational and infrastructure assets whose individual demands are aggregated to obtain the total facility demand. This representation enables the model to capture both local building-level behavior and campus-wide demand characteristics.

\subsection{Facility Load Hierarchy}

The facility load profile is developed through a hierarchical decomposition process. At the highest level, the total facility demand is separated into IT demand and non-IT demand which is presented in equation (\ref{facility_total}). 

\begin{equation}
P^{Facility}_{[t]}=P^{IT}_{[t]}+P^{NonIT}_{[t]} .
\label{facility_total}
\end{equation}
\begin{equation}
\begin{aligned}
P^{IT}_{[t]}=&P^{Crit}_{[t]}+P^{Interactive}_{[t]}+P^{Inference}_{[t]} +P^{Training}_{[t]}\\
&+P^{Batch}_{[t]}+P^{Storage}_{[t]} + P^{Network}_{[t]} + P^{GPUTran}_{[t]} 
\end{aligned}
\label{IT_load}
\end{equation}
\begin{equation}
P^{NonIT}_{[t]}=P^{Loss}_{[t]}+P^{Cooling}_{[t]}+P^{Aux}_{[t]} .
\label{Non-IT_Load}
\end{equation}

The IT load shown in equation (\ref{IT_load}) is represented as the aggregation of several computational workload categories commonly observed in hyperscale AI campuses. These categories include critical cloud services ($P^{Crit}_{[t]}$), interactive services ($P^{Interactive}_{[t]}$), AI inference workloads ($P^{Inference}_{[t]}$), AI training workloads ($P^{Training}_{[t]}$), batch-processing activities ($P^{Batch}_{[t]}$), storage infrastructure ($P^{Storage}_{[t]}$), networking equipment ($P^{Network}_{[t]}$), and high-frequency GPU transient behavior ($P^{GPUTran}_{[t]}$). Critical cloud services represent always-on computational resources required to maintain continuous service availability. Interactive services represent latency-sensitive user-facing applications whose utilization follows human activity patterns. AI inference workloads represent model-serving activities that process prompts and generate output tokens. AI training workloads represent large-scale optimization processes executed on GPU clusters. Batch-processing workloads represent flexible computational tasks that can often be shifted in time. Storage and networking systems provide supporting data-management and communication services whose operation is strongly coupled to both training and inference activities. The aggregate IT demand is obtained by combining these workload categories according to their installed capacities, utilization levels, and operational states. The resulting IT profile therefore reflects both the diversity of computational services and the interactions among them.

The non-IT load is further decomposed into electrical losses ($P^{Loss}_{[t]}$), cooling-system demand ($P^{Cooling}_{[t]}$), and auxiliary facility services ($P^{Aux}_{[t]}$) which is shown in equation (\ref{Non-IT_Load}). Electrical losses capture power dissipated in UPS systems, transformers, power-distribution equipment, and substations. Cooling demand represents the electricity consumed by chillers, pumps, fans, and cooling-support equipment required to remove heat generated by computing hardware. Auxiliary demand represents building-support systems such as controls, monitoring systems, communications equipment, security systems, lighting, and balance-of-plant loads. The following subsections describe the proposed data center load modeling framework and present the associated mathematical formulations and equations used to characterize its behavior. Please note that, unlike the optimization indices, parameters, and variables, the data center modeling parameters are intentionally defined within the body of the paper. Given the large number of model-specific parameters, this approach improves readability and allows each parameter to be introduced in the context of its corresponding equation, making the formulations easier to follow and interpret.
\subsection{Bounded Smoothed Stochastic Processes}
Slow stochastic variability is represented using a moving-average filtered Gaussian process. For a generic independent sequence $\epsilon_k\sim\mathcal{N}(0,\sigma^2)$ and smoothing window $L$, the filtered disturbance is
\begin{equation}
\widetilde{\epsilon}_k=\frac{1}{L}\sum_{j=0}^{L-1}\epsilon_{k-j}.
\label{eq:smooth_noise}
\end{equation}
Boundary handling is performed by finite convolution over the available samples. A general bounded process is written as
\begin{equation}
x_k=\operatorname{clip}\!\left(\widehat{x}_k,x^{\min},x^{\max}\right),
\label{eq:clip}
\end{equation}
where $\widehat{x}_k$ is the unbounded process and
\begin{equation}
\operatorname{clip}(x,a,b)=\min\{\max\{x,a\},b\}.
\end{equation}
Clipping prevents synthetic operating indices from producing physically inadmissible utilization, efficiency, power-factor, temperature, or power values.

\subsection{Annual, Monthly, Weekly, and Daily Modulation}

\subsubsection{Seasonal index}

A smooth summer-intensity index is defined from the day of year $n_d$:
\begin{equation}
s_d=\operatorname{clip}\!\left[\frac{1}{2}+\frac{1}{2}\sin\!\left(\frac{2\pi(n_d-n_{\mathrm{pk}})}{N_y}\right),0,1\right],
\label{eq:summer_index}
\end{equation}
where $n_{\mathrm{pk}}$ controls the seasonal phase and $N_y$ is the number of days in the annual cycle. The daily value $s_d$ is repeated over all minutes of day $d$ to obtain $s_m$.

\subsubsection{Monthly demand and temperature factors}

Let $\mu(m)$ return the calendar month associated with minute $m$. The monthly AI-demand multiplier and monthly ambient-temperature bias are lookup processes:
\begin{align}
f_{\mathrm{AI},m}^{\mathrm{mon}} &= F_{\mathrm{AI}}[\mu(m)], \\
\Delta T_m^{\mathrm{mon}} &= F_T[\mu(m)].
\end{align}
This formulation allows calibrated monthly factors to be replaced without changing the governing model.

\subsubsection{Weekend and annual growth factors}
To account for temporal variations in data center operation, a weekend adjustment factor and a gradual growth factor are incorporated into the synthetic load model. The weekend modifier captures differences in workload behavior between weekdays and weekends, while the growth factor represents progressive changes in facility utilization and demand over the simulation horizon, enabling the model to reflect evolving operational conditions more realistically. The weekend modifier is
\begin{equation}
f_{d}^{\mathrm{wk}}=
\begin{cases}
f_{\mathrm{weekend}}, & d\text{ is Saturday or Sunday},\\
1, & \text{otherwise}.
\end{cases}
\end{equation}
A linear within-horizon growth factor is used to represent gradual capacity utilization or demand growth:
\begin{equation}
f_d^{\mathrm{gr}}=f_{\mathrm{gr},0}+
\left(f_{\mathrm{gr},1}-f_{\mathrm{gr},0}\right)
\frac{d-1}{\max(1,N_d-1)}.
\label{eq:growth}
\end{equation}

\subsubsection{Daily stochastic factors}

To capture day-to-day variability in hyperscale data center operation, stochastic scaling factors are introduced for overall workload intensity, AI training activity, AI inference activity, and ambient temperature conditions. These factors are sampled from normal distributions characterized by user-defined means and variances, thereby enabling realistic representation of daily fluctuations while preserving the statistical characteristics of the modeled workloads. A fixed random seed is employed to ensure reproducibility of the dataset. 

\begin{align}
f_d^{\mathrm{work}} &\sim \mathcal{N}(\mu_{\mathrm{work}},\sigma_{\mathrm{work}}^2), \\
f_d^{\mathrm{tr}} &\sim \mathcal{N}(\mu_{\mathrm{tr}},\sigma_{\mathrm{tr}}^2), \\
f_d^{\mathrm{inf}} &\sim \mathcal{N}(\mu_{\mathrm{inf}},\sigma_{\mathrm{inf}}^2), \\
\Delta T_d^{\mathrm{day}} &\sim \mathcal{N}(\mu_T,\sigma_T^2).
\end{align}

Each day is assigned a category $z_d\in\mathcal{Z}$, where the modeled categories are normal operation, weekend operation, product launch, maintenance, and training sprint. A categorical multiplier is
\begin{equation}
f_d^{\mathrm{type}}=F_z[z_d].
\end{equation}
Special day types are sampled without replacement and may overwrite the default normal or weekend classification. The category influences both the low-frequency workload and the probability, duration, and amplitude of fast transient events.

\subsection{Ambient-Temperature Model}

The minute-level ambient temperature combines a reference temperature, monthly bias, annual seasonality, diurnal variation, daily weather offset, and smoothed weather noise:
\begin{align}
\widehat{T}_{a,m}={}&T_{\mathrm{ref}}+\Delta T_m^{\mathrm{mon}}+A_{\mathrm{sea}}s_m \\
&+A_{\mathrm{day}}\sin\!\left(\frac{2\pi(h_m-\phi_T)}{H_d}\right)
+\Delta T_{d_m}^{\mathrm{day}}+\widetilde{\epsilon}_{T,m},
\end{align}
\begin{equation}
T_{a,m}=\operatorname{clip}\!\left(\widehat{T}_{a,m},T_a^{\min},T_a^{\max}\right).
\label{eq:ambient_temp}
\end{equation}
This temperature series affects cooling coefficient of performance, economizer availability, transformer losses, auxiliary demand, and seasonal thermal loading.

\subsection{High Resolution IT Workload Indices}

Let the set of IT components be
\begin{equation}
\begin{aligned}
    \mathcal{C}=\{\mathrm{critical},\mathrm{interactive},\mathrm{training},\mathrm{inference},\mathrm{batch},\\ 
\mathrm{storage},\mathrm{network}\}
\end{aligned}
\end{equation}
Each component is first represented by a dimensionless utilization or workload index. Numerical coefficients are intentionally separated from the equations and are reported in the parameter tables.

\subsubsection{Critical computing}

The critical-service index represents persistent computing and control services with small diurnal and seasonal variation:
\begin{align}
\widehat{W}_{m}^{\mathrm{crit}}={}&b_{\mathrm{crit}}+a_{\mathrm{crit},1}
\sin\!\left(\frac{2\pi(h_m-\phi_{\mathrm{crit}})}{H_d}\right)\\ & \nonumber
+a_{\mathrm{crit},2}s_m+\widetilde{\epsilon}_{\mathrm{crit},m},\\
W_m^{\mathrm{crit}}={}&\operatorname{clip}\!\left(
\widehat{W}_{m}^{\mathrm{crit}}f_{\mathrm{AI},m}^{\mathrm{mon}}
f_{d_m}^{\mathrm{work}}f_{d_m}^{\mathrm{gr}}f_{d_m}^{\mathrm{type}},
W_{\mathrm{crit}}^{\min},W_{\mathrm{crit}}^{\max}\right).
\end{align}

\subsubsection{Interactive services}

Interactive demand contains a daily cycle and a shorter intraday harmonic:
\begin{align}
\widehat{W}_{m}^{\mathrm{int}}={}&b_{\mathrm{int}}+a_{\mathrm{int},1}
\sin\!\left(\frac{2\pi(h_m-\phi_{\mathrm{int}})}{H_d}\right) \\ &
+a_{\mathrm{int},2}\sin\!\left(\frac{2\pi h_m}{H_{\mathrm{int}}}\right) 
+a_{\mathrm{int},3}s_m+\widetilde{\epsilon}_{\mathrm{int},m}, \nonumber
\end{align}
\begin{equation}
W_m^{\mathrm{int}}=\operatorname{clip}\!\left(
\widehat{W}_{m}^{\mathrm{int}}f_{\mathrm{AI},m}^{\mathrm{mon}}
f_{d_m}^{\mathrm{wk}}f_{d_m}^{\mathrm{work}}f_{d_m}^{\mathrm{gr}}
f_{z,m}^{\mathrm{int}},W_{\mathrm{int}}^{\min},W_{\mathrm{int}}^{\max}\right),
\end{equation}
where $f_{z,m}^{\mathrm{int}}$ provides additional event-day scaling for product launches and maintenance periods.

\subsubsection{Prompt and decode token activity}

Prompt-processing activity is modeled by
\begin{align}
\widehat{I}_{m}^{\mathrm{prompt}}={}&b_p+a_{p,1}
\sin\!\left(\frac{2\pi(h_m-\phi_p)}{H_d}\right) \\ &
+a_{p,2}\sin\!\left(\frac{2\pi h_m}{H_p}\right) \nonumber
+a_{p,3}s_m+\widetilde{\epsilon}_{p,m},\\
I_{m}^{\mathrm{prompt,base}}={}&\widehat{I}_{m}^{\mathrm{prompt}}
f_{\mathrm{AI},m}^{\mathrm{mon}}f_{d_m}^{\mathrm{inf}}f_{d_m}^{\mathrm{gr}}.
\end{align}
Short prompt bursts are superimposed using Gaussian pulses. For burst $r$ with center $c_r$, width $\omega_r$, and amplitude $A_r$,
\begin{equation}
g_{r,m}=A_r\exp\!\left[-\frac{1}{2}\left(\frac{m-c_r}{\omega_r}\right)^2\right].
\label{eq:gaussian_pulse}
\end{equation}
The prompt index is then
\begin{equation}
I_m^{\mathrm{prompt}}=\operatorname{clip}\!\left(
I_m^{\mathrm{prompt,base}}+\sum_{r\in\mathcal{R}_{d_m}}g_{r,m},
I_p^{\min},I_p^{\max}\right)
\end{equation}
The number, amplitude, and timing of bursts depend on the day type. Burst amplitudes are additionally scaled by monthly AI demand and seasonal intensity.

Decode activity is correlated with prompt activity while retaining an independent short-cycle oscillation and stochastic term:
\begin{align}
I_m^{\mathrm{decode}}={}&\operatorname{clip}\!\left[
\rho_{dp}I_m^{\mathrm{prompt}}
+a_d\sin\!\left(\frac{2\pi m}{H_d^{\mathrm{dec}}}\right)\right. \notag\\
&\left.+\widetilde{\epsilon}_{d,m},
I_d^{\min},I_d^{\max}\right].
\label{eq:decode_index}
\end{align}

The aggregate inference index is a token-work-weighted average:
\begin{equation}
W_m^{\mathrm{inf}}=
\operatorname{clip}\!\left(
\frac{w_pI_m^{\mathrm{prompt}}+w_dI_m^{\mathrm{decode}}}{w_p+w_d},
W_{\mathrm{inf}}^{\min},W_{\mathrm{inf}}^{\max}\right)
\label{eq:inference_index}
\end{equation}

\subsubsection{Accelerator training and inference requests}

The baseline training request is modelled as follows: 
\begin{align}
\widehat{G}_m^{\mathrm{tr}}={}&b_{\mathrm{tr}}+a_{\mathrm{tr},1}
\sin\!\left(\frac{2\pi(h_m-\phi_{\mathrm{tr}})}{H_d}\right)
+a_{\mathrm{tr},2}s_m \\ \nonumber
&+a_{\mathrm{tr},3}\sin\!\left(\frac{2\pi n_m}{D_{\mathrm{tr}}}\right),\\
G_m^{\mathrm{tr,base}}={}&\widehat{G}_m^{\mathrm{tr}}
f_{\mathrm{AI},m}^{\mathrm{mon}}f_{d_m}^{\mathrm{tr}}f_{d_m}^{\mathrm{gr}},
\end{align}
where $n_m$ is the minute-level day-of-year index. The inference request is

\begin{equation}
\begin{aligned}
G_m^{\mathrm{inf,base}}={}&\left[b_{\mathrm{ginf}}+a_{\mathrm{ginf},1}W_m^{\mathrm{inf}}\right.\\
&\left.+a_{\mathrm{ginf},2}\sin\!\left(\frac{2\pi(h_m-\phi_{\mathrm{ginf}})}{H_d}\right)\right]
 f_{d_m}^{\mathrm{inf}}f_{\mathrm{AI},m}^{\mathrm{mon}}.
\end{aligned}
\label{eq:inference_gpu_base}
\end{equation}

Longer workload events are introduced using a trapezoidal function. For event $r$ with start $s_r$, end $e_r$, ramp duration $L_r$, and signed amplitude $A_r$, define
\begin{equation}
\psi_r(m)=
\begin{cases}
0, & m<s_r,\\
A_r\dfrac{m-s_r}{L_r}, & s_r\le m<s_r+L_r,\\
A_r, & s_r+L_r\le m<e_r-L_r,\\
A_r\dfrac{e_r-m}{L_r}, & e_r-L_r\le m<e_r,\\
0, & m\ge e_r.
\end{cases}
\label{eq:trapezoid}
\end{equation}
Training sprints add positive training events; product launches increase inference request while reducing schedulable training; maintenance reduces training request; and multi-day campaigns add persistent positive training demand. Thus,
\begin{align}
G_m^{\mathrm{tr}}={}&\operatorname{clip}\!\left(G_m^{\mathrm{tr,base}}+\sum_{r\in\mathcal{R}_{\mathrm{tr}}}\psi_r(m)+\widetilde{\epsilon}_{\mathrm{tr},m},G_{\mathrm{tr}}^{\min},\right. \notag\\
&\left.G_{\mathrm{tr}}^{\max}\right),
\label{eq:training_gpu}
\end{align}
\begin{align}
G_m^{\mathrm{inf}}={}&\operatorname{clip}\!\left(G_m^{\mathrm{inf,base}}+\sum_{r\in\mathcal{R}_{\mathrm{inf}}}\psi_r(m)+\widetilde{\epsilon}_{\mathrm{inf},m},G_{\mathrm{inf}}^{\min},\right. \notag\\
&\left.G_{\mathrm{inf}}^{\max}\right).
\label{eq:inference_gpu}
\end{align}

\subsubsection{Batch, storage, and network services}

Batch processing is preferentially shifted to off-peak nighttime periods and is partially displaced by training demand:
\begin{align}
\widehat{W}_m^{\mathrm{batch}}={}&b_{\mathrm{batch}}
+a_{\mathrm{early}}\mathbb{I}\!\left(h_m\in\mathcal{H}_{\mathrm{early}}\right)
+\notag\\ a_{\mathrm{late}}\mathbb{I}\!\left(h_m\in\mathcal{H}_{\mathrm{late}}\right) 
&-a_{\mathrm{disp}}G_m^{\mathrm{tr}}
+a_{\mathrm{wk}}\mathbb{I}\!\left(d_m\text{ is weekend}\right)
\notag\\
&+\widetilde{\epsilon}_{\mathrm{batch},m},
\label{eq:batch_unbounded}\\
W_m^{\mathrm{batch}}={}&\operatorname{clip}\!\left[
\widehat{W}_m^{\mathrm{batch}}
\left(b_{\mathrm{sea}}+a_{\mathrm{sea}}s_m\right),\right.
\notag\\
&\left.W_{\mathrm{batch}}^{\min},
W_{\mathrm{batch}}^{\max}\right].
\label{eq:batch_index}
\end{align}

where $\mathbb{I}(\cdot)$ is the indicator function.

Storage and network activity are represented as
\begin{align}
\widehat{W}_m^{\mathrm{stor}}={}&b_{\mathrm{stor}}+a_{\mathrm{stor},1}
\sin\!\left(\frac{2\pi(h_m-\phi_{\mathrm{stor}})}{H_d}\right)
+a_{\mathrm{stor},2}G_m^{\mathrm{tr}}\notag\\
&+a_{\mathrm{stor},3}s_m
+\widetilde{\epsilon}_{\mathrm{stor},m},\\
W_m^{\mathrm{stor}}={}&\operatorname{clip}\!\left(
\widehat{W}_m^{\mathrm{stor}}f_{\mathrm{AI},m}^{\mathrm{mon}},
W_{\mathrm{stor}}^{\min},W_{\mathrm{stor}}^{\max}\right),\\
\widehat{W}_m^{\mathrm{net}}={}&b_{\mathrm{net}}+a_{\mathrm{net},1}
\sin\!\left(\frac{2\pi(h_m-\phi_{\mathrm{net}})}{H_d}\right)
+a_{\mathrm{net},2}W_m^{\mathrm{inf}} \notag\\
&+a_{\mathrm{net},3}G_m^{\mathrm{tr}}
+a_{\mathrm{net},4}s_m+\widetilde{\epsilon}_{\mathrm{net},m},\\
W_m^{\mathrm{net}}={}&\operatorname{clip}\!\left(
\widehat{W}_m^{\mathrm{net}}f_{\mathrm{AI},m}^{\mathrm{mon}},
W_{\mathrm{net}}^{\min},W_{\mathrm{net}}^{\max}\right).
\end{align}

\subsection{IT Background Envelope and Component Allocation}

The minute-level campus IT envelope is expressed as a fraction of full IT capacity:

\begin{align}
\widehat{u}_{\mathrm{IT},m}={}&b_{\mathrm{IT}}+a_{\mathrm{IT},1}
\sin\!\left(\frac{2\pi(h_m-\phi_{\mathrm{IT}})}{H_d}\right)
\notag\\
&+a_{\mathrm{IT},2}\sin\!\left(\frac{2\pi h_m}{H_{\mathrm{IT}}}\right)
+a_{\mathrm{IT},3}(f_{\mathrm{AI},m}^{\mathrm{mon}}-1)
\notag\\
&+a_{\mathrm{IT},4}s_m+a_{\mathrm{IT},5}(f_{d_m}^{\mathrm{work}}-1)
+\widetilde{\epsilon}_{\mathrm{IT},m},\\
u_{\mathrm{IT},m}={}&\operatorname{clip}\!\left(
\widehat{u}_{\mathrm{IT},m},u_{\mathrm{IT}}^{\min},u_{\mathrm{IT}}^{\max}\right),\\
P_{\mathrm{IT},m}^{\mathrm{env}}={}&P_{\mathrm{IT}}^{\max}u_{\mathrm{IT},m}.
\label{eq:it_envelope}
\end{align}

Let $\alpha_c^{\mathrm{N}}$ denote the normalized background share of component $c$, satisfying
\begin{equation}
\alpha_c^{\mathrm{N}}\ge0,\qquad \sum_{c\in\mathcal{C}}\alpha_c^{\mathrm{N}}=1.
\end{equation}
Let $X_{c,m}$ denote the corresponding component index:
\begin{equation}
X_{c,m}=\begin{cases}
W_m^{\mathrm{crit}}, & c=\mathrm{critical},\\
W_m^{\mathrm{int}}, & c=\mathrm{interactive},\\
G_m^{\mathrm{tr}}, & c=\mathrm{training},\\
W_m^{\mathrm{inf}}, & c=\mathrm{inference},\\
W_m^{\mathrm{batch}}, & c=\mathrm{batch},\\
W_m^{\mathrm{stor}}, & c=\mathrm{storage},\\
W_m^{\mathrm{net}}, & c=\mathrm{network}.
\end{cases}
\end{equation}
The preliminary component power is
\begin{equation}
\widetilde{P}_{c,m}=P_{\mathrm{IT},m}^{\mathrm{env}}\alpha_c^{\mathrm{N}}
\frac{X_{c,m}}{\overline{X}_c},
\label{eq:preliminary_component}
\end{equation}
where $\overline{X}_c$ is the full-horizon mean of $X_{c,m}$. Mean normalization preserves the intended long-run component shares before the instantaneous balance correction.

Because independent normalized indices do not necessarily sum to the prescribed IT envelope at every minute, a balancing factor is applied:
\begin{align}
\kappa_m&=\frac{P_{\mathrm{IT},m}^{\mathrm{env}}}
{\max\!\left(\sum_{c\in\mathcal{C}}\widetilde{P}_{c,m},\varepsilon\right)},\\
P_{c,m}&=\kappa_m\widetilde{P}_{c,m}.
\label{eq:component_correction}
\end{align}
Consequently,
\begin{equation}
\sum_{c\in\mathcal{C}}P_{c,m}=P_{\mathrm{IT},m}^{\mathrm{env}},
\end{equation}
subject only to numerical tolerance $\varepsilon$.

\subsection{Expansion to One-Second Resolution}

The corrected minute-level component powers, total IT power, and ambient temperature are mapped to the one-second grid using zero-order hold:
\begin{align}
P_{c,t}^{(0)} &= P_{c,\lfloor t/S_m\rfloor},\\
P_{\mathrm{IT},t}^{(0)} &= P_{\mathrm{IT},\lfloor t/S_m\rfloor}^{\mathrm{env}},\\
T_{a,t} &= T_{a,\lfloor t/S_m\rfloor}.
\end{align}
The superscript $(0)$ denotes the background profile before second-scale transient events are imposed. Zero-order hold intentionally avoids introducing artificial sub-minute ramps; all rapid ramps are generated explicitly by the transient-event model.

\subsection{Second-Scale Transient Event Model}

Three transient families are included. Their occurrence probabilities depend on day type, and their start times are randomized around nominal daily centers. Let $p_{j,z}$ be the probability of event type $j\in\{1,2,3\}$ on a day of category $z$. The occurrence indicator is
\begin{equation}
\delta_{j,d}\sim\operatorname{Bernoulli}(p_{j,z_d}).
\end{equation}
If an event is selected, its actual start time is
\begin{equation}
t_{j,d}^{\mathrm{start}}=t_{d}^{0}+t_j^{\mathrm{base}}+\xi_{j,d},
\end{equation}
where $t_d^0$ is the first second of day $d$ and $\xi_{j,d}$ is a bounded random timing offset.

During a full-campus event, a prescribed IT loading fraction $f_{j,t}$ is imposed:
\begin{equation}
P_{\mathrm{IT},t}=P_{\mathrm{IT}}^{\max}f_{j,t}.
\label{eq:event_total_it}
\end{equation}
The total event power is decomposed using event-specific shares $\alpha_c^{\mathrm{E}}$:
\begin{equation}
P_{c,t}=\alpha_c^{\mathrm{E}}P_{\mathrm{IT},t},\qquad
\sum_{c\in\mathcal{C}}\alpha_c^{\mathrm{E}}=1.
\label{eq:event_component_shares}
\end{equation}
The event shares place greater weight on accelerator training than the normal background shares.

\subsubsection{Sustained synchronized ramp event}

This event represents a synchronized transition from a low IT loading fraction $f_1^{\mathrm{L}}$ to a high fraction $f_1^{\mathrm{H}}$, followed by a sustained hold and controlled recovery. For relative event time $\theta=t-t_1^{\mathrm{start}}$,

\begin{equation}
\Delta f_1=f_1^{\mathrm{H}}-f_1^{\mathrm{L}}
\end{equation}

\begin{equation}
\begin{aligned}
f_1(\theta)=
\begin{cases}
f_1^{\mathrm{L}}, & \theta<0,\\
f_1^{\mathrm{L}}
+\Delta f_1\dfrac{\theta}{T_1^{\mathrm{r}}},
& 0\le\theta\le T_1^{\mathrm{r}},\\
f_1^{\mathrm{H}},
& T_1^{\mathrm{r}}<\theta<T_1^{\mathrm{r}}+T_1^{\mathrm{h}},\\
f_1^{\mathrm{H}}
-\Delta f_1
\dfrac{\theta-T_1^{\mathrm{r}}-T_1^{\mathrm{h}}}
{T_1^{\mathrm{rec}}},
& T_1^{\mathrm{r}}+T_1^{\mathrm{h}}\le\theta\\
&\le T_1^{\mathrm{r}}+T_1^{\mathrm{h}}+T_1^{\mathrm{rec}},\\
f_1^{\mathrm{L}}, & \text{otherwise}.
\end{cases}
\end{aligned}
\label{eq:th1}
\end{equation}

The theoretical campus IT step and average imposed ramp are
\begin{align}
\Delta P_{1}^{\mathrm{IT}}&=P_{\mathrm{IT}}^{\max}(f_1^{\mathrm{H}}-f_1^{\mathrm{L}}),\\
R_1^{\mathrm{IT}}&=\frac{\Delta P_1^{\mathrm{IT}}}{T_1^{\mathrm{r}}}.
\end{align}
For an equally allocated building, replace $P_{\mathrm{IT}}^{\max}$ with $P_{\mathrm{IT},b}^{\max}$.

\subsubsection{Stochastic repeated synchronized burst event}

This event consists of $K_2$ ramp-hold-recovery bursts distributed over a fixed event window $T_2^{\mathrm{win}}$. The nominal burst spacing is
\begin{equation}
\Delta T_2=\left\lfloor\frac{T_2^{\mathrm{win}}}{K_2}\right\rfloor.
\end{equation}
For burst $k$, the start time is
\begin{equation}
t_{2,k}^{\mathrm{start}}=t_2^{\mathrm{start}}+k\Delta T_2,
\qquad k=0,\ldots,K_2-1.
\end{equation}
Each burst follows the same piecewise form as Eq.~\eqref{eq:th1}, using TH2-specific low and high fractions and ramp, hold, and recovery durations. The individual-burst theoretical change and ramp are
\begin{align}
\Delta P_{2}^{\mathrm{IT}}&=P_{\mathrm{IT}}^{\max}(f_2^{\mathrm{H}}-f_2^{\mathrm{L}}),\\
R_2^{\mathrm{IT}}&=\frac{\Delta P_2^{\mathrm{IT}}}{T_2^{\mathrm{r}}}.
\end{align}

\subsubsection{Second-resolution transition spike}

This event represents a severe synchronized short-timescale IT stress event. In the full-campus mode, the campus alternates between high and low loading fractions at the one-second sampling limit. For $K_3$ spikes,
\begin{equation}
f_3(\theta)=
\begin{cases}
f_3^{\mathrm{H}}, & \theta=2k,\\
f_3^{\mathrm{L}}, & \theta=2k+1,
\end{cases}
\qquad k=0,\ldots,K_3-1.
\label{eq:th3_full}
\end{equation}
with low-state conditioning immediately before and after the spike train. The one-sample upward change is
\begin{equation}
\Delta P_3^{\mathrm{IT}}=P_{\mathrm{IT}}^{\max}(f_3^{\mathrm{H}}-f_3^{\mathrm{L}}),
\end{equation}
and the corresponding discrete ramp at one-second resolution is numerically equal to $\Delta P_3^{\mathrm{IT}}$ in MW/s.

An alternative diversified mode modifies only a participating fraction $\gamma_3$ of the pre-event training load. Let $P_{\mathrm{tr},t}^{0}$ be the baseline training demand and let $P_{\mathrm{nontr},t}$ be the sum of the remaining IT components. The modified training demand is
\begin{align}
P_{\mathrm{tr},t}^{\mathrm{new}}={}&(1-\gamma_3)P_{\mathrm{tr},t}^{0}
+\gamma_3P_{\mathrm{tr},t}^{0}f_{3,t},\\
P_{\mathrm{tr},t}={}&\min\!\left[P_{\mathrm{tr},t}^{\mathrm{new}},
\max(P_{\mathrm{IT}}^{\max}-P_{\mathrm{nontr},t},0)\right],\\
P_{\mathrm{IT},t}={}&P_{\mathrm{nontr},t}+P_{\mathrm{tr},t}.
\label{eq:th3_training_only}
\end{align}
This cap prevents the component total from exceeding campus IT capacity.

After all transient events are applied, total IT load is reconstructed as
\begin{equation}
P_{\mathrm{IT},t}=\operatorname{clip}\!\left(
\sum_{c\in\mathcal{C}}P_{c,t},\varepsilon,P_{\mathrm{IT}}^{\max}\right).
\label{eq:final_it}
\end{equation}

\subsection{Power Factor and Electrical Conversion Losses}

The fractions of IT demand associated with training and inference are
\begin{align}
r_{\mathrm{tr},t}&=\frac{P_{\mathrm{tr},t}}{\max(P_{\mathrm{IT},t},\varepsilon)},\\
r_{\mathrm{inf},t}&=\frac{P_{\mathrm{inf},t}}{\max(P_{\mathrm{IT},t},\varepsilon)}.
\end{align}
The second-level power factor is
\begin{align}
\widehat{\mathrm{PF}}_t={}&\mathrm{PF}_0-k_{\mathrm{PF,tr}}r_{\mathrm{tr},t}
-k_{\mathrm{PF,inf}}r_{\mathrm{inf},t} \\
&-a_{\mathrm{PF}}\sin\!\left(\frac{2\pi h_t}{H_d}\right)
+\widetilde{\epsilon}_{\mathrm{PF},t},\\
\mathrm{PF}_t={}&\operatorname{clip}\!\left(
\widehat{\mathrm{PF}}_t,\mathrm{PF}^{\min},\mathrm{PF}^{\max}\right).
\label{eq:pf}
\end{align}

Define the normalized IT loading ratio
\begin{equation}
\lambda_t=\operatorname{clip}\!\left(
\frac{P_{\mathrm{IT},t}}{P_{\mathrm{IT}}^{\max}},
\lambda^{\min},\lambda^{\max}\right).
\end{equation}
The UPS efficiency is represented by a bounded concave quadratic:
\begin{align}
\widehat{\eta}_{\mathrm{UPS},t}&=\eta_{\mathrm{UPS},0}
+a_{\eta,1}\lambda_t-a_{\eta,2}\lambda_t^2,\\
\eta_{\mathrm{UPS},t}&=\operatorname{clip}\!\left(
\widehat{\eta}_{\mathrm{UPS},t},\eta_{\mathrm{UPS}}^{\min},
\eta_{\mathrm{UPS}}^{\max}\right).
\end{align}
The UPS loss is
\begin{equation}
P_{\mathrm{UPS},t}^{\mathrm{loss}}=P_{\mathrm{IT},t}
\left(\frac{1}{\eta_{\mathrm{UPS},t}}-1\right).
\label{eq:ups_loss}
\end{equation}

Transformer losses include fixed core loss and load-dependent copper loss with a high-temperature correction:
\begin{align}
f_{T,t}^{\mathrm{tr}}&=1+k_T^{\mathrm{tr}}\max(0,T_{a,t}-T_{\mathrm{tr}}^{\mathrm{ref}}),\\
P_{\mathrm{tr},t}^{\mathrm{loss}}&=P_{\mathrm{tr}}^{\mathrm{fix}}
+k_{\mathrm{tr}}^{\mathrm{cu}}\lambda_t^2f_{T,t}^{\mathrm{tr}}.
\label{eq:transformer_loss}
\end{align}
PDU losses contain linear and quadratic terms:
\begin{equation}
P_{\mathrm{PDU},t}^{\mathrm{loss}}=
 k_{\mathrm{PDU},1}P_{\mathrm{IT},t}
+k_{\mathrm{PDU},2}P_{\mathrm{IT},t}^2.
\label{eq:pdu_loss}
\end{equation}
The aggregate electrical loss is
\begin{equation}
P_{\mathrm{elec},t}^{\mathrm{loss}}=
P_{\mathrm{UPS},t}^{\mathrm{loss}}+P_{\mathrm{tr},t}^{\mathrm{loss}}
+P_{\mathrm{PDU},t}^{\mathrm{loss}}.
\label{eq:electrical_loss}
\end{equation}

\subsection{Thermal Response and Cooling Demand}
The heat presented to the cooling system is approximated as the sum of IT power and electrical conversion losses:
\begin{equation}
Q_t^{\mathrm{heat}}=P_{\mathrm{IT},t}+P_{\mathrm{elec},t}^{\mathrm{loss}}.
\end{equation}
A first-order discrete thermal lag prevents the cooling plant from responding instantaneously to IT transients:
\begin{align}
\alpha_{\mathrm{th}}&=\frac{\Delta t}{\max(\tau_{\mathrm{th}},\Delta t)},\\
Q_t^{\mathrm{th}}&=\alpha_{\mathrm{th}}Q_t^{\mathrm{heat}}
+(1-\alpha_{\mathrm{th}})Q_{t-1}^{\mathrm{th}},
\label{eq:thermal_lag}
\end{align}
where $\Delta t=1$ s and $\tau_{\mathrm{th}}$ is the effective thermal time constant. The implementation initializes the filtered trajectory to reduce startup bias.

For multiple cooling technologies, effective reference COP and ambient sensitivity are weighted by their modeled campus shares $w_q$:
\begin{align}
\mathrm{COP}_{\mathrm{ref}}^{\mathrm{eff}}&=\sum_{q\in\mathcal{Q}}w_q\mathrm{COP}_{\mathrm{ref},q},\\
k_{a}^{\mathrm{eff}}&=\sum_{q\in\mathcal{Q}}w_qk_{a,q},\\
\sum_{q\in\mathcal{Q}}w_q&=1.
\end{align}
The temperature-dependent COP is
\begin{align}
\widehat{\mathrm{COP}}_t&=\mathrm{COP}_{\mathrm{ref}}^{\mathrm{eff}}
-k_a^{\mathrm{eff}}\max(0,T_{a,t}-T_{\mathrm{COP}}^{\mathrm{ref}}),\\
\mathrm{COP}_t&=\operatorname{clip}\!\left(
\widehat{\mathrm{COP}}_t,\mathrm{COP}^{\min},\mathrm{COP}^{\max}\right).
\label{eq:cop}
\end{align}

The economizer multiplier is a piecewise function of ambient temperature:
\begin{equation}
e(T_{a,t})=
\begin{cases}
e_1, & T_{a,t}\le \theta_1,\\
e_2, & \theta_1<T_{a,t}\le\theta_2,\\
e_3, & \theta_2<T_{a,t}\le\theta_3,\\
e_4, & T_{a,t}>\theta_3.
\end{cases}
\label{eq:economizer}
\end{equation}
A value below unity represents reduced compressor demand under favorable ambient conditions.

The normalized cooling load is
\begin{equation}
\lambda_t^{\mathrm{cool}}=\operatorname{clip}\!\left(
\frac{Q_t^{\mathrm{th}}}{P_{\mathrm{IT}}^{\max}},
\lambda_{\mathrm{cool}}^{\min},\lambda_{\mathrm{cool}}^{\max}\right).
\end{equation}
The cooling subsystems are
\begin{align}
P_{\mathrm{ch},t}&=\frac{Q_t^{\mathrm{th}}}{\mathrm{COP}_t}e(T_{a,t}),\\
P_{\mathrm{fan},t}&=k_{\mathrm{fan}}(\lambda_t^{\mathrm{cool}})^3,\\
P_{\mathrm{pump},t}&=k_{\mathrm{pump}}(\lambda_t^{\mathrm{cool}})^3.
\end{align}
Cooling staging and controls are represented by

\begin{equation}
\begin{aligned}
P_{\mathrm{stage},t}={}&a_{s,1}\sin\!\left(\frac{2\pi t}{T_{s,1}}\right)
+a_{s,2}\sin\!\left(\frac{2\pi t}{T_{s,2}}\right)\\
&+A_{s,1}\mathbb{I}(h_t\in\mathcal{H}_{s,1})
+A_{s,2}\mathbb{I}(h_t\in\mathcal{H}_{s,2})\\
&+k_{s,T}\max(0,T_{a,t}-T_s^{\mathrm{ref}}).
\end{aligned}
\end{equation}

The continuous cooling demand is
\begin{equation}
P_{\mathrm{cool},t}^{\mathrm{cont}}=\max\!\left(
P_{\mathrm{ch},t}+P_{\mathrm{fan},t}+P_{\mathrm{pump},t}
+P_{\mathrm{stage},t},0\right).
\end{equation}
To represent discrete equipment blocks of rating $P_{\mathrm{blk}}$, the final cooling load is
\begin{equation}
P_{\mathrm{cool},t}=P_{\mathrm{blk}}
\left\lceil\frac{P_{\mathrm{cool},t}^{\mathrm{cont}}}{P_{\mathrm{blk}}}\right\rceil.
\label{eq:cooling_discrete}
\end{equation}

\subsection{Auxiliary and Miscellaneous Demand}

Auxiliary demand represents lighting, controls, security, support systems, and other load-dependent balance-of-plant services:
\begin{align}
\widehat{P}_{\mathrm{aux},t}={}&a_{\mathrm{aux},0}P_{\mathrm{IT}}^{\max}
+a_{\mathrm{aux},1}P_{\mathrm{IT},t}
+a_{\mathrm{aux},2}\sin\!\left(\frac{2\pi h_t}{H_d}\right) \\
&+a_{\mathrm{aux},3}s_{d_t}+\widetilde{\epsilon}_{\mathrm{aux},t},\\
P_{\mathrm{aux},t}={}&\operatorname{clip}\!\left(
\widehat{P}_{\mathrm{aux},t},a_{\mathrm{aux}}^{\min}P_{\mathrm{IT}}^{\max},
 a_{\mathrm{aux}}^{\max}P_{\mathrm{IT}}^{\max}\right).
\end{align}
Miscellaneous demand is

\begin{equation}
\begin{aligned}
P_{\mathrm{misc},t}=\operatorname{clip}\!\Bigl({}&
 a_{\mathrm{misc},0}P_{\mathrm{IT}}^{\max}
 +\widetilde{\epsilon}_{\mathrm{misc},t},a_{\mathrm{misc}}^{\min}P_{\mathrm{IT}}^{\max},\\
& a_{\mathrm{misc}}^{\max}P_{\mathrm{IT}}^{\max}\Bigr).
\end{aligned}
\end{equation}

\subsection{Non-IT Calibration and Facility Demand}

Before calibration, non-IT demand is
\begin{equation}
P_{\mathrm{NIT},t}^{\mathrm{raw}}=P_{\mathrm{elec},t}^{\mathrm{loss}}
+P_{\mathrm{cool},t}+P_{\mathrm{aux},t}+P_{\mathrm{misc},t}.
\end{equation}
Let $\mathcal{T}_{\mathrm{hi}}$ denote seconds for which IT demand exceeds a specified fraction $\beta_{\mathrm{hi}}$ of full IT capacity:
\begin{equation}
\mathcal{T}_{\mathrm{hi}}=\{t:P_{\mathrm{IT},t}\ge\beta_{\mathrm{hi}}P_{\mathrm{IT}}^{\max}\}.
\end{equation}
The reference non-IT load is
\begin{equation}
P_{\mathrm{NIT}}^{\mathrm{ref}}=
\begin{cases}
\dfrac{1}{|\mathcal{T}_{\mathrm{hi}}|}\displaystyle\sum_{t\in\mathcal{T}_{\mathrm{hi}}}
P_{\mathrm{NIT},t}^{\mathrm{raw}}, & |\mathcal{T}_{\mathrm{hi}}|>0,\\
\max_tP_{\mathrm{NIT},t}^{\mathrm{raw}}, & |\mathcal{T}_{\mathrm{hi}}|=0.
\end{cases}
\end{equation}
A common calibration factor aligns the non-IT model with the nominal full-load non-IT target:
\begin{equation}
k_{\mathrm{NIT}}=\frac{P_{\mathrm{NIT}}^{\max}}
{\max(P_{\mathrm{NIT}}^{\mathrm{ref}},\varepsilon)}.
\label{eq:nonit_calibration}
\end{equation}
Each non-IT subcomponent is multiplied by $k_{\mathrm{NIT}}$. Therefore,
\begin{align}
P_{\mathrm{elec},t}^{\mathrm{cal}}&=k_{\mathrm{NIT}}P_{\mathrm{elec},t}^{\mathrm{loss}},\\
P_{\mathrm{cool},t}^{\mathrm{cal}}&=k_{\mathrm{NIT}}P_{\mathrm{cool},t},\\
P_{\mathrm{aux},t}^{\mathrm{cal}}&=k_{\mathrm{NIT}}P_{\mathrm{aux},t},\\
P_{\mathrm{misc},t}^{\mathrm{cal}}&=k_{\mathrm{NIT}}P_{\mathrm{misc},t},\\
P_{\mathrm{NIT},t}&=P_{\mathrm{elec},t}^{\mathrm{cal}}
+P_{\mathrm{cool},t}^{\mathrm{cal}}+P_{\mathrm{aux},t}^{\mathrm{cal}}
+P_{\mathrm{misc},t}^{\mathrm{cal}},\\
P_{\mathrm{fac},t}&=P_{\mathrm{IT},t}+P_{\mathrm{NIT},t}.
\label{eq:facility_load}
\end{align}
The common factor preserves the relative temporal behavior and relative subcomponent composition of the raw non-IT model while matching its high-IT operating level to the selected campus target.

\subsection{Reactive Power, Apparent Power, PUE, and Ramps}

Facility reactive and apparent power are calculated from real power and power factor:
\begin{align}
Q_{\mathrm{fac},t}&=P_{\mathrm{fac},t}\tan\!\left[\cos^{-1}(\mathrm{PF}_t)\right],\\
S_{\mathrm{fac},t}&=\frac{P_{\mathrm{fac},t}}{\max(\mathrm{PF}_t,\varepsilon)}.
\end{align}
The second-level power usage effectiveness is
\begin{equation}
\mathrm{PUE}_t=\frac{P_{\mathrm{fac},t}}
{\max(P_{\mathrm{IT},t},\varepsilon)}.
\label{eq:pue}
\end{equation}
The discrete one-second facility and IT ramps are
\begin{align}
R_{\mathrm{fac},t}^{(s)}&=\frac{P_{\mathrm{fac},t}-P_{\mathrm{fac},t-1}}{\Delta t_s},\\
R_{\mathrm{IT},t}^{(s)}&=\frac{P_{\mathrm{IT},t}-P_{\mathrm{IT},t-1}}{\Delta t_s},
\end{align}
where $\Delta t_s=1$ s. Accordingly, the numerical differences are expressed directly in MW/s.

\subsection{Aggregation to Minute Resolution}

Any second-level variable $x_t$ is aggregated to a minute mean as
\begin{equation}
\overline{x}_m=\frac{1}{S_m}\sum_{j=0}^{S_m-1}x_{mS_m+j}.
\label{eq:minute_aggregation}
\end{equation}
Minute-level reactive and apparent power are recalculated from the aggregated real power and power factor:
\begin{align}
Q_{\mathrm{fac},m}&=P_{\mathrm{fac},m}
\tan\!\left[\cos^{-1}(\mathrm{PF}_m)\right],\\
S_{\mathrm{fac},m}&=\frac{P_{\mathrm{fac},m}}{\max(\mathrm{PF}_m,\varepsilon)}.
\end{align}
The minute-to-minute changes are
\begin{align}
R_{\mathrm{fac},m}^{(\mathrm{min})}&=P_{\mathrm{fac},m}-P_{\mathrm{fac},m-1},\\
R_{\mathrm{IT},m}^{(\mathrm{min})}&=P_{\mathrm{IT},m}-P_{\mathrm{IT},m-1}.
\end{align}
These quantities are reported in MW per minute interval. Longer aggregation intervals may be constructed from the second-level series by replacing $S_m$ in Eq.~\eqref{eq:minute_aggregation} with the desired number of seconds per interval.


\section{Data Center Load Smoothing Optimization and SSTI Mitigation using Load Side BESS}\label{Problem_formulation_loadSmoothing}
The proposed multi-phase optimization framework comprises two distinct phases. The first phase focuses on BESS design and load smoothing optimization at the load side , while the second phase addresses resilient design of BESS and operation scheduling optimization of BTM generation assets at the generation side of the microgrid. The load-smoothing optimization problem, defined by equation (\ref{load_smooth_OF})--(\ref{Net_acceptable_ramp}), minimizes the combined cost of load shedding and net-load variability through the coordinated operation of a load-side BESS. The objective function presented in equation (\ref{load_smooth_OF}) consists of two components: a load shedding cost term, represented by the shed load multiplied by the Value of Lost Load (VOLL), and a ramping rate penalty term that penalizes sharp increases and decreases in the net load profile. By simultaneously minimizing load curtailment and rapid load fluctuations, the optimization seeks to produce a smoother and more operationally manageable demand profile.
\begin{align}\label{load_smooth_OF} 
\min \quad
&  \underbrace{\sum_{t} P_{[t]}^{\text{shed}} \cdot \text{VOLL}}_{\text{Load Shedding Cost}} 
 + \underbrace{\sum_{t} (P_{[t]}^{Net} - P_{[t]}^{Net})^{2} \cdot RRP_{}}_{\text{Ramp Rate Penalty Cost for Smoothing}} 
 \end{align}
\begin{equation}\label{Load_smooth_balance1} 
P_{[t]}^{Net} - P_{[t]}^{Load} - P_{[t]}^{ch} + P_{[t]}^{dis} + P_{[t]}^{Shed} = 0 
\end{equation}
\begin{equation}\label{Load_shedding_smoothing}
    0 \le  P_{[t]}^{Shed}  \le \overline {P_{[t]}^{Load}}  ,\quad \forall\; t \in \mathcal{T}
\end{equation}
\begin{equation}\label{BES_charge_smoothing}
    0 \le P_{[t]}^{ch} \le \overline{P^{ch}} \;.\; \phi_{[t]},\quad \forall\; t \in \mathcal{T}
\end{equation}
\begin{equation}\label{BES_discharge_smoothing}
    0 \le P_{[t]}^{dis} \le \overline {P^{dis}} \;.\; (1-\phi_{[t]}),\quad \forall\; t \in \mathcal{T}
\end{equation}
\begin{equation}\label{BES_SOC_smoothing}
    E_{[t]}=E_{[t-1]} + (P_{[t]}^{ch}. \eta^{ch} - P_{[t]}^{dis}/\eta^{dis}).\Delta t ,\; \forall\; t \in \mathcal{T}
\end{equation}
\begin{equation}\label{BES_SOC_limit_smoothing}
    \underline {SOC^{}}\;.\; \overline {E} \le E_{[t]} \le \overline{SOC^{}}\;.\; \overline {E},\quad \forall\; t \in \mathcal{T}
\end{equation}
\begin{equation}\label{BES_MaxEnergy_limit_smoothing}
    \overline {E}= E^{R}.f^{EOL}. f^{Temp,E}.f^{Res}.f^{Ava}., \forall\; t \in \mathcal{T}
\end{equation}
\begin{equation}\label{BES_MaxPower_limit_smoothing}
    \overline {P^{}} = P^{rated}. f^{PCS}. f^{Temp,P}.f^{Res}.f^{Ava}, \forall\; t \in \mathcal{T}
\end{equation}
\begin{equation}\label{SOC_load_smoothing}
    SOC_{[t]} = E_{[t]}\;/\ \overline {E}_{[t]},\quad \forall\; t \in \mathcal{T}
\end{equation}
\begin{equation}\label{SOC_initial_BESS_Load}
    E_{[t]} = SOC^{init}\;.\; \overline{E},\quad \forall\; t = t_0
\end{equation}
\begin{equation}\label{SOC_final_BESS_Load}
    E_{[t]} = SOC^{end}\;.\; \overline{E},\quad \forall\; t = T
\end{equation}
\begin{equation}\label{Net_acceptable_ramp}
     -R^{allowed}_{}\le P_{[t]}^{Net} - P_{[t-1]}^{Net} \le R^{allowed}_{} ,\quad \forall\; t \in \mathcal{T}
\end{equation}
The power balance equation is enforced by equation (\ref{Load_smooth_balance1}), while the load shedding limits are defined in equation (\ref{Load_shedding_smoothing}). The BESS design and operational constraints are represented in equations (\ref{BES_charge_smoothing})-(\ref{SOC_final_BESS_Load}), where equations (\ref{BES_charge_smoothing}) and (\ref{BES_discharge_smoothing}) refer to charging and discharging limits of BESS. Equations (\ref{BES_SOC_smoothing}) and (\ref{BES_SOC_limit_smoothing}) show the BESS energy balance and energy limits constraints \cite{haggi2025real}. BESS max energy capacity is shown in equation (\ref{BES_MaxEnergy_limit_smoothing}). Apart from efficiency and depth of discharge and other parameters,  BESS size is affected by loss factors such as end of life capacity (degradation over lifetime), temperature derates, availability factor, and additional reserve. Temperature derate, availability factor, power conversion system (PCS) and reserve factor affect BESS power rating which is shown in (\ref{BES_MaxPower_limit_smoothing}). Additionally, equations (\ref{SOC_load_smoothing})-(\ref{SOC_final_BESS_Load}) show SOC, initial and final energy levels, respectively. Finally, equation (\ref{Net_acceptable_ramp}) shows the acceptable ramping limits (allowed ramping threshold).

\section{Resilient BESS Design and BTM Generation Assets Techno-Economic Optimization}\label{Problem_formulation_loadSmoothing}
\subsection{Objective Function}
The BTM techno-economic optimization and resilience planning framework is formulated through (\ref{OF_genside})--(\ref{Res_req_Ren}), where the BESS size as well as optimal operational setpoints are determined while accounting for economic performance, technical and physical equipment constraints, and system reliability requirements.

\begin{align}\label{OF_genside}
\min \quad
& \underbrace{\sum_{t} P_{[t]}^{\text{Imp}} \cdot LMP_{[t]}}_{\text{Grid Import Cost}} 
 + \underbrace{\sum_{t} P_{[t]}^{\text{Exp}} \cdot PPA_{[t]}}_{\text{Grid Export Revenue}} \\
& + \underbrace{\sum_{t,i} P_{[t]}^{\text{PV}} \cdot C_{[i]}^{\text{PV}}}_{\text{PV Cost}} \nonumber
 + \underbrace{\sum_{t,i} P_{[t]}^{\text{Wind}} \cdot C_{[i]}^{\text{Wind}}}_{\text{Wind Cost}} \\
& + \underbrace{\sum_{t,i} P_{[t,i]}^{\text{GT}} \cdot C_{[i]}^{\text{GT,SRMC}} + U_{[t,i]}^{} \cdot C_{[i]}^{\text{GT, NL}} }_{\text{Gas Turbine Short Run Marginal Cost and No-Load Cost}}  \nonumber \\
& + \underbrace{\sum_{t,i} SU_{[t,i]} \cdot C_{[i]}^{\text{SU,GT}}}_{\text{Gas Turbines' Startup Cost}} 
+ \underbrace{\sum_{t,i} SD_{[t,i]} \cdot \nonumber C_{[i]}^{\text{SD,GT}}}_{\text{Gas Turbines' Shutdown Cost}} \\
 &+ \underbrace{\sum_{t} P_{[t]}^{\text{shed}} \cdot \text{VOLL}}_{\text{Load Shedding Cost}} 
+ \underbrace{\sum_{t,i} \rho^{Crtl} \cdot \nonumber P_{[t]}^{PV_{Crtl}}}_{\text{Curtailed PV Penalty Cost}} \\
& + \underbrace{\sum_{t} \xi^{Crtl} \cdot  P_{[t]}^{Wind_{Crtl}}}_{\text{Curtailed Wind Penalty Cost}}
 + \underbrace{\sum_{t,i} \gamma^{Crtl} \cdot \nonumber P_{[t,i]}^{GT}}_{\text{Curtailed Turbines Penalty Cost}} \\ \nonumber
& + \underbrace{\sum_{t} \mu^{Res} \cdot  R_{[t]}^{Slack}}_{\text{Slack Reserve Penalty Cost for Feasibility}}  + \underbrace{\sum_{t} P_{[t]}^{\text{FC}} \cdot  C_{[t]}^{NG}}_{\text{Natural Gas FC Cost}}\\ 
& + \underbrace{ \overline{P^{ch}} \cdot  IC_{Power}^{}}_{\text{Investment Cost on BESS Power}} + \underbrace{\overline{SOC^{\text{}}} \cdot  IC_{Energy}}_{\text{Investment Cost on BESS Energy}} \nonumber
\end{align}
 \begin{equation}\label{GT_SRMC_calcs}
    C_{[i]}^{\text{GT,SRMC}} = GT_{[i]}^{HR}\;.\; C_{}^{\text{fuel}} \; + \; V^{O\&M}_{[i]}
\end{equation}
 \begin{equation}\label{GT_No_load_calcs}
    C_{[i]}^{\text{GT,NL}} = GT_{[i]}^{HR, NL}\;.\; C_{}^{\text{fuel}} \;.\; \underline{P^{GT}_{[i]}}
\end{equation}
\begin{equation}\label{IC_energy}
    IC_{Energy} = c_{en} \: \frac{r(1+r)^h}{(1+r)^h - 1} \cdot \frac{N_{\mathrm{Opt. Time}}}{N_{\mathrm{year}}}
\end{equation}
\begin{equation}\label{IC_power}
    IC_{Power} = c_{pow} \: \frac{r(1+r)^h}{(1+r)^h - 1} \cdot \frac{N_{\mathrm{Opt. Time}}}{N_{\mathrm{year}}}
\end{equation}
The objective function is formulated to minimize the total operational cost of BTM generation assets and investment cost of BESS. The operational cost components include grid import costs based on locational marginal prices (LMPs) and grid export revenues for surplus energy supplied back to the grid. Although this study focuses exclusively on BTM data center planning and operation, grid import and export formulations are included to provide a generalized framework applicable to projects with potential future grid interconnection. Additional cost components include solar and wind energy annualized operation and maintenance costs, gas turbine short-run marginal and no-load costs, and gas turbine startup and shutdown costs to capture unit commitment and operational flexibility requirements. Load shedding is penalized using the Value of Lost Load (VOLL), while additional penalty terms are assigned to curtailed PV generation, wind generation. A reserve shortfall penalty term is also incorporated within the framework to penalize reserve violations and maintain optimization model feasibility under stressed operating conditions where reserve requirements cannot be fully satisfied. Curtailed gas turbine penalty cost is also considered for grid connection mode in case if the goal is still maximizing the utilization of BTM gas turbine assets (this is optional in grid connection mode). Furthermore, the objective function includes the operating costs of natural gas powered FC units. Finally, investment cost terms associated with both the power and energy capacities of the BESS are included (which can be calculated by equations (\ref{IC_energy}) and (\ref{IC_power})), enabling the simultaneous optimization of long-term resource planning and operational performance.
\subsection{Constraints}
The following constraints need to be considered for reliable and more realistic operation of BTM data center generation assets. 
\begin{align}\label{Power_balance1} 
P_{[t]}^{Imp} - P_{[t]}^{Exp} + P_{[t]}^{PV} + P_{[t]}^{Wind} + \sum_i{P_{[t,i]}^{GT}}
- P_{[t]}^{Load} \\ - P_{[t]}^{ch} + P_{[t]}^{dis} + P_{[t]}^{FC} + P_{[t]}^{Shed} = 0 \nonumber
\end{align}
\begin{equation}\label{Load_shedding}
    0 \le  P_{[t]}^{Shed}  \le \overline {P_{[t]}^{Load}}  ,\quad \forall\; t \in \mathcal{T}
\end{equation}
\begin{equation}\label{Grid_import}
    \underline {P_{[t]}^{Imp}} \;.\; \delta_{[t]} \;.\; \theta_{[t]}^{G}\le  P_{[t]}^{Imp}  \le \overline {P_{[t]}^{Imp}} \;.\; \delta_{[t]}^{G} \;.\; \theta_{[t]}^{G},\quad \forall\; t \in \mathcal{T}
\end{equation}
\begin{equation}\label{Grid_export}
    \underline {P_{[t]}^{Exp}} \;.\; (1-\delta_{[t]})\;.\; \theta_{[t]}^{G}\le  P_{[t]}^{Exp} \le \overline {P_{[t]}^{Exp}} \;.\; (1-\delta_{[t]}) \;.\; \theta_{[t]}^{G}
\end{equation}
\begin{equation}\label{Grid_import_ramp}
     -R^{G,Dn}_{}\le P_{[t]}^{Imp} - P_{[t-1]}^{Imp} \le R^{G,Up}_{} ,\quad \forall\; t \in \mathcal{T}
\end{equation}
\begin{equation}\label{Grid_export_ramp}
     -R^{G,Dn}_{}\le P_{[t]}^{Exp} - P_{[t-1]}^{Exp} \le R^{G,Up}_{} ,\quad \forall\; t \in \mathcal{T}
\end{equation}
\begin{equation}\label{GT_limit}
   \underline{P^{GT}_{[i]}}.U_{[t,i]} . \theta_{[t,i]}^{GT}\le P_{[t,i]}^{GT} \le \overline{P^{GT}_{[i]}} . U_{[t]}. \theta_{[t,i]}^{GT}, \;\;\forall\; t,i \in \mathcal{T,I}
\end{equation}
\begin{equation}\label{GT_Binary_transition}
   U_{[t,i]} - U_{[t-1,i]} = SU_{[t,i]} - SD_{[t,i]},\quad \forall\; t \in \mathcal{T}, i \in \mathcal{I}
\end{equation}
\begin{equation}\label{GT_SUSD}
   SU_{[t,i]} + SD_{[t,i]} \le 1 ,\quad \forall\; t \in \mathcal{T}, i \in \mathcal{I}
\end{equation}
\begin{equation}\label{GT_ramp_up}
\begin{aligned}
      P_{[t,i]}^{GT} - P_{[t-1,i]}^{GT} \le R^{GT,Up}_{i}\;.\; U_{[t-1,i]} \;+\; R^{GT,SU}_{i}\;.\; SU_{[t,i]} 
    \\
      \quad \forall\; t \in \mathcal{T}, i \in \mathcal{I}
\end{aligned}
\end{equation}
\begin{equation}\label{GT_ramp_down}
\begin{aligned}
      P_{[t-1,i]}^{GT} - P_{[t,i]}^{GT} \le R^{GT,Dn}_{i}\;.\; U_{[t,i]} \;+\; R^{GT,SD}_{i}\;.\; SD_{[t,i]} 
    \\
      \quad \forall\; t \in \mathcal{T}, i \in \mathcal{I}
\end{aligned}
\end{equation}
\begin{equation}\label{min_UT}
     \sum_{\tau = t}^{t + UT_i - 1} U_{[\tau,i]} \;\ge\; UT_{[i]} \cdot SU_{[t,i]} ,\quad \forall\; t \in \mathcal{T}, i \in \mathcal{I}
\end{equation}
\begin{equation}\label{min_DT}
      \sum_{\tau = t}^{t + DT_i - 1} (1 - U_{[\tau,i]}) \;\ge\; DT_i \cdot SD_{[t,i]} ,\quad \forall\; t \in \mathcal{T}, i \in \mathcal{I}
\end{equation}
\begin{equation}\label{GT_reserve}
0 \le P_{[t,i]}^{GT,Res}  \;\leq\; \overline{P^{GT}_{[i]}} \;.\; U_{[t]} \;.\; \theta_{[t,i]}^{GT}-  P_{[t,i]}^{GT},\;\;\forall\; t \in \mathcal{T}, i \in \mathcal{I}
\end{equation}
\begin{equation}\label{GT_reserve_deliver}
P_{[t,i]}^{GT,Res} \;\leq\; R^{GT,Up}_{i}\;.\; U_{[t,i]}\;.\ \tau^{Delivery},\quad \forall\; t \in \mathcal{T}, i \in \mathcal{I}
\end{equation}
\begin{equation}\label{PV_limit}
    0 \le  P_{[t]}^{PV}  \le \overline {P_{[t]}^{PV}}\;.\; \theta_{[t]}^{PV}  ,\quad \forall\; t \in \mathcal{T}
\end{equation}
\begin{equation}\label{PV_crtl}
     P_{[t]}^{PV} + P_{[t]}^{PV_{Crtl}} - \overline {P_{[t]}^{PV}} = 0  ,\quad \forall\; t \in \mathcal{T}
\end{equation}
\begin{equation}\label{Wind_limit}
    0 \le  P_{[t]}^{Wind}  \le \overline {P_{[t]}^{Wind}}\;.\; \theta_{[t]}^{Wind}  ,\quad \forall\; t \in \mathcal{T}
\end{equation}
\begin{equation}\label{Wind_crtl}
     P_{[t]}^{Wind} + P_{[t]}^{Wind_{Crtl}} - \overline {P_{[t]}^{Wind}} = 0  ,\quad \forall\; t \in \mathcal{T}
\end{equation}
\begin{equation}\label{BES_charge_Gen}
    0 \le P_{[t]}^{ch} \le \overline{P^{ch}} \;.\; \phi_{[t]}\;.\; \theta_{[t]}^{BESS},\quad \forall\; t \in \mathcal{T}
\end{equation}
\begin{equation}\label{BES_discharge_Gen}
    0 \le P_{[t]}^{dis} \le \overline {P^{dis}} \;.\; (1-\phi_{[t]}) \;.\; \theta_{[t]}^{BESS},\quad \forall\; t \in \mathcal{T}
\end{equation}
\begin{equation}\label{BESS_reserve}
0 \le P_{[t]}^{BESS,Res}  \;\leq\; \overline {P^{dis}} . (1-\phi_{[t]})\;.\; \theta_{[t]}^{BESS} -  P_{[t]}^{dis}, \forall\; t \in \mathcal{T}
\end{equation}
\begin{equation}\label{BES_SOC_Gen}
    E_{[t]}=E_{[t-1]} + (P_{[t]}^{ch}. \eta^{ch} - P_{[t]}^{dis}/\eta^{dis}).\Delta t ,\; \forall\; t \in \mathcal{T}
\end{equation}
\begin{align}\label{BES_SOC_limit_Gen} 
\underline {SOC^{}}\;.\; \overline {E} + E_{[t]}^{Brdg} + E_{[t]}^{Resi} + E_{[t]}^{BS}  \\ \le E_{[t]} \le \overline{SOC^{}}\;.\;\overline {E} ,\quad \forall\; t \in \mathcal{T} \nonumber
\end{align}
\begin{equation}\label{BES_MaxEnergy_limit_Gen}
    \overline {E}= E^{R}.f^{EOL}.f^{Temp,E}.f^{Avail},\forall\; t \in \mathcal{T}
\end{equation}
\begin{equation}\label{Bridge_SOC}
    E_{[t]}^{Brdg} = (P_{[t]}^{GT} \;.\; \tau^{Brdg})/\eta^{dis}, \forall\; t \in \mathcal{T}
\end{equation}
\begin{equation}\label{Resilience_SOC}
    E_{[t]}^{Resi} = (R^{Largest,Gen} \;.\; \tau^{Resi})/\eta^{dis}, \forall\; t \in \mathcal{T}
\end{equation}
\begin{equation}\label{Blackstart_SOC}
    E_{[t]}^{BS} = (P_{[t]}^{BS} \;.\; \tau^{BS})/\eta^{dis}, \forall\; t \in \mathcal{T}
\end{equation}
\begin{equation}\label{BES_MaxPower_Gen}
    \overline {P^{}} = P^{rated}. f^{PCS}. f^{Temp,P}.f^{Ava}, \forall\; t \in \mathcal{T}
\end{equation}
\begin{equation}\label{SOC_Gen}
    SOC_{[t]} = E_{[t]}\;/\ \overline {E}_{[t]},\quad \forall\; t \in \mathcal{T}
\end{equation}
\begin{equation}\label{SOC_initial_BESS_Gen}
    E_{[t]} = SOC^{init}\;.\; \overline{E},\quad \forall\; t = t_0
\end{equation}
\begin{equation}\label{SOC_final_BESS_Gen}
    E_{[t]} = SOC^{end}\;.\; \overline{E},\quad \forall\; t = T
\end{equation}
\begin{equation}\label{FC_limits}
    \underline {P^{FC}}\;.\; \theta_{[t]}^{FC} \le P_{[t]}^{FC} \le \overline {P^{FC}} \;.\; \theta_{[t]}^{FC},\quad \forall\; t \in \mathcal{T}
\end{equation}
\begin{equation}\label{FC_power}
    E_{[t]}^{FC} = {FC}^{HR} \;.\; P_{[t]}^{FC},\quad \forall\; t \in \mathcal{T}
\end{equation}
\begin{equation}\label{FC_ramp}
     -R^{FC,Dn}_{}\le P_{[t]}^{FC} - P_{[t-1]}^{FC} \le R^{FC,Up}_{} ,\quad \forall\; t \in \mathcal{T}
\end{equation}
\begin{equation}\label{FC_reserve}
0 \le P_{[t]}^{FC,Res}  \;\leq\; \overline{P^{FC}_{}}\;.\; \theta_{[t]}^{FC} -  P_{[t]}^{FC},\quad \forall\; t \in \mathcal{T}
\end{equation}
\begin{equation}\label{FC_reserve_deliver}
P_{[t,i]}^{FC,Res} \;\leq\; R^{FC,Up}_{i}\;.\ \tau^{Delivery},\quad \forall\; t \in \mathcal{T}, i \in \mathcal{I}
\end{equation}
\begin{align}\label{min_rserve} 
\sum_{t,i} P_{[t,i]}^{GT,Res} + P_{[t]}^{FC,Res} + P_{[t]}^{BESS,Res}\\ + R^{Slack}_{[t]}\;\geq\; {R_{[t]}^{Req}},\quad \forall\; t \in \mathcal{T}, i \in \mathcal{I} \nonumber
\end{align}
\begin{equation}\label{Slack_reserve}
R^{Slack}_{[t]} \geq 0
\quad \forall t \in T
\end{equation}
\begin{equation}\label{Res_req_dicompose}
{R_{[t]}^{Req}} = {R_{[t]}^{Largest,Gen}} + {R_{[t]}^{DC,Load}} + {R_{[t]}^{Ren}}
\quad \forall t \in T
\end{equation}
\begin{equation}\label{Res_req_largestGen}
{R_{[t]}^{Largest,Gen}} = Max (\overline{P_{[i]}^{GT})}, \quad \forall t \in T, i \in \mathcal{I} 
\end{equation}
\begin{equation}\label{Res_req_DCload}
{R_{[t]}^{DC,Load}} = \alpha_L \;.\; P_{[t]}^{Load}
\quad \forall t \in T
\end{equation}
\begin{equation}\label{Res_req_Ren}
{R_{[t]}^{Ren}} = \alpha_P \;.\; \overline {P_{[t]}^{PV}} + \alpha_W \;.\; \overline {P_{[t]}^{Wind}}
\quad \forall t \in T
\end{equation}

Where equations (\ref{Power_balance1}) and (\ref{Load_shedding}) show the system level power balance and load shedding minimum and maximum allowable limits, respectively. While Grid import and export limits are also shown in equations (\ref{Grid_import}) and (\ref{Grid_export}), grid import and export ramp limits are presented in equations (\ref{Grid_import_ramp}) and (\ref{Grid_export_ramp}). These constraints ensure that imported and exported power from data centers would not result in stability issues, area control error (ACE) signal changes, etc. in case of potential grid connection, The gas turbine constraints model the operational behavior and commitment status of thermal generation units within the microgrid. Equations (\ref{GT_limit}) limits the output power of each gas turbine between its minimum and maximum generation capacities based on its on/off commitment status and its availability. (\ref{GT_Binary_transition}) and (\ref{GT_SUSD}) constraints refer to the startup and shutdown transitions of the units while preventing simultaneous startup and shutdown actions within the same time interval. Ramp rate constraints are shown in equations (\ref{GT_ramp_up}) and (\ref{GT_ramp_down}) limits the change in power output between consecutive time intervals to ensure realistic and stable turbine operation. Minimum uptime and minimum downtime of gas turbines are shown in equations (\ref{min_UT}) and (\ref{min_DT}), respectively. Equations (\ref{GT_reserve}) and (\ref{GT_reserve_deliver}) show the reserve headroom of gas turbines which is crucial for joint energy and reserve optimization, and the response time to deliver that reserve, respectively. 


The Solar PV constraints define the operational limits and curtailment behavior of these systems within the microgrid. The equation (\ref{PV_limit}) limits the PV power output based on the available solar generation at each time interval. Constraint (\ref{PV_crtl}) models PV curtailment by representing the difference between the available and utilized solar power. Wind constraints are also shown in equations (\ref{Wind_limit})-(\ref{Wind_crtl}) with explanations similar to solar constraints.

The BESS constraints, presented in equations (\ref{BES_charge_Gen})-(\ref{SOC_final_BESS_Gen}), show the charging, discharging, and behavior of the BESS within the BTM data center microgrid. The charging and discharging constraints, presented in equations (\ref{BES_charge_Gen}) and (\ref{BES_discharge_Gen}), limit the BESS power based on its maximum operating capacities while preventing simultaneous charging and discharging through a binary variable. Equation (\ref{BESS_reserve}) shows the BESS power reserve for joint energy and reserve dispatch scenarios, and the energy capacity balance equation (\ref{BES_SOC_Gen}) updates the stored energy level by considering charging and discharging efficiencies over time. In addition, in equation (\ref{BES_SOC_limit_Gen}), SOC limits are imposed to maintain BESS energy within its allowable energy capacity range and ensure reliable operation by storing energy for reserve. Equations (\ref{BES_MaxEnergy_limit_Gen}) and (\ref{BES_MaxPower_Gen}) show the factors that affect both energy and power rating of BESS to have a reliable operation till end life of BESS. Equations (\ref{SOC_Gen}) - (\ref{SOC_final_BESS_Gen}) refer to SOC calculation and initial and end states of BESS SOC. 

FC constraints are presented in equations (\ref{FC_limits}) through (\ref{FC_reserve_deliver}) where show FC power generation limits, energy calculation based on heat rate, ramping limits, and reserve headroom, reserve delivery over delivery response time, respectively. Final system level reserve requirement constraints are presented in equations (\ref{min_rserve})-(\ref{Res_req_Ren}) where the reserve requirement calculated based on largest generator capacity, data center load error, and renewables uncertainty error with appropriate weights for both solar and wind(obtained from a complete historical data analysis). 

\begin{figure*}[]
\center
\footnotesize
\captionsetup{font={footnotesize}}
\begin{tabular}{cc}
\subfloat[Annual ERCOT LMP prices] {\includegraphics[height=1.8in,width=2.1in]{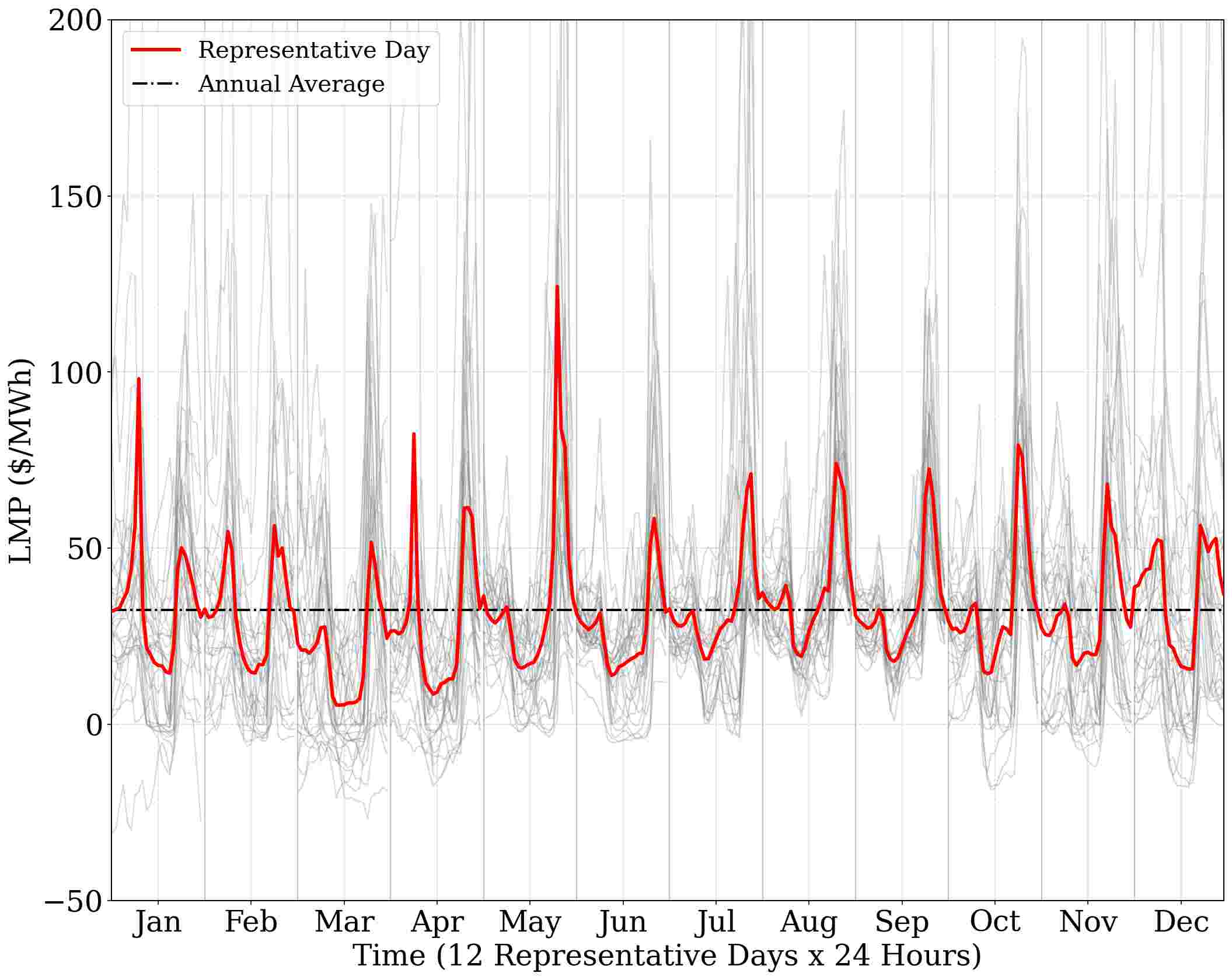}}\label{LMP_profilee}
\subfloat[ERCOT Solar power profile]{\includegraphics[height=1.8in,width=2.1in]{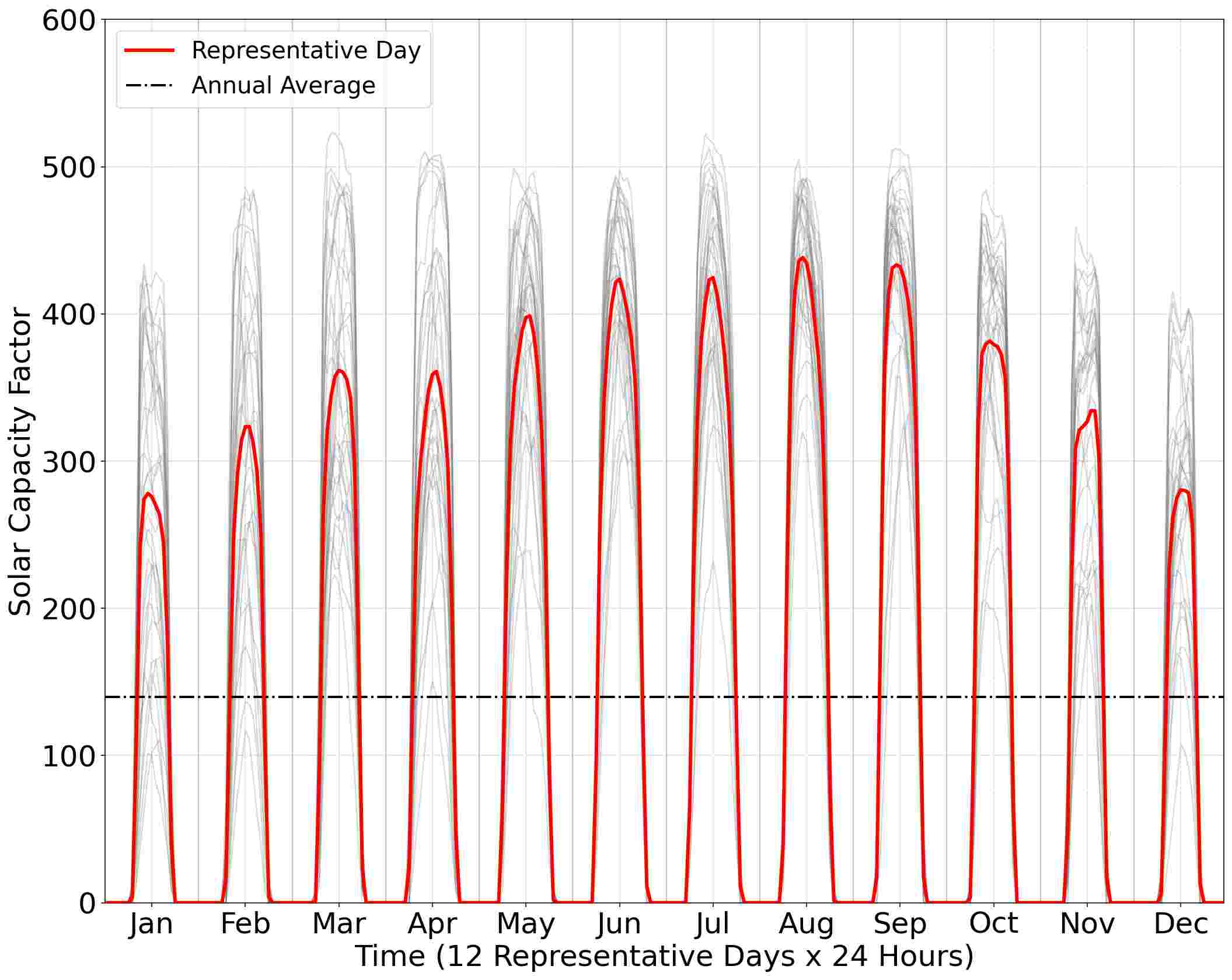}}\label{Solar_profile} 
\subfloat[ERCOT Wind power profile] {\includegraphics[height=1.8in,width=2.1in]{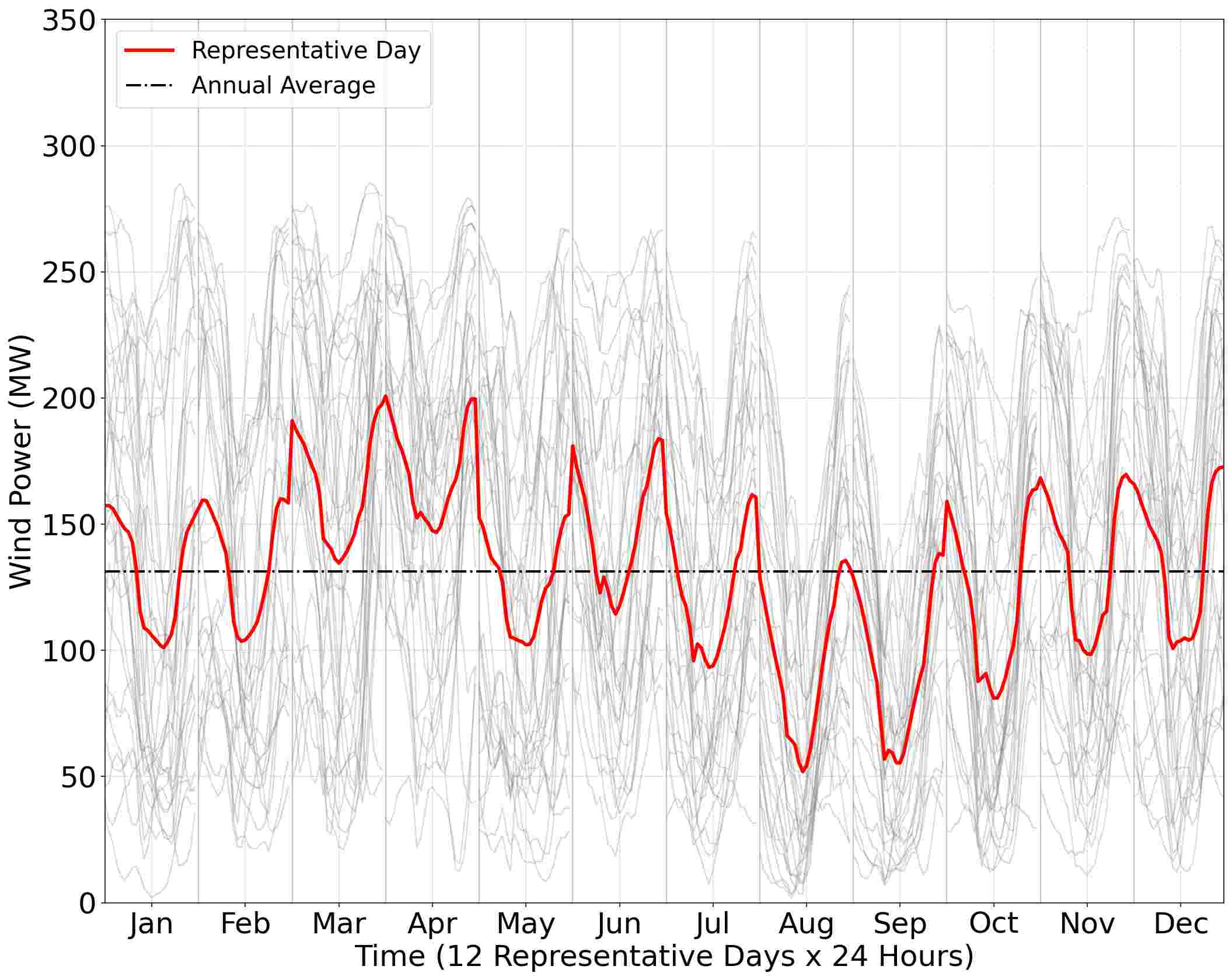}}\label{Wind_profile} 
\end{tabular}
\caption{Year 2025 ERCOT LMP, solar, and wind data and final annual representative profile.}
\label{System_input_profiles}
\end{figure*}

\vspace{1.5mm}
\section{Simulation Results and Numerical Analysis}\label{Results}


\begin{table}[!t]
\captionsetup{justification=raggedright,singlelinecheck=false,font={footnotesize}}
\caption{Techno-Economic Parameters of the BESS.}
\label{BESS_Parameters_Table}
\centering
\scriptsize
\setlength{\tabcolsep}{2.2pt}
\renewcommand{\arraystretch}{1.08}

\begin{tabularx}{\columnwidth}{
>{\centering\arraybackslash}p{0.16\columnwidth}
>{\centering\arraybackslash}X
>{\centering\arraybackslash}p{0.18\columnwidth}
>{\centering\arraybackslash}p{0.12\columnwidth}
}
\toprule
\textbf{Category} & \textbf{Parameter} & \textbf{Unit} & \textbf{Value} \\
\midrule

\multirow{13}{*}{\rotatebox[origin=c]{0}{Technical}}
& Charge/Discharge Efficiency & p.u. & 0.924 \\
& PCS factor & p.u. & 0.980 \\
& AC loss factor & p.u. & 0.980 \\
& Temperature derate factor & p.u. & 0.995 \\
& Minimum SOC & p.u. & 0.100 \\
& Maximum SOC & p.u. & 0.900 \\
& Initial SOC & p.u. & 0.100 \\
& Final SOC & p.u. & 0.100 \\
& Storage duration & h & 2, 4 \\
& Lifetime & years & 20 \\
& End-of-life factor & p.u. & 0.649 \\
& Availability factor & p.u. & 0.990 \\
& Reserve margin & \% & 10 \\
\midrule

\multirow{4}{*}{\rotatebox[origin=c]{0}{Economic}}
& Power capital cost & \$/kW & 312 \\
& Energy capital cost & \$/kWh & 297 \\
& Fixed O\&M cost & \$/kW-year & 34 \\
& Discount rate & \% & 6 \\

\bottomrule
\end{tabularx}
\end{table}

\begin{table}[]
\captionsetup{justification=raggedright, singlelinecheck=false, font={footnotesize}}
\caption{Techno-Economic Parameters of the Gas Turbines.}
\label{GTs_TE_Parameters}
\centering
\scriptsize
\setlength{\tabcolsep}{1.4pt}
\renewcommand{\arraystretch}{1.08}
\begin{tabularx}{\columnwidth}{>{\centering\arraybackslash}p{0.14\columnwidth} >{\centering\arraybackslash}X >{\centering\arraybackslash}p{0.16\columnwidth} >{\centering\arraybackslash}p{0.10\columnwidth} >{\centering\arraybackslash}p{0.10\columnwidth} >{\centering\arraybackslash}p{0.10\columnwidth}}
\toprule
\textbf{Category} & \textbf{Parameter} & \textbf{Unit} & \multicolumn{3}{c}{\textbf{Value}} \\
\cmidrule(lr){4-6}
& & & \textbf{GT 1} & \textbf{GT 2} & \textbf{GT 3} \\
\midrule
\multirow{11}{*}{Technical}
& Maximum power capacity & MW & 100 & 180 & 225 \\
& Minimum stable generation & MW & 35 & 75 & 90 \\
& Ramping rate & MW/min & 40 & 80 & 100 \\
& Startup ramp rate & MW/min & 40 & 80 & 100 \\
& Shutdown ramp rate & MW/min & 40 & 80 & 100 \\
& Minimum uptime & h & 1 & 2 & 4 \\
& Minimum downtime & h & 1 & 2 & 4 \\
& Heat rate & Btu/kWh & 11500 & 8400 & 6800 \\
& Availability factor & p.u. & 0.98 & 0.98 & 0.97 \\
\midrule
\multirow{4}{*}{Economic}
& Startup cost & \$/start & 4500 & 10000 & 20000 \\
& Shutdown cost & \$/shutdown & 400 & 1000 & 2000 \\
& Fixed O\&M cost & \$/kW-year & 10 & 15 & 18 \\
& Variable O\&M cost & \$/MWh & 5.5 & 5 & 4.5 \\
& Fuel price & \$/MMBtu & 3.5 & 3.5 & 3.5 \\
\bottomrule
\end{tabularx}
\end{table}

Given that a substantial proportion of future hyperscale data centers are expected to be deployed in Texas, West Texas was selected as the representative study region. Consequently, regional climatic and geographic characteristics, including ambient temperature conditions, were incorporated into the analysis \cite{West_texas_temp}. The proposed data center microgrid consists of a 600-MW solar field, a 400-MW wind farm, five gas turbine generators (one SCGT of type GT1, three SCGTs of type GT2, and one CCGT of type GT3), a FC power plant including 26 modules (each 12.5 MW) and total capacity of 325 MW and 10\% ramp rate limit, and two BESS systems, with one installed on the generation side and the other on the load side. To accurately represent renewable energy variability, ERCOT 2025 historical generation data were used to develop the solar and wind production profiles \cite{ercot_generation_2025}. $\alpha_L$, $\alpha_P$, and $\alpha_W$ were considered 5\%, 5\% and 3\% respectively to account for uncertainty error on total reserve amount. Considering BESS charging and discharging equations, the product of a continuous variable and a binary variable results in a non-convex formulation. To maintain computational tractability, this bilinear term is linearized using the Big-M reformulation method presented in \cite{haggi2022proactive} and \cite{wu2016exact}.
The techno-economic optimization inputs for the gas turbines and BESSs are (based on engineering experiences and validated by information provided by NREL ATB \cite{nrel2025atb}) summarized in Table \ref{GTs_TE_Parameters} and Table \ref{BESS_Parameters_Table}, respectively. The proposed framework is primarily intended for BTM data center microgrids. Nevertheless, grid-interaction variables, including power import/export and market-based pricing signals, are incorporated to maintain a generalized formulation and support future grid-connected studies. For the BTM case studies presented in this paper, the corresponding constraints and objective function terms are deactivated. Presented in Figure (\ref{System_input_profiles}a) representative ERCOT LMP profiles are included to establish a baseline for future market participation analyses and to compare the economic value of BTM generation against actual grid prices. The proposed formulation can therefore be readily extended to grid-integrated data center microgrids without modification to the underlying optimization framework.

Since the generation-side dispatch optimization is performed at an hourly resolution, the Sample Average Approximation (SAA) method \cite{kim2014guide}\cite{haggi2022hydrogen} was employed to convert the original annual time-series data into representative monthly profiles. Specifically, a single 24-hour representative profile was generated for each month, resulting in a reduced optimization horizon of 288 hours, lowering the computational burden, while preserving the seasonal changes of the underlying data. The resulting representative profiles are shown in Figures (\ref{System_input_profiles}b), and (\ref{System_input_profiles}c), where the gray curves represent the original daily profiles and the red curves denote the corresponding monthly representative profiles used in the optimization framework. 

To ensure reliable data center operation, several penalty terms were considered in the objective function of the optimizations to reflect operational priorities. Given the critical nature of BTM data center operation, load shedding is assigned a value of lost load (VOLL) of \$,1000,000 /MWh and reserve penalty of \$500,000 /MWh to strongly discourage unserved energy. Solar and wind curtailments are also penalized at \$1,000/MWh to maximize the usage of low-cost renewable resources. Furthermore, a ramp-rate penalty of \$10,000/MWh is introduced to discourage rapid power fluctuations and promote smoother system operation. Collectively, these penalty factors prioritize reliability, renewable energy utilization, and load-smoothing performance within the proposed microgrid framework.

\begin{figure}[]
\centering
\footnotesize
\captionsetup{justification=raggedright, singlelinecheck=false, font={footnotesize}}
\includegraphics[width=3.4in]{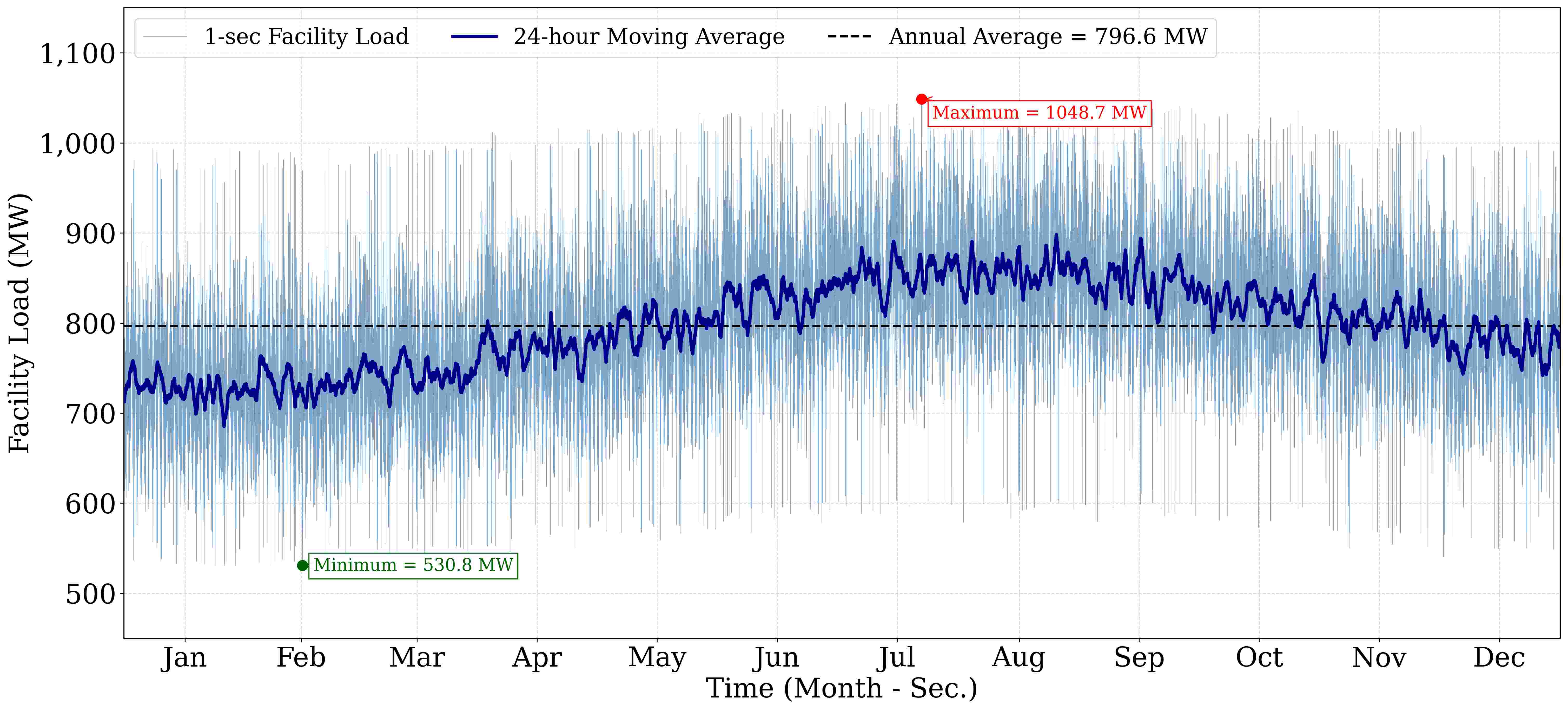}
\caption{Annual data center facility load with stochastic bursts and seconds resolution.}
\label{DC_load_sec}
\end{figure}

The proposed optimization frameworks (load smoothing, and operational planning of BESS) are formulated as mixed integer linear programs (MILP). Simulations were carried out on a system with an Intel(R) Core (TM) Ultra 7 CPU of 2.2 GHz, and 32 GB RAM. The proposed framework was coded on Python \cite{PythonRef} and the optimization problems were solved using the Gurobi solver with a termination gap of 0.01\% \cite{gurobi}.

\subsection{Data Center Load Modeling Results:}

\begin{figure*}[]
\centering
\footnotesize
\captionsetup{font={footnotesize}}
\setlength{\tabcolsep}{2pt}
\renewcommand{\arraystretch}{1.0}
\begin{tabular}{ccc}
\subfloat[Data center facility load\label{facility_decom}]{\includegraphics[width=2\textwidth,height=1.8in,keepaspectratio]{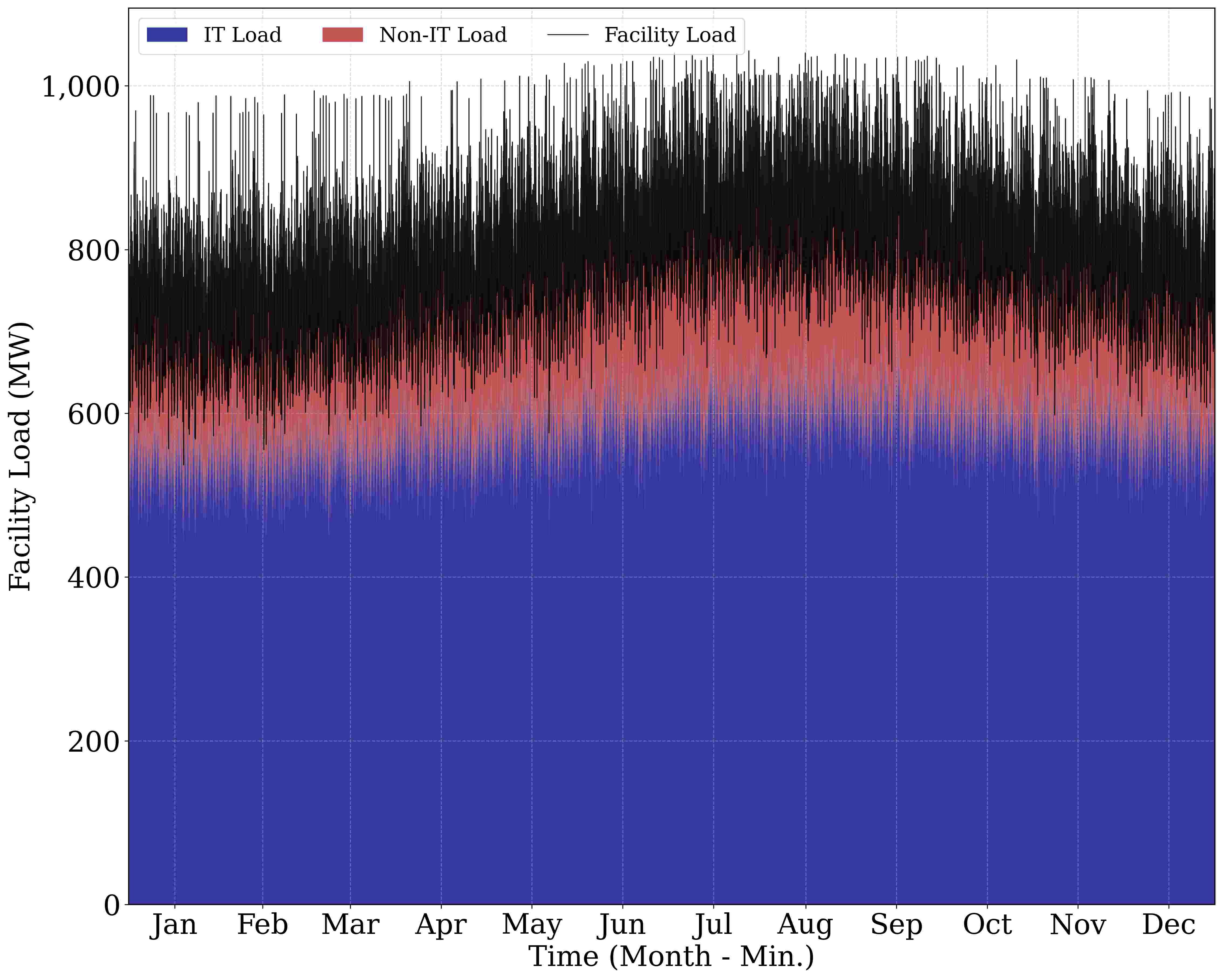}} &
\subfloat[IT load decomposition\label{IT_decom}]{\includegraphics[width=2\textwidth,height=1.8in,keepaspectratio]{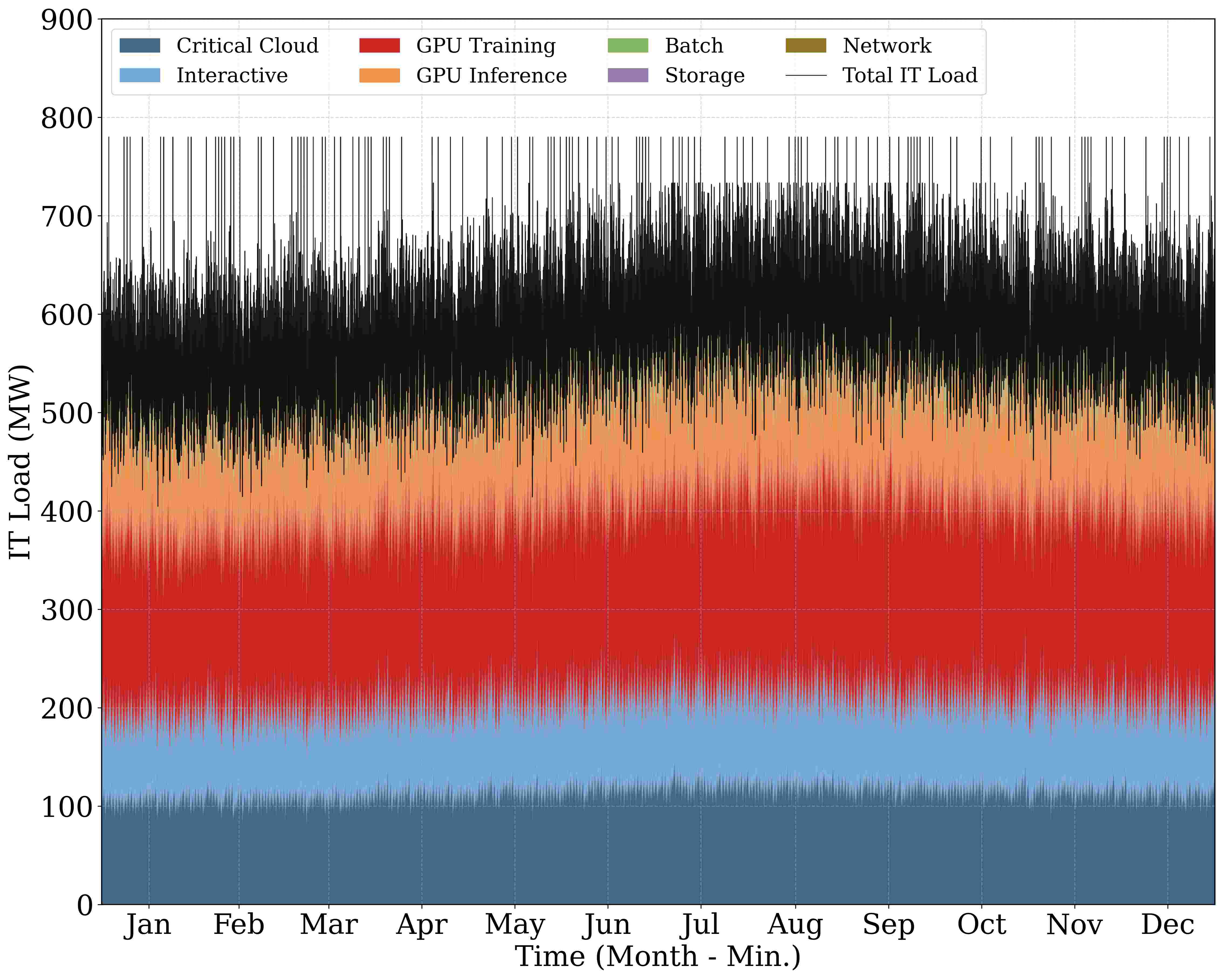}} &
\subfloat[Non-IT load decomposition\label{Non_IT_decom}]{\includegraphics[width=2\textwidth,height=1.8in,keepaspectratio]{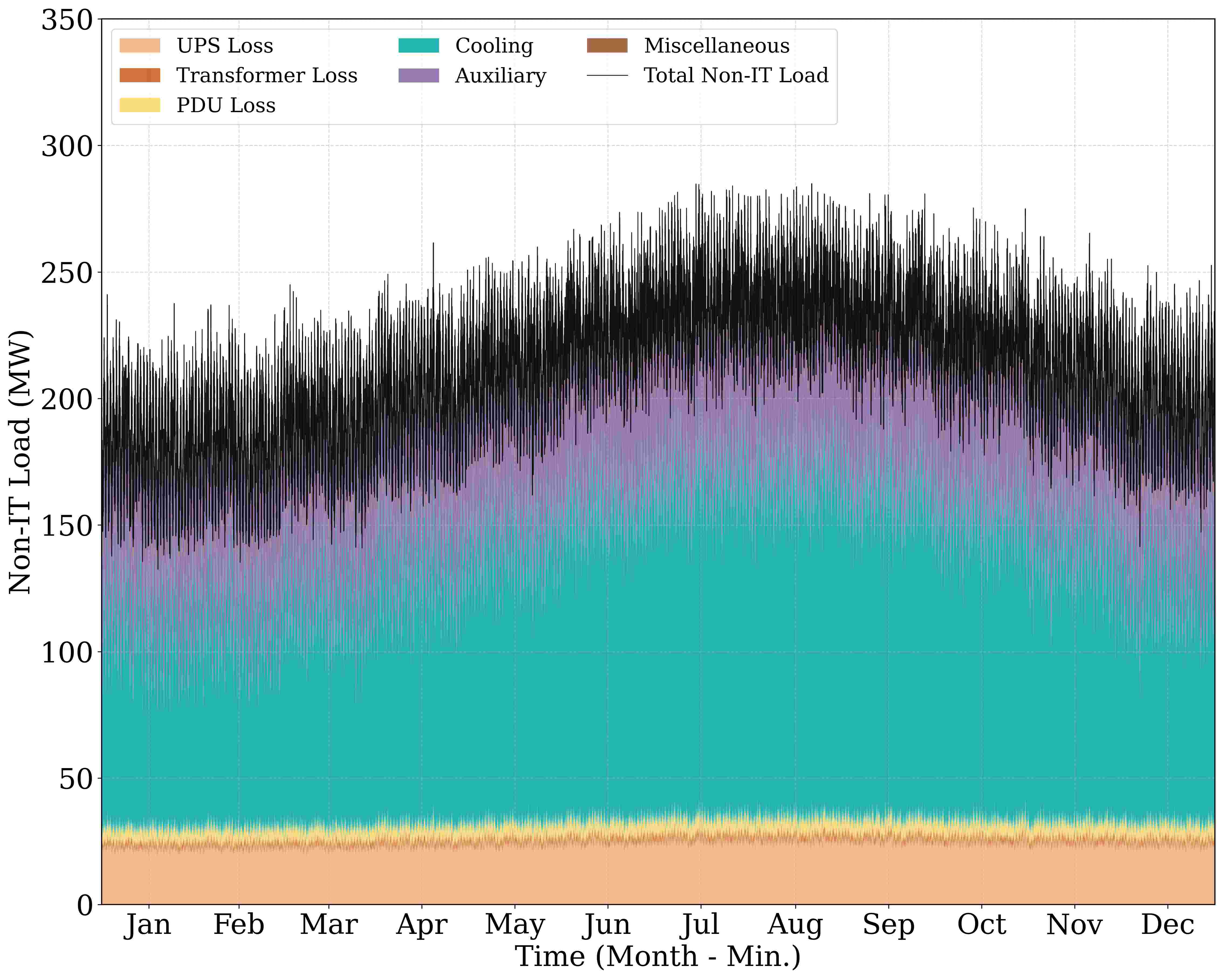}} \\
\subfloat[Data center building's workload\label{Building_decom}]{\includegraphics[width=2\textwidth,height=1.8in,keepaspectratio]{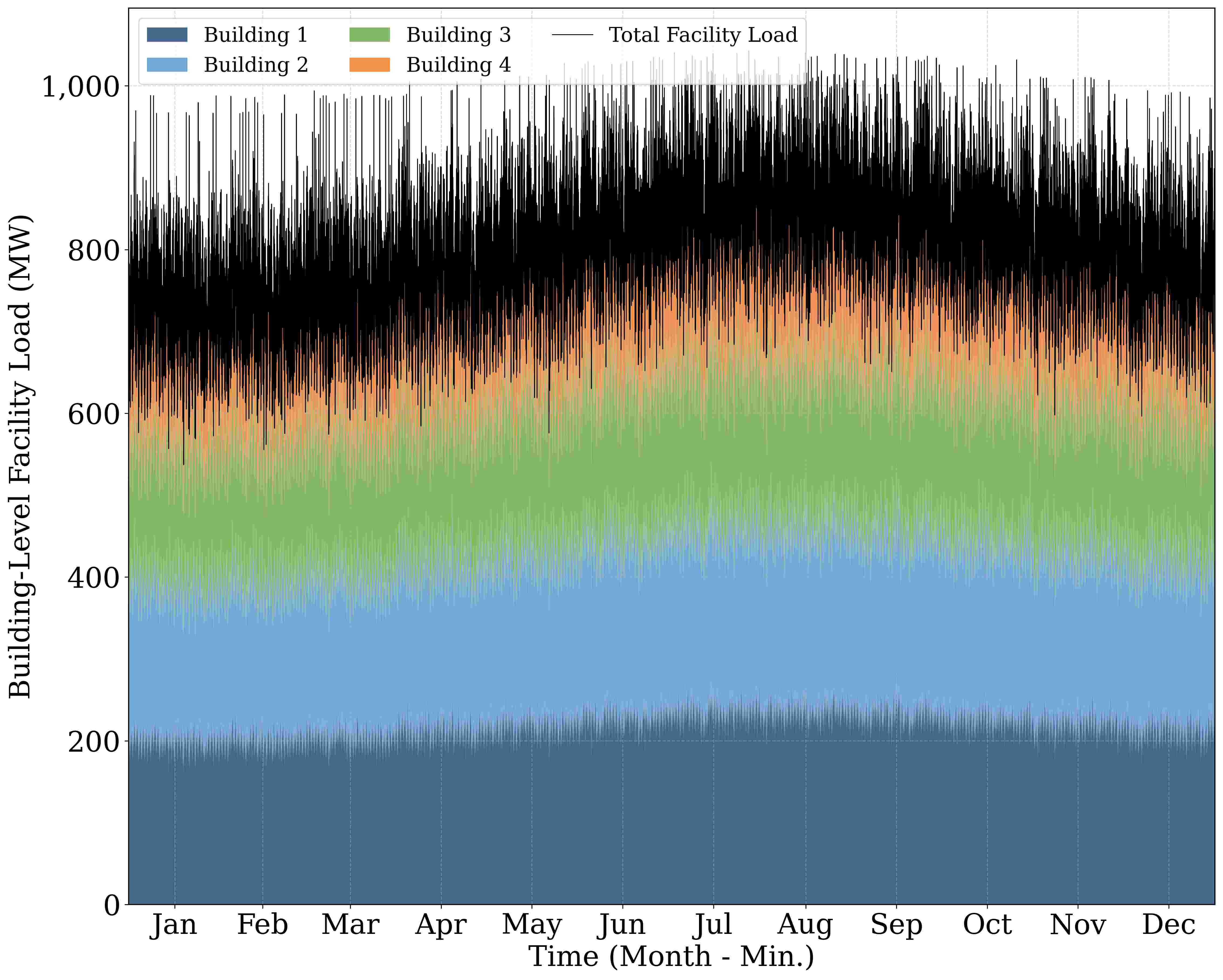}} &
\subfloat[IT load workload profiles\label{IT_profile}]{\includegraphics[width=2\textwidth,height=1.8in,keepaspectratio]{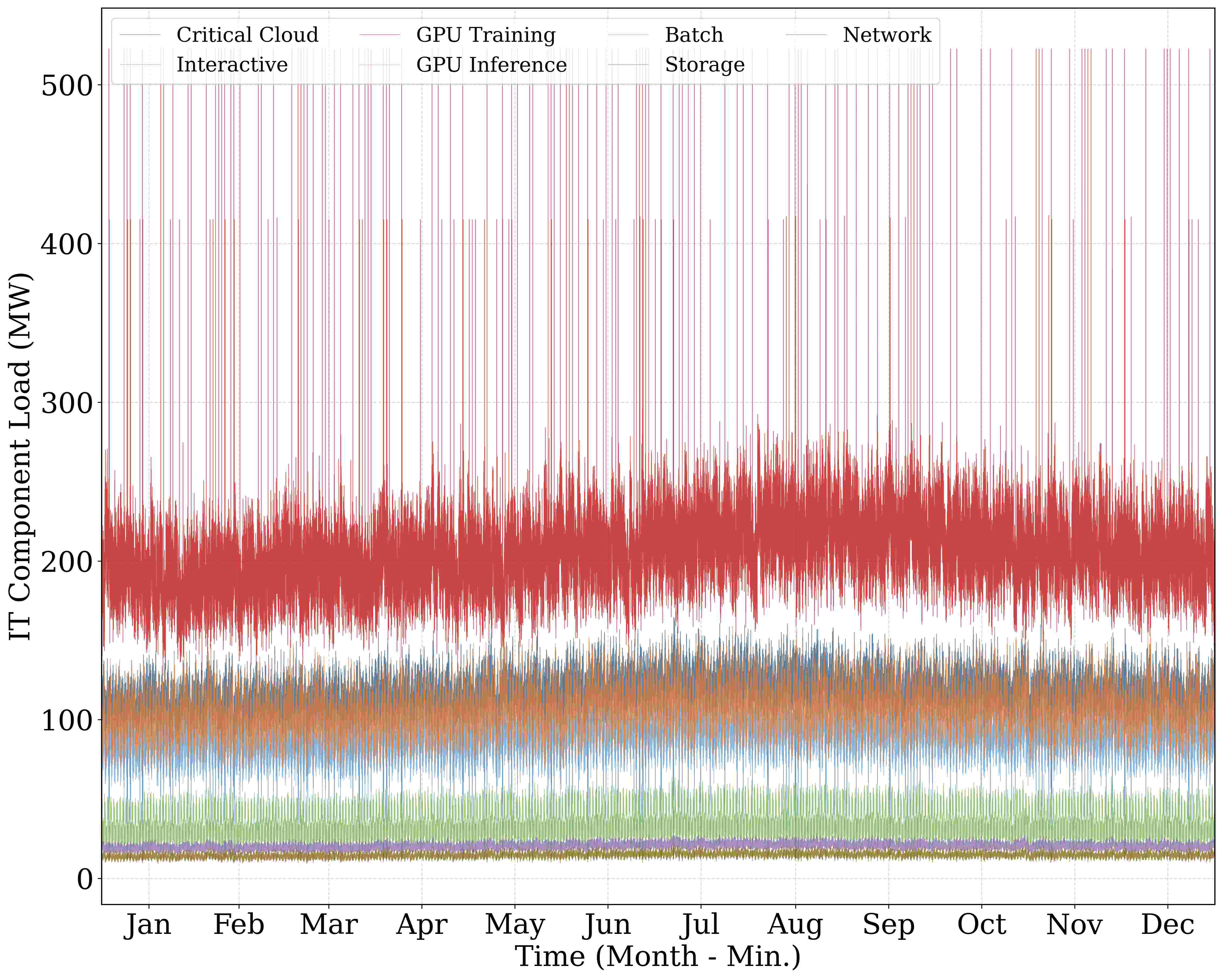}} &
\subfloat[Non-IT load workload profiles\label{Non_IT_profile6}]{\includegraphics[width=2\textwidth,height=1.8in,keepaspectratio]{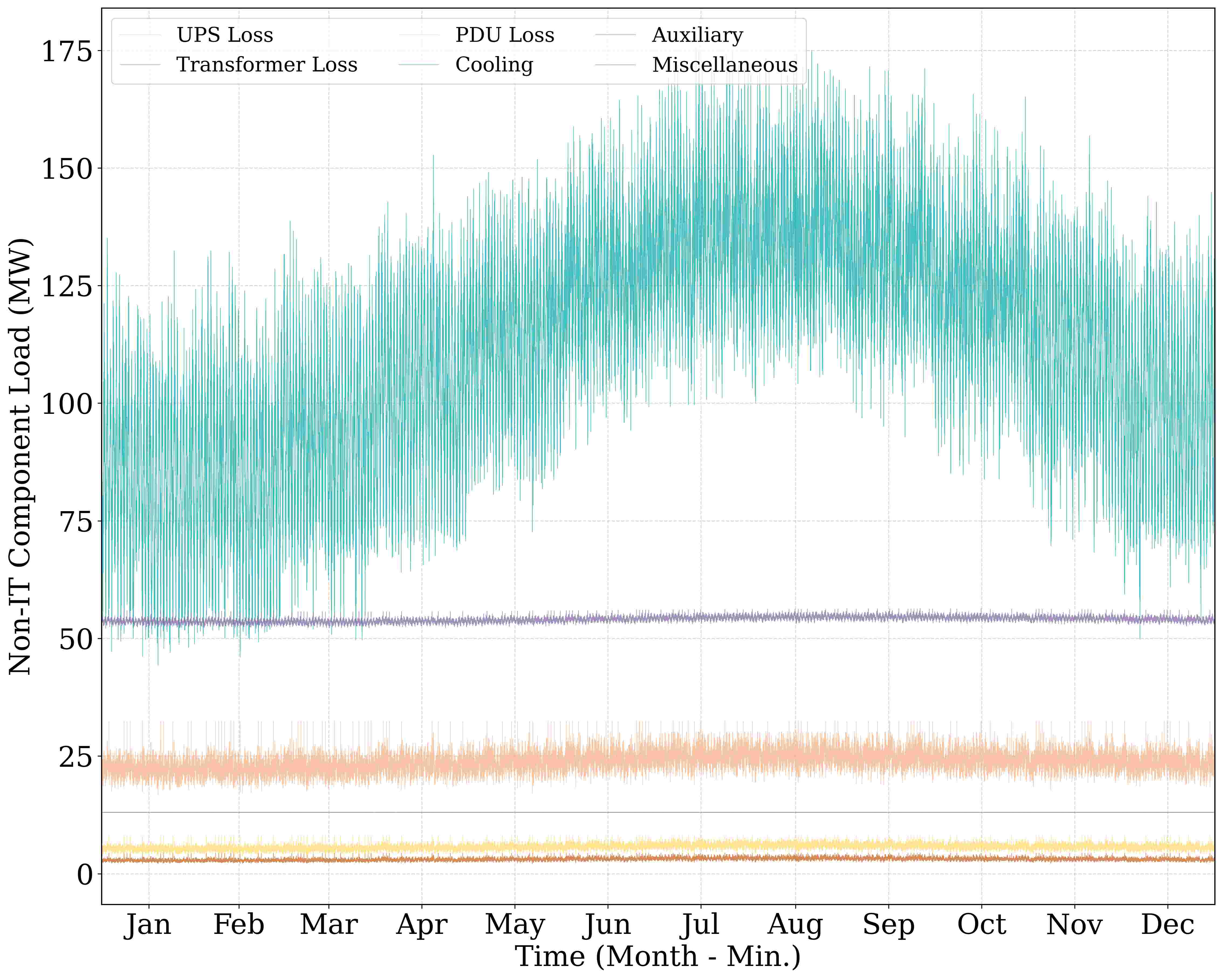}} \\
\subfloat[Monthly PUE range\label{PUE_boxplot7}]{\includegraphics[width=2\textwidth,height=1.8in,keepaspectratio]{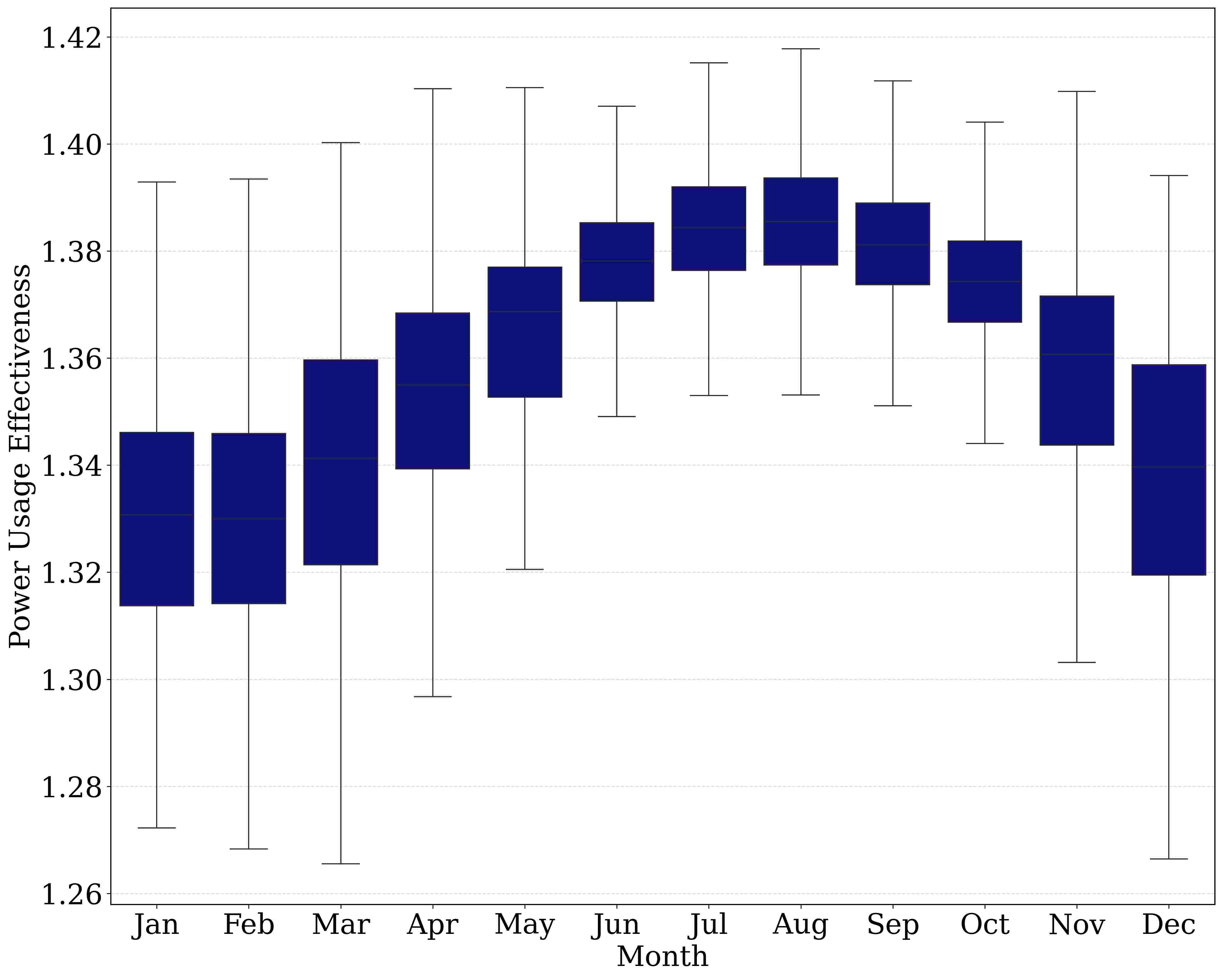}} &
\subfloat[Monthly average IT load components\label{Monthly_IT}]{\includegraphics[width=2\textwidth,height=1.8in,keepaspectratio]{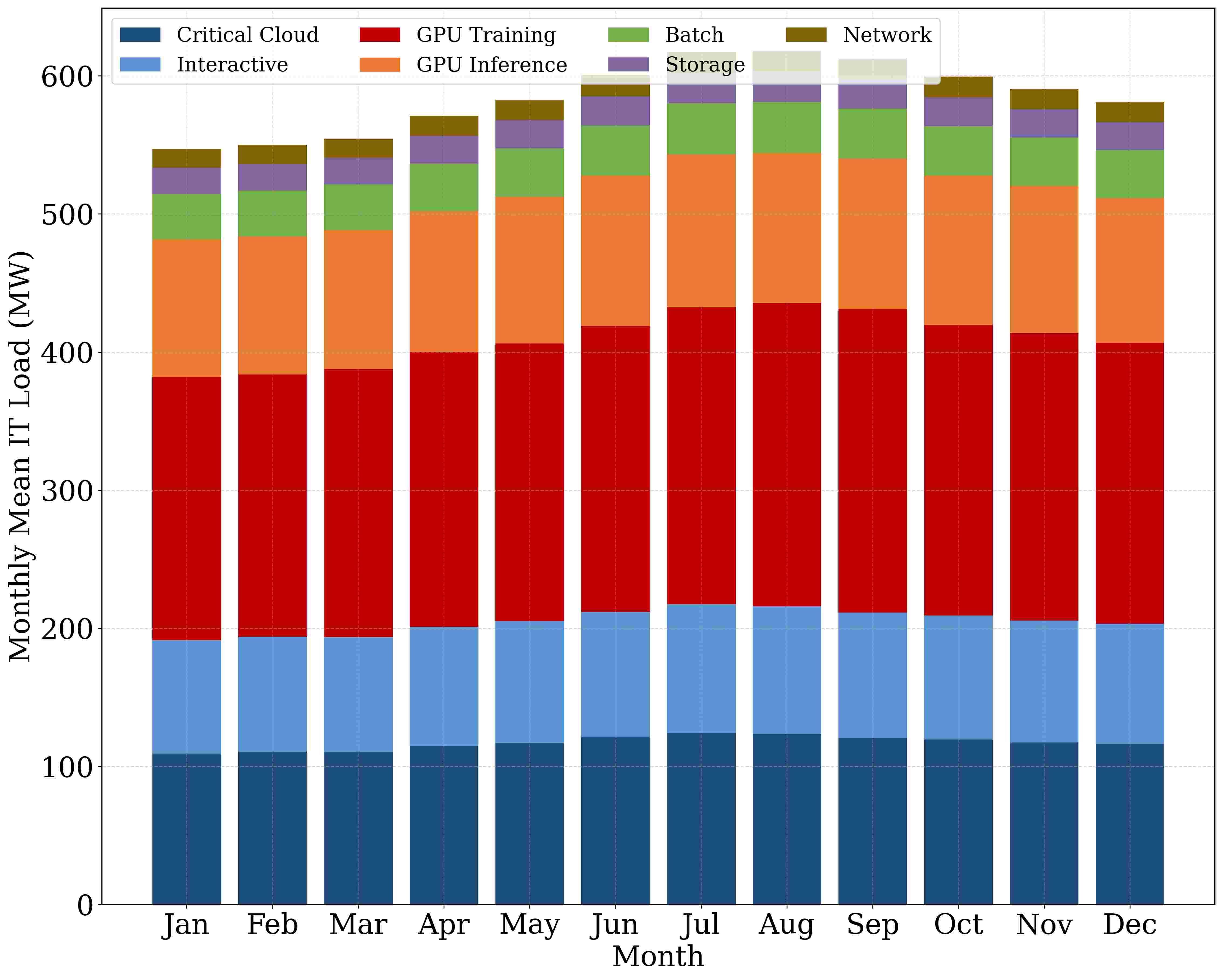}} &
\subfloat[Monthly average Non-IT load components\label{Monthly_NonIT}]{\includegraphics[width=2\textwidth,height=1.8in,keepaspectratio]{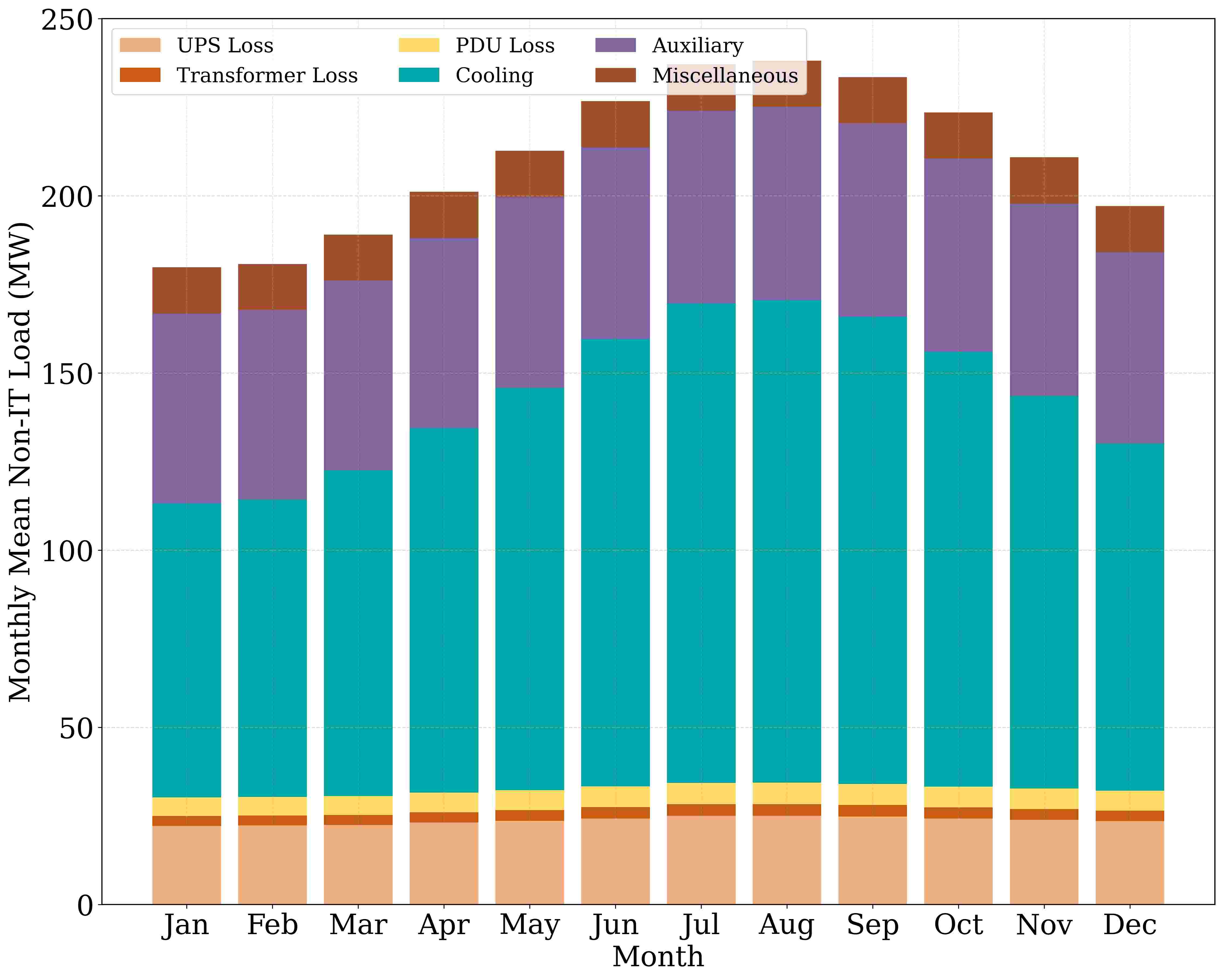}} \\
\subfloat[Annual energy by workload category\label{Energy_decom}]{\includegraphics[width=2\textwidth,height=1.8in,keepaspectratio]{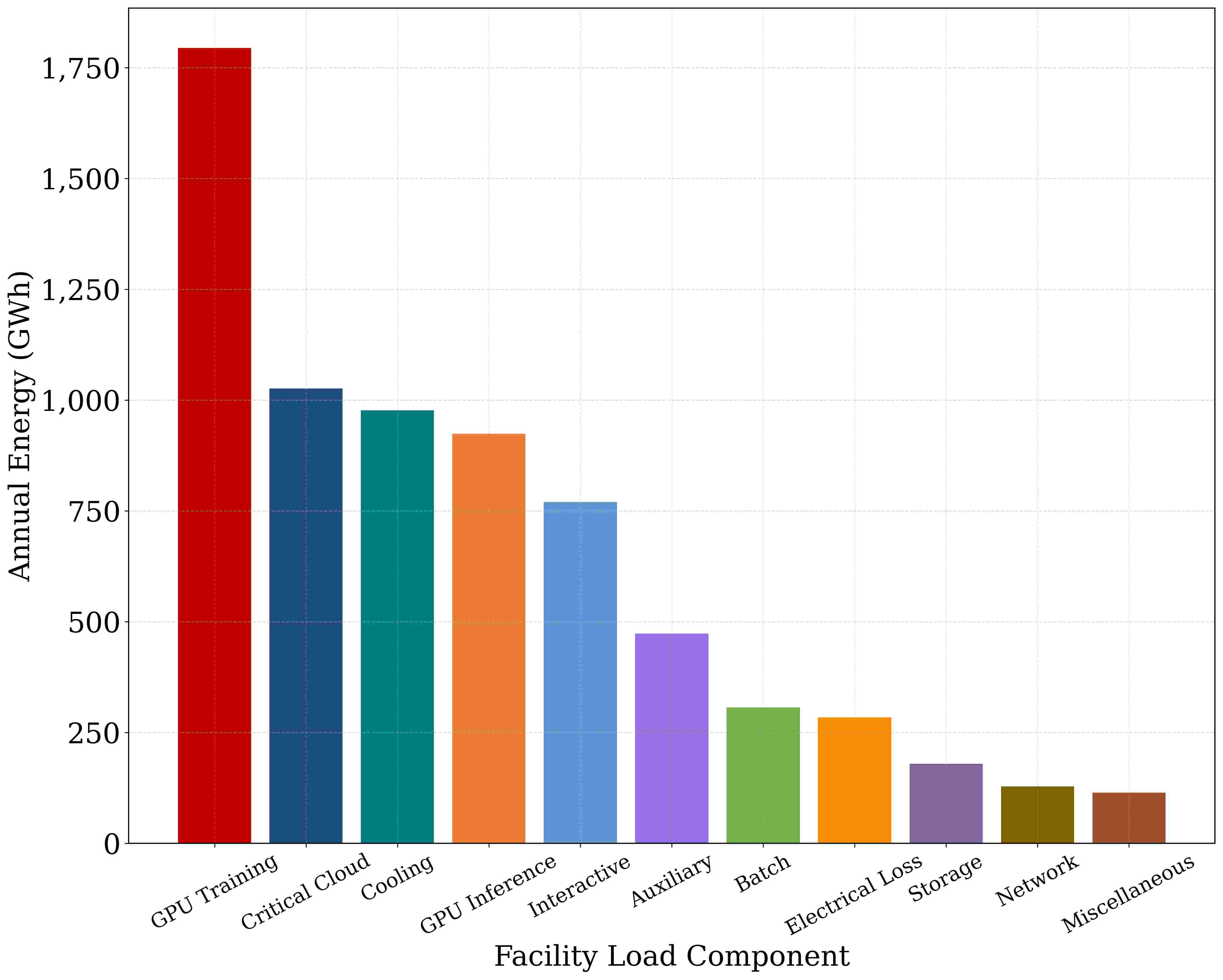}} &
\subfloat[PUE vs IT load and temperature\label{PUE_colorbar}]{\includegraphics[width=2\textwidth,height=1.8in,keepaspectratio]{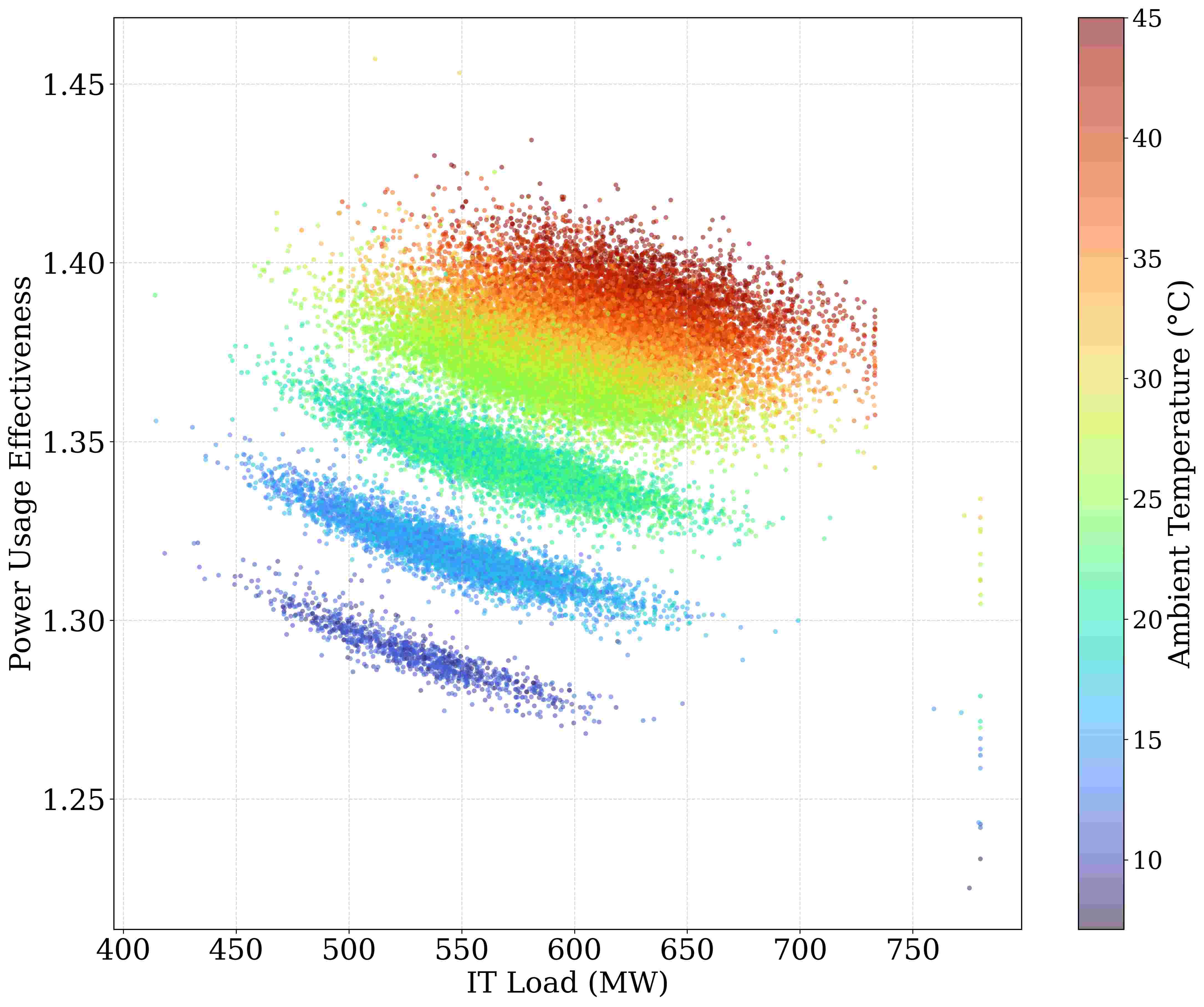}} &
\subfloat[Cooling load vs temperature\label{Cooling_colorbar}]{\includegraphics[width=2\textwidth,height=1.8in,keepaspectratio]{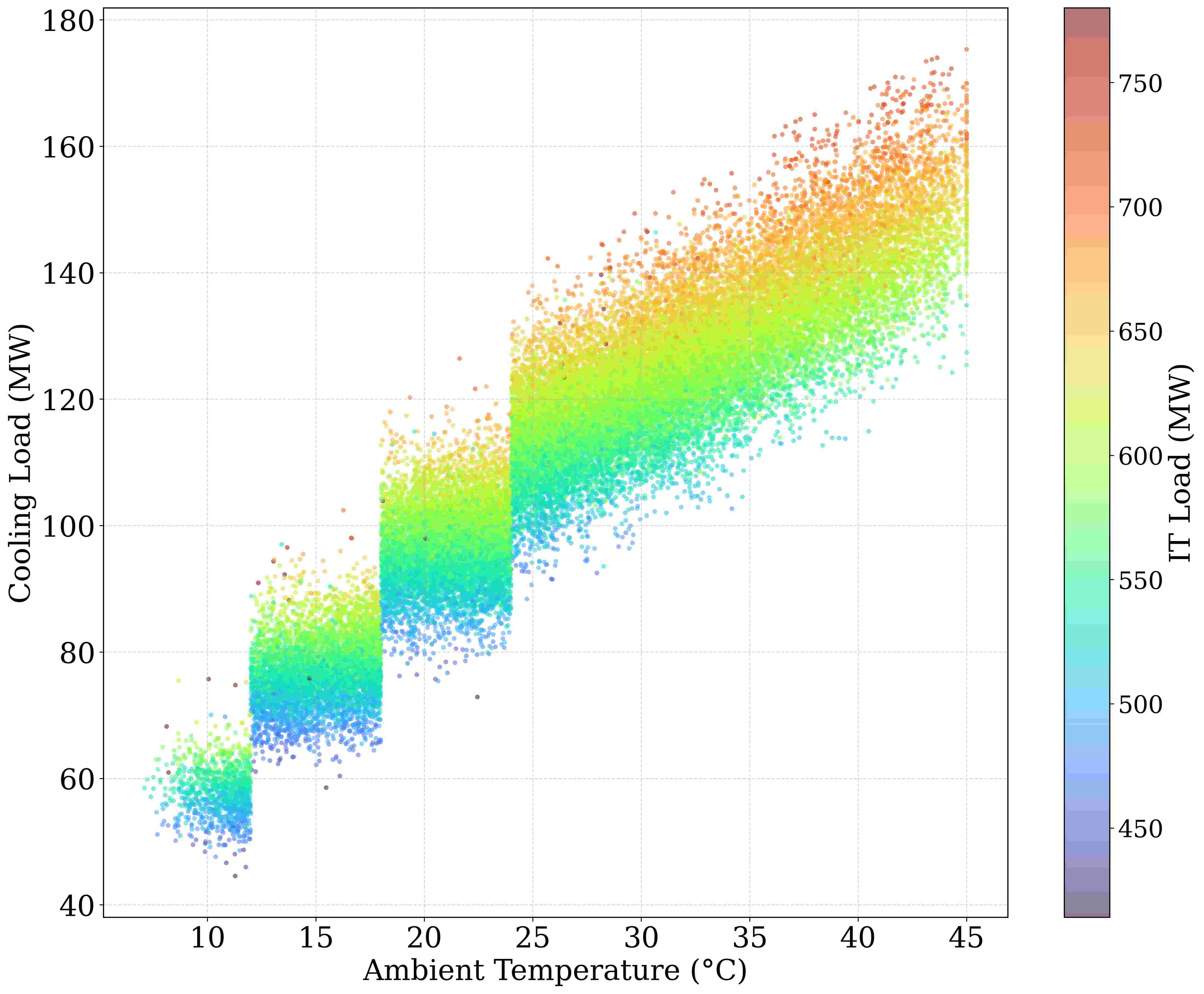}}
\end{tabular}
\caption{Data center load modeling results.}
\label{DC_load_figures}
\end{figure*}

Figure (\ref{DC_load_sec}) presents the annual facility load profile at 1-second resolution, demonstrating the ability of the proposed digital twin to simultaneously capture long-term load evolution and short-duration stochastic variability. The simulated facility exhibits an annual average demand of 796.6 MW, with loading conditions ranging from 530.8 MW to 1048.7 MW. The 24-hour moving average reveals a clear seasonal pattern, with facility demand increasing from winter toward summer, reaching its highest sustained operating levels between June and September before gradually declining. This trend is consistent with the model formulation, where AI workload intensity increases during mid-year periods while higher ambient temperatures elevate cooling-system consumption and infrastructure losses. In contrast to the smooth seasonal trajectory, the second-level profile preserves significant high-frequency fluctuations throughout the year, reflecting the stochastic workload-generation framework embedded within the digital twin. These variations are produced by the combined effects of mean-reverting workload dynamics, random job arrivals, workload migrations, GPU scheduler activities, cluster-shedding events, and second-scale autoregressive noise. The resulting profile demonstrates that future AI-oriented hyperscale facilities may operate across a wide dynamic range rather than around a single steady-state operating point. The coexistence of seasonal trends, stochastic variability, and transient peak events indicates that both energy adequacy and short-term flexibility must be considered when evaluating grid interconnections, BESS sizing, and microgrid operation for large-scale data centers.

Data center load modeling detailed results are shown in Figures (\ref{facility_decom})-(\ref{Cooling_colorbar}) with minute-level resolution (for better visualization compared to second based resolution). These figures highlights several important characteristics of the proposed data center digital twin beyond simple load profile. First, the model preserves the expected hierarchy between IT and non-IT demand, where IT loads dominate total facility consumption while non-IT systems govern a large portion of the seasonal variability. As shown in Figures (\ref{facility_decom})-(\ref{Non_IT_decom}), annual fluctuations in total facility demand are primarily associated with changes in cooling requirements rather than changes in the underlying workload composition. This behavior emerges naturally from the model structure, where cooling demand depends on both IT-generated heat and ambient-temperature-dependent COP degradation. Consequently, the facility load envelope broadens during summer periods even though the IT subsystem remains the dominant consumer. On the other hand, GPU-centric workloads represent the primary source of computational variability. Figures (\ref{IT_decom}), (\ref{IT_profile}), and (\ref{Energy_decom}) show that GPU training consistently accounts for the largest share of IT demand and annual energy consumption. This is a direct consequence of the workload allocation assumptions, where GPU training receives the highest nominal share and is further influenced by stochastic job arrivals, synchronized scheduler events, workload bursts, and transient stress scenarios. In contrast, cloud, storage, and network workloads form a comparatively stable base load. The resulting behavior resembles modern AI-oriented hyperscale facilities where a relatively predictable service layer is superimposed with highly dynamic GPU clusters that drive peak demand and short-term variability.

The results also demonstrate that non-IT demand cannot be accurately represented using a fixed percentage of IT load. Figures (\ref{Non_IT_decom}), (\ref{Non_IT_profile6}), and (\ref{Monthly_NonIT}) indicate that cooling demand exhibits a substantially stronger seasonal response than any other infrastructure component. This behavior arises from the combined effects of ambient-temperature variation, nonlinear cooling-system efficiency, economizer operation, thermal inertia, and load-dependent auxiliary equipment. Consequently, a modest increase in IT utilization can produce a disproportionately larger increase in cooling demand during warm operating periods. The distinct operating regions visible in the cooling-load distribution further indicate transitions between different cooling-efficiency regimes rather than a simple linear relationship between IT power and cooling power. In addition, Figures (\ref{PUE_boxplot7} ) and (\ref{PUE_colorbar}) show that PUE follows a clear seasonal trend, with higher values occurring during warmer periods when cooling-system performance deteriorates. Although increasing IT utilization improves the use of fixed infrastructure assets, the scatter patterns suggest that ambient temperature introduces a larger shift in PUE than workload magnitude alone. The temperature-dependent clustering observed in Figure (\ref{PUE_colorbar}) reflects the combined effects of operation, COP degradation, and thermal-response dynamics embedded within the model. This result emphasizes that facility efficiency is fundamentally a coupled workload-environment phenomenon rather than a fixed design parameter.

Finally, the consistency across annual, monthly, and subsystem-level representations demonstrates the internal coherence of the digital twin framework. Building-level aggregation, workload decomposition, annual energy allocation, and temperature-dependent cooling behavior all preserve the same underlying trends without introducing contradictory operating patterns. Figures (\ref{Monthly_IT})-(\ref{Cooling_colorbar}) show that GPU training and cooling emerge as the dominant contributors to energy consumption, while the strong correlation between ambient temperature, cooling demand, and PUE confirms that the simulated facility responds realistically to both computational and environmental drivers. Collectively, these results indicate that the proposed framework captures not only the stochastic behavior of hyperscale AI workloads, but also the thermodynamic and infrastructure interactions required for high-fidelity power-system, microgrid, and BESS-planning studies.

\begin{figure}[]
\centering
\footnotesize
\captionsetup{justification=raggedright, singlelinecheck=false, font={footnotesize}}
\includegraphics[width=3.4in]{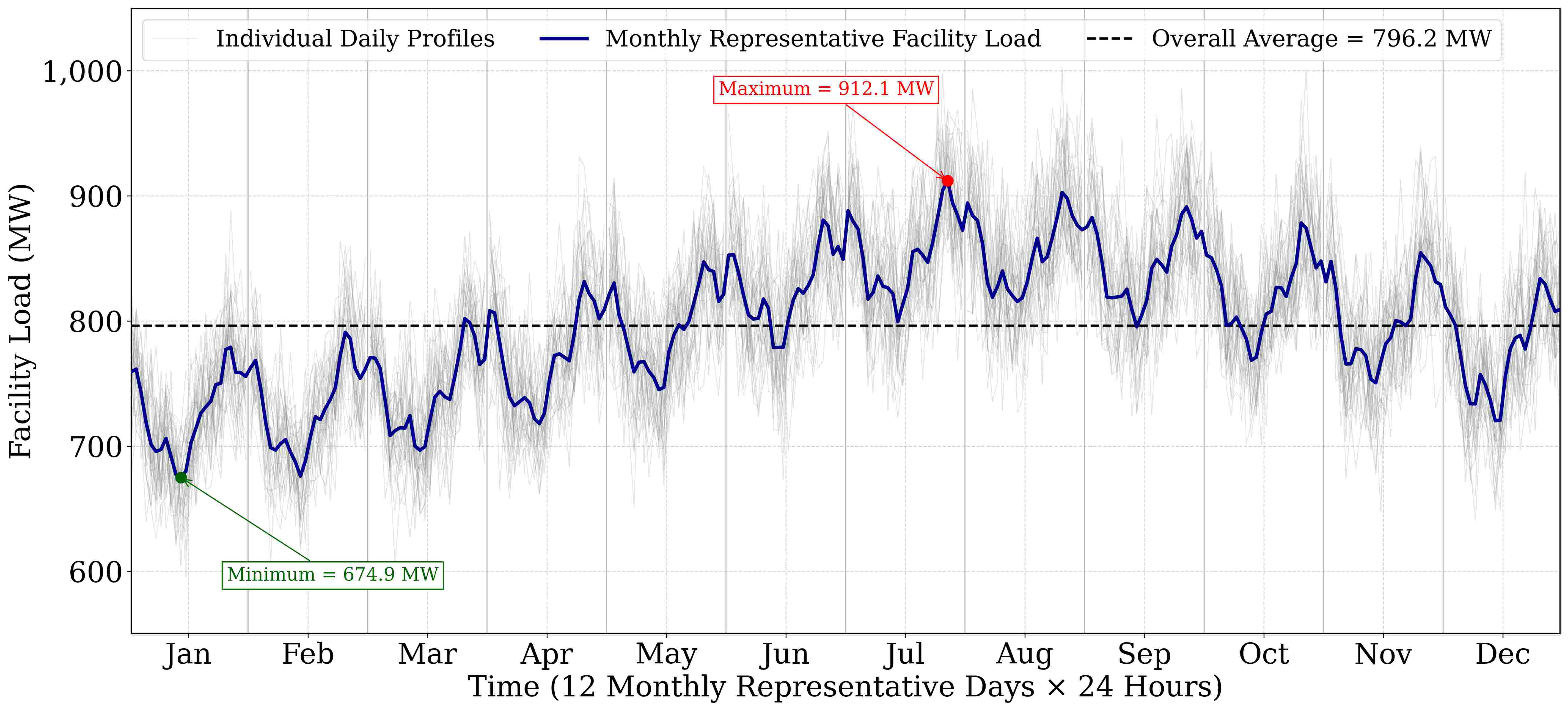}
\caption{Representative load profile for data center with hourly resolution.}
\label{DC_load_rep_hr}
\end{figure}

\begin{figure}[]
\centering
\footnotesize
\captionsetup{justification=raggedright, singlelinecheck=false, font={footnotesize}}
\includegraphics[width=3.4in]{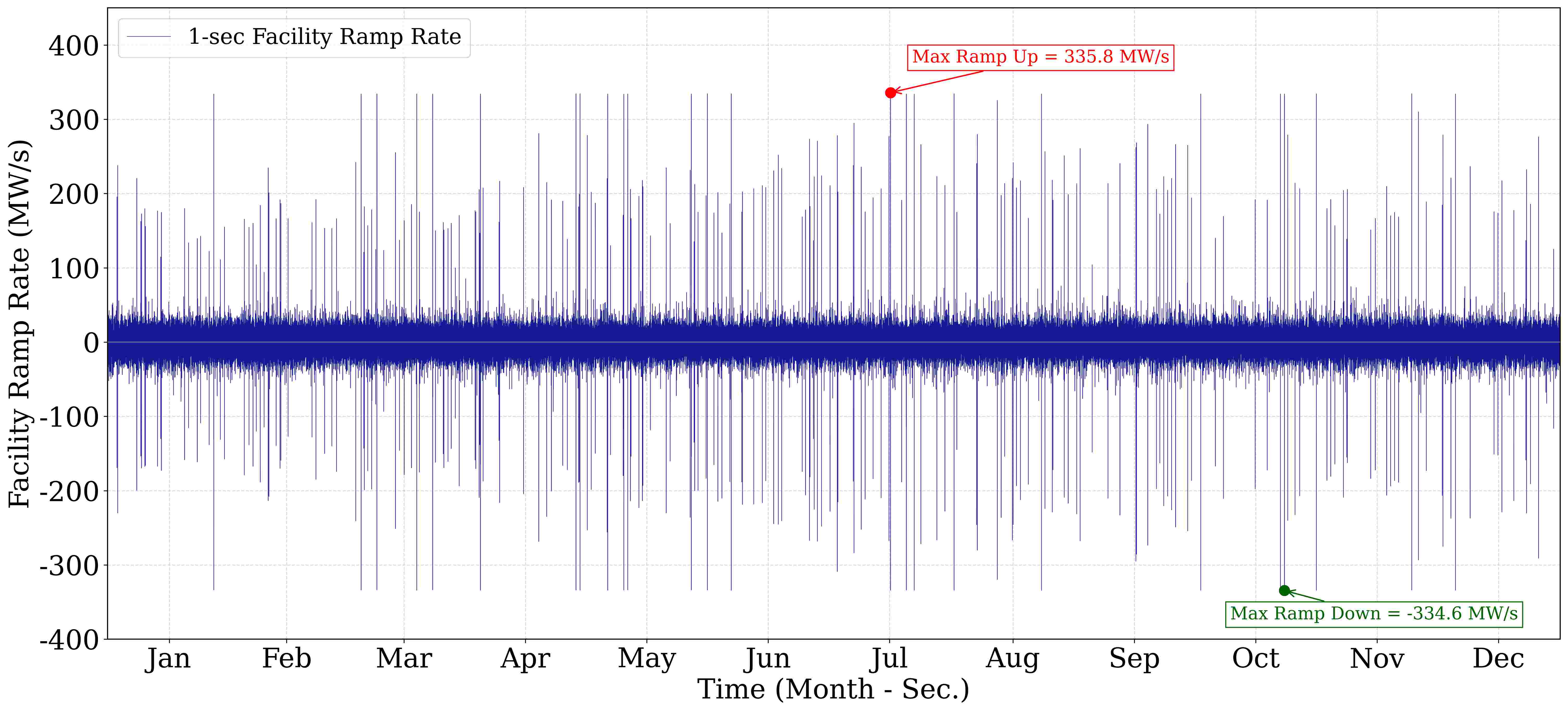}
\caption{Annual data center facility load ramping variation with second resolution.}
\label{DC_load_ramp_sec}
\end{figure}

\subsection{Load Smoothing Optimization Results:}
\begin{figure*}[t]
\centering
\footnotesize
\captionsetup{font={footnotesize}}
\begin{tabular}{ccc}
\subfloat[Smoothing with 50 MW BESS]{\includegraphics[height=1.6in,width=2in]{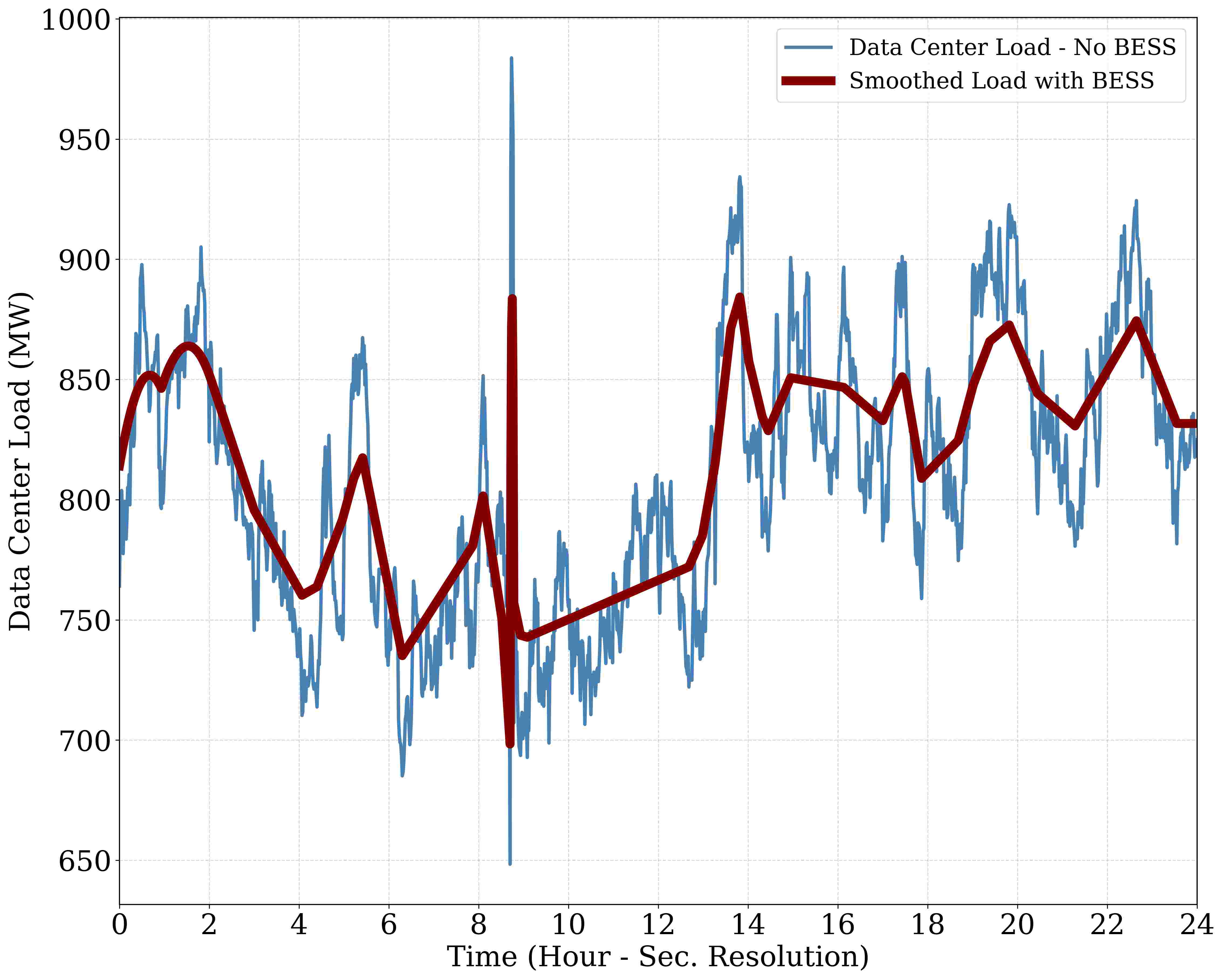}\label{DC_load_50MW_BESS}} &
\subfloat[Smoothing with 150 MW BESS]{\includegraphics[height=1.6in,width=2in]{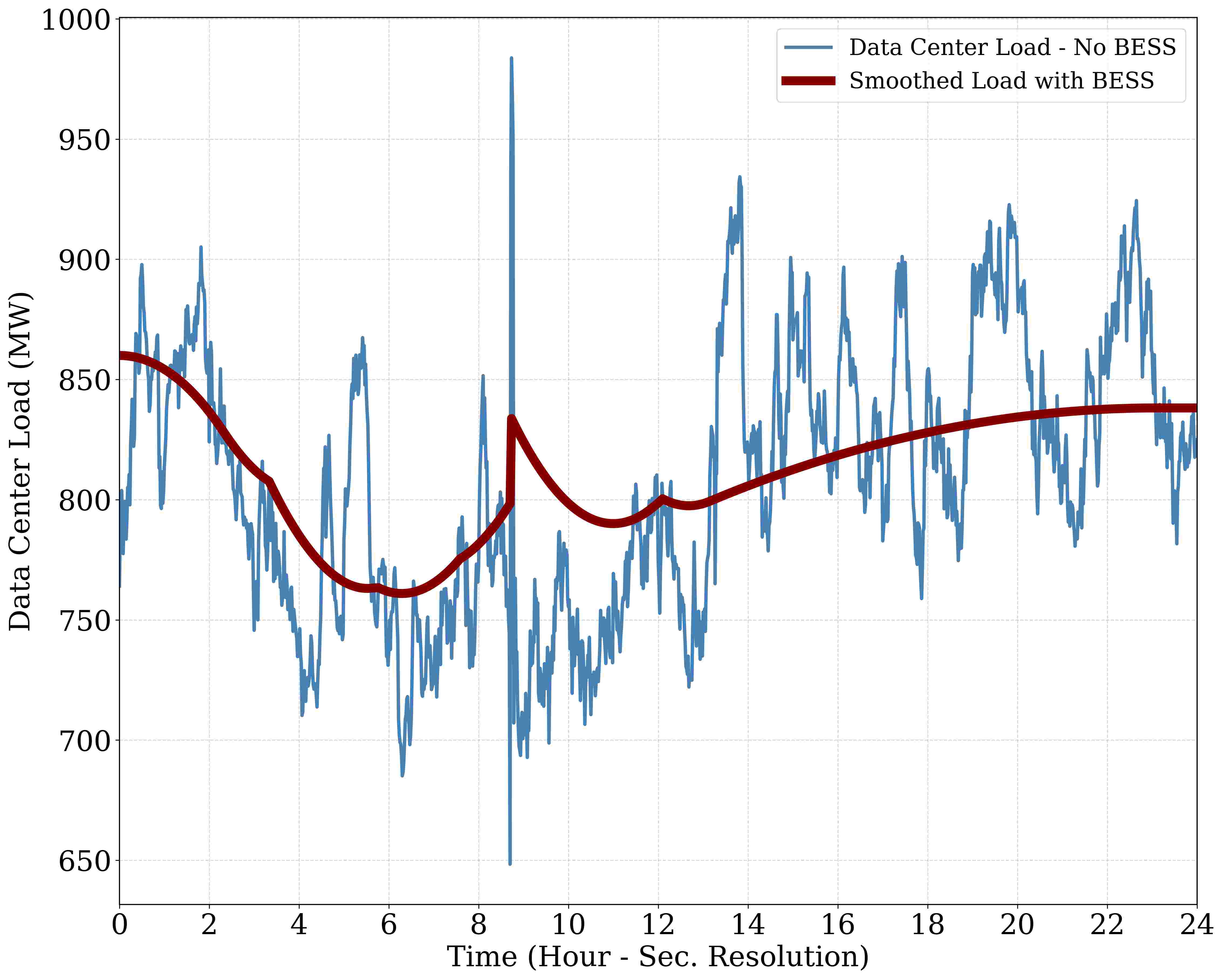}\label{DC_load_150MW_BESS}} &
\subfloat[Smoothing with 383 MW BESS]{\includegraphics[height=1.6in,width=2in]{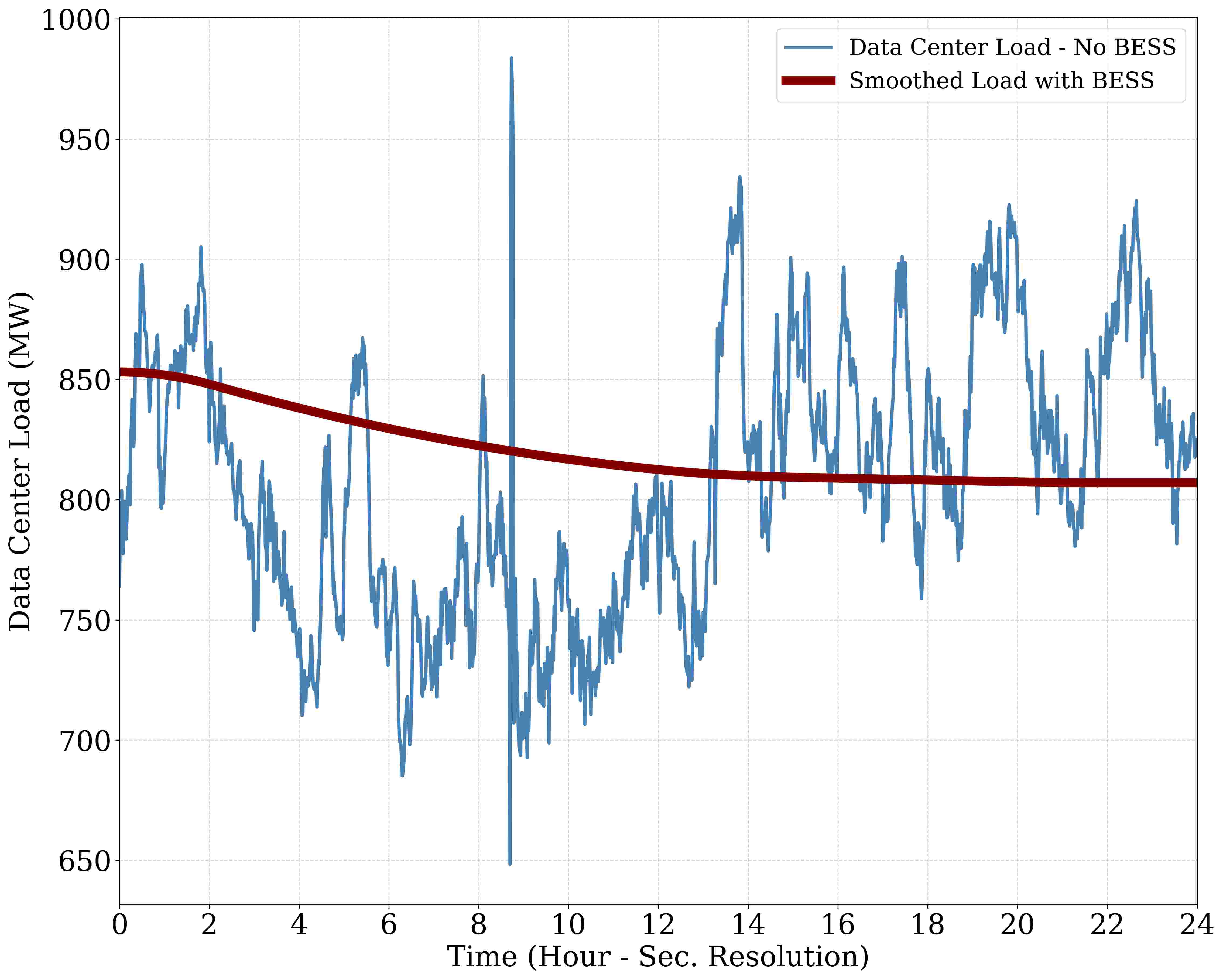}\label{DC_load_383MW_BESS}} \\
\subfloat[50 MW BESS Scheduling]{\includegraphics[height=1.6in,width=2in]{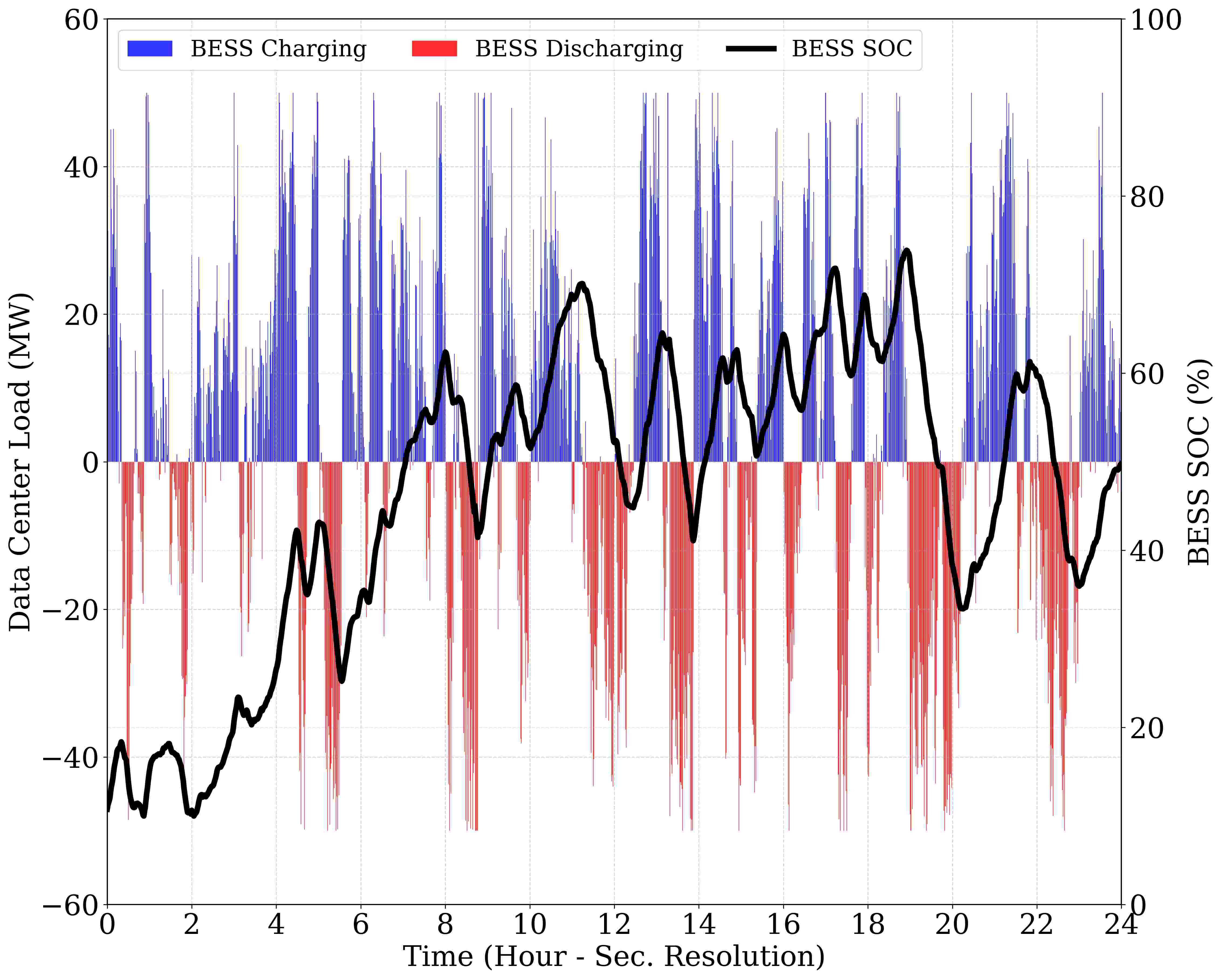}\label{SOC_50MW_BESS}} &
\subfloat[150 MW BESS Scheduling]{\includegraphics[height=1.6in,width=2in]{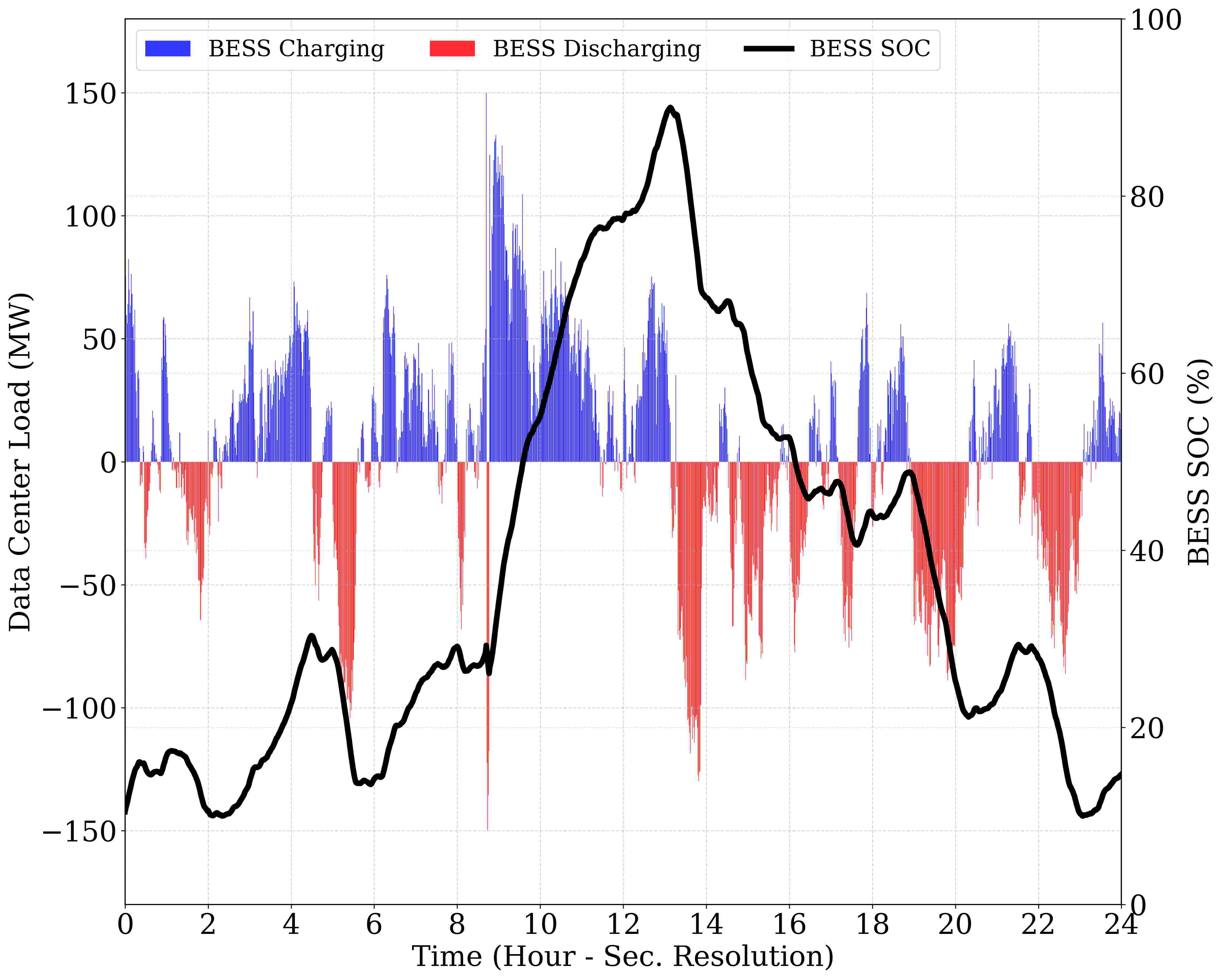}\label{SOC_load_150MW_BESS}} &
\subfloat[383 MW BESS Scheduling]{\includegraphics[height=1.6in,width=2in]{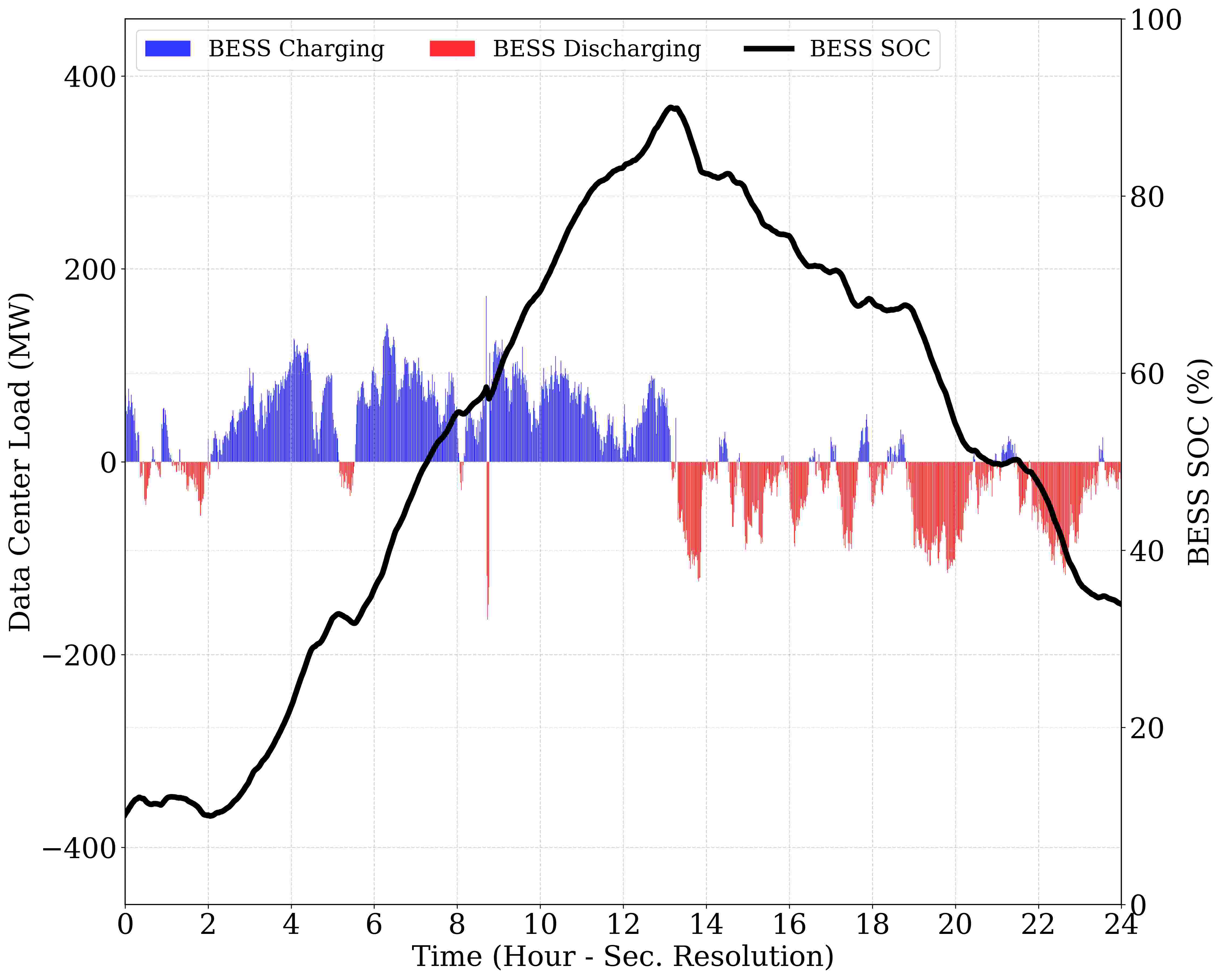}\label{SOC_load_383MW_BESS}}
\end{tabular}
\caption{Load smoothing optimization results with and without BESS deployment for different sizes, together with the corresponding BESS scheduling (charging and discharging) and SOC curves.}
\label{Load_Smoothing_and_BESS_SOC_resultss}
\end{figure*}

Considering Figure (\ref{DC_load_ramp_sec}), which illustrates the annual data center load ramping variations at one-second resolution, the maximum observed load transient reaches 335.8 MW/s. To prevent these rapid fluctuations from transferring to the generation side, the load-side BESS should be designed to absorb such shocks and provide a smoother and more predictable demand profile. Accordingly, the BESS sizing methodology presented in Section \ref{Problem_formulation_loadSmoothing} incorporates efficiency losses, temperature derating, availability, end-of-life degradation, PCS losses, and a 10\% design margin to ensure reliable performance throughout the asset lifetime. Based on the parameters summarized in Table \ref{BESS_Parameters_Table}, the resulting optimal BESS size is 383 MW with a two-hour duration, providing sufficient capacity to reliably mitigate the most severe load transients and support system stability.

Figures (\ref{DC_load_50MW_BESS})-(\ref{DC_load_383MW_BESS}) compares the load-smoothing performance achieved with 50 MW, 150 MW, and 383 MW BESS configurations. The 50 MW BESS provides limited smoothing capability, as its charge/discharge power is insufficient to fully mitigate the rapid second-by-second load fluctuations. Increasing the BESS size to 150 MW significantly improves the smoothing performance and reduces the magnitude of load variability observed by the generation system. However, several sharp spikes remain, particularly during extreme ramping events associated with the maximum load variation of 335.8 MW. The 383 MW BESS, which represents the reliable size after accounting for temperature derating, availability, PCS losses, and a 10\% reserve margin, delivers the smoothest load profile and effectively eliminates the remaining high-frequency spikes. Consequently, the generation system is exposed to a more stable and predictable demand profile, facilitating reliable generation scheduling and dispatch.

Figures (\ref{SOC_50MW_BESS})-(\ref{SOC_load_383MW_BESS})  presents the optimal BESS charging/discharging schedules and corresponding SOC profiles for the 50 MW, 150 MW, and 383 MW BESS configurations. For the 50 MW case, the limited power capacity requires frequent charging and discharging actions to mitigate load fluctuations, which is reflected by the corresponding variations in the SoC profile. Increasing the BESS rating to 150 MW provides greater operational flexibility, allowing the BESS to respond more effectively to rapid load variations and absorb a larger portion of the transient ramping events. The 383 MW BESS exhibits the highest smoothing capability, with sufficient charging and discharging capacity to absorb the most severe load transients and maintain a significantly smoother demand profile. Although the maximum second-to-second load variation reaches 335.8 MW, the full BESS power rating is not utilized during this specific event because it occurs when the baseline load is approximately 850 MW, allowing the optimal controller to balance charging and discharging actions. Consequently, the required BESS output depends not only on the magnitude of the ramp but also on the timing at which the transient occurs. The 383 MW rating already loss factors and a 10\% reserve margin, ensuring reliable mitigation of worst-case load transients and potential issues related to SSTI for generation side. By preventing high-frequency load spikes from propagating to the generation side, the BESS reduces the risk of SSTI on generation side and operational stress on generation assets and enhances overall system stability.

It should be noted that the adopted load-smoothing formulation penalizes all rapid load-ramping events through the ramp-rate penalty term, encouraging the BESS to mitigate both small and large transient load variations. In practice, allowable ramping limits can be enforced at the point of interconnection (POI) \cite{ercot_lcl_2026} where ERCOT plan to set this limit as 10 MW/sec, such that only load variations exceeding the prescribed threshold are penalized while smaller fluctuations are intentionally tolerated. Under this approach, the generation system would be exposed only to moderate load ramps, while the load-side BESS continues to absorb high-frequency transients and sharp load spikes. Although generation-side flexibility resources can accommodate part of these variations, mitigating rapid load fluctuations at the load side remains desirable to reduce operational stress on generation assets and minimize potential SSTI concerns. Furthermore, the presented results are based on a deterministic single-horizon optimization framework. Future work will investigate rolling-horizon optimization approaches to better capture forecast uncertainty, intertemporal operating decisions, and real-world dispatch conditions.

\subsection{BTM Generation Mix and Resilient BESS Sizing Optimization Results:}
\begin{figure}[]
\centering
\footnotesize
\captionsetup{justification=raggedright, singlelinecheck=false, font={footnotesize}}
\includegraphics[width=3.4in]{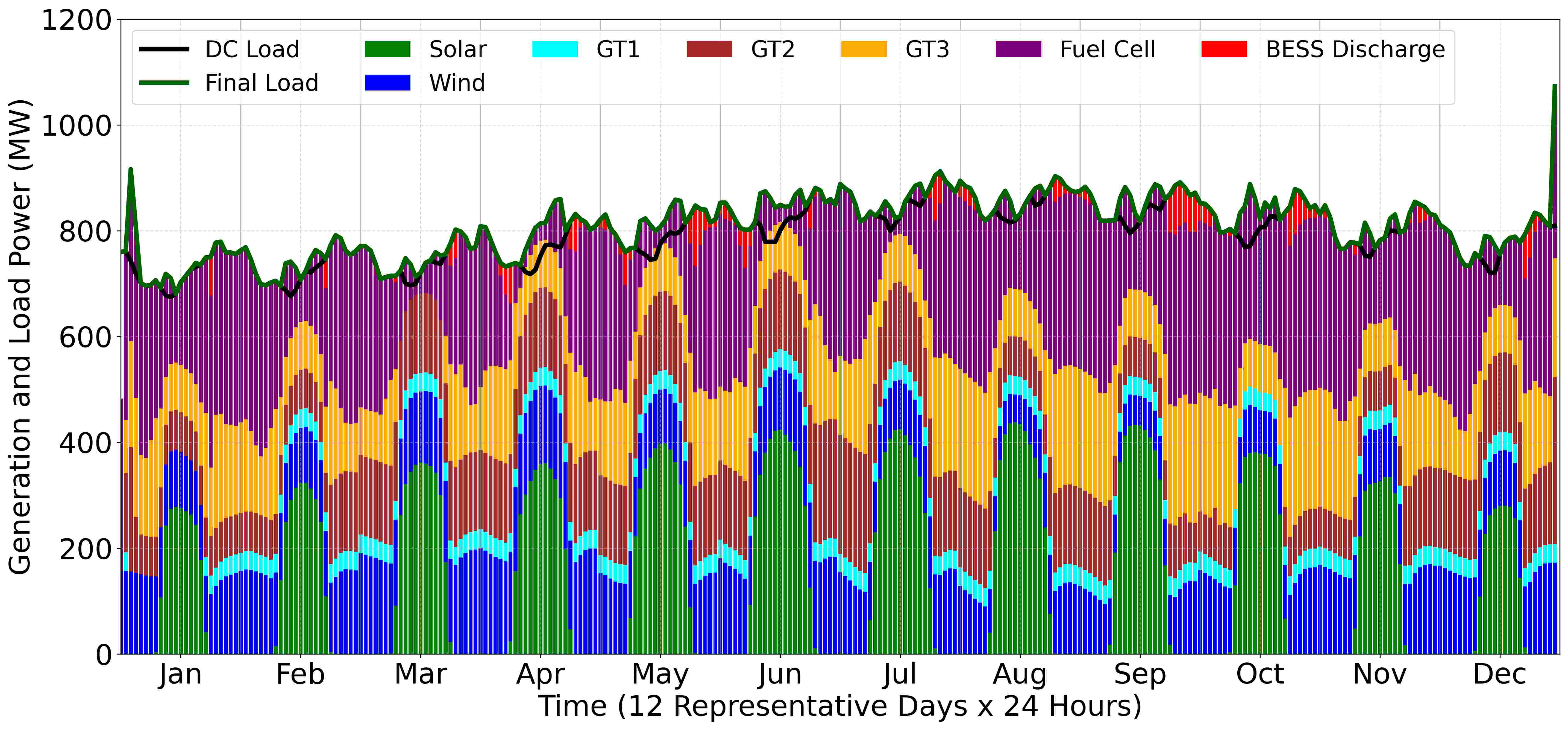}
\caption{Generation dispatch by resource type for BTM data center load supply.}
\label{Gen_vs_load}
\end{figure}

\begin{figure}[]
\centering
\footnotesize
\captionsetup{justification=raggedright, singlelinecheck=false, font={footnotesize}}
\includegraphics[width=3.4in]{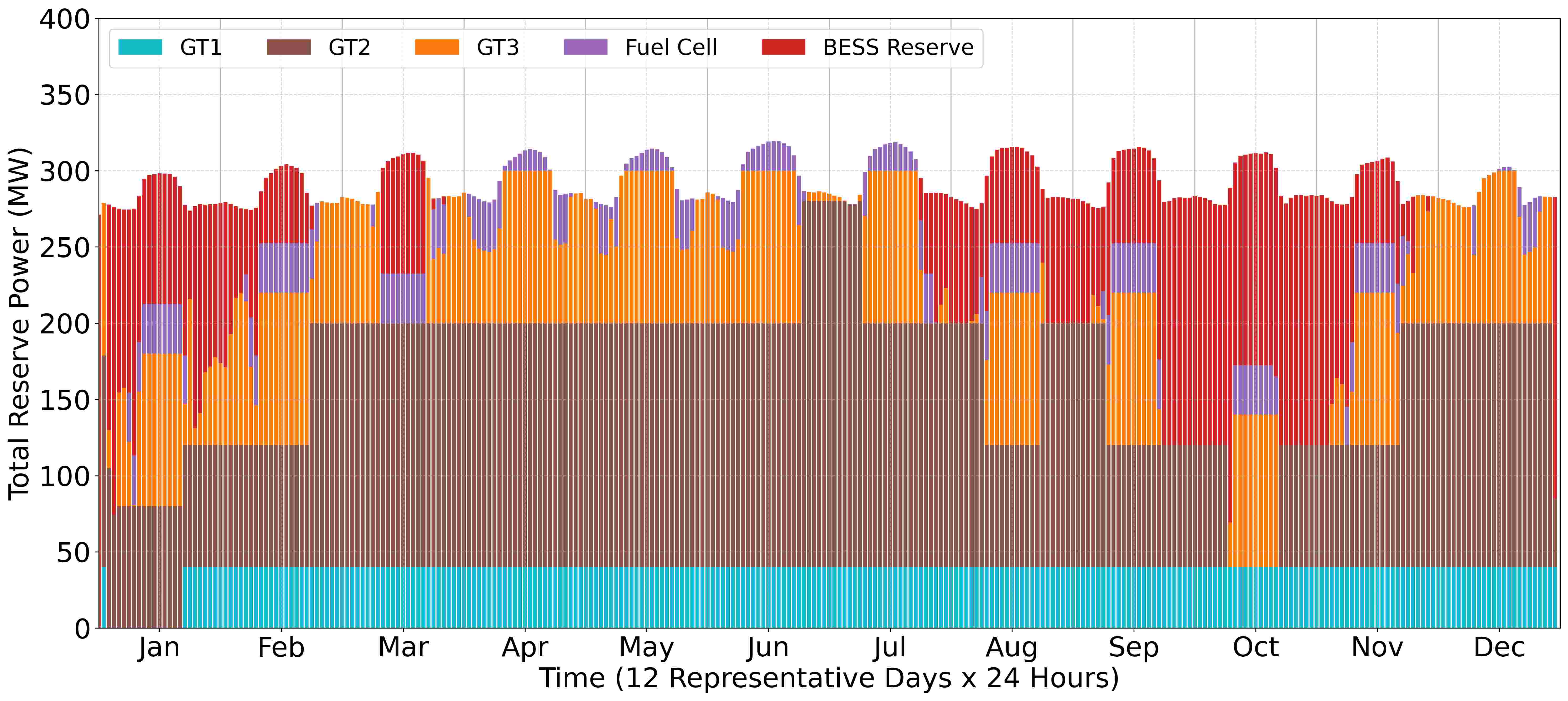}
\caption{Total reserve decomposition for GTs, FC, and BESS. }
\label{System_reserve_pltt}
\end{figure}

\subsubsection{Resilient BESS Size and Optimal Generation Dispatch} After solving the optimization problem defined in equations (\ref{OF_genside})–(\ref{Res_req_Ren}), the optimal generation-side BESS size is determined to be 124 units (based on info presented in Table \ref{BESS_Parameters_Table}), corresponding to a total power installed capacity of 271.1 MW with a four-hour energy duration which is equivalent to 1084.4 MWh energy. The selected size results from the trade-off between capital investment, operating cost minimization, reliability requirements, and resiliency considerations. In addition to economic objectives, the optimization explicitly enforces N-1 contingency criteria, renewable generation uncertainty, load uncertainty, and equipment availability constraints, ensuring that the resulting solution can reliably support data center operation under both normal and contingency conditions. Figure (\ref{Gen_vs_load}) illustrates the optimal generation dispatch used to supply the data center load. As expected, the dispatch follows the merit-order principle, whereby lower-cost resources are utilized before higher-cost units. Renewable generation is dispatched first due to its near-zero marginal operating cost, followed by GT3, FCs, GT2, and GT1. The BESS participates in the optimal dispatch whenever it is economically beneficial or required to satisfy reserve and reliability constraints. Consequently, the resulting generation portfolio minimizes total system cost while maintaining operational reliability and resiliency. The corresponding reserve allocation is also presented in Figure (\ref{System_reserve_pltt}), which shows the contribution of each resource toward the total system spinning reserve requirement. The reserve requirement accounts for the largest generator outage together with renewable generation and load uncertainties. The results demonstrate how reserve responsibilities are shared among the gas turbines, FCs, and BESS to ensure sufficient contingency response capability. Such reserves are essential for maintaining reliable operation following the loss of the online generating unit. 

\begin{figure}[]
\centering
\footnotesize
\captionsetup{justification=raggedright, singlelinecheck=false, font={footnotesize}}
\includegraphics[width=3.4in]{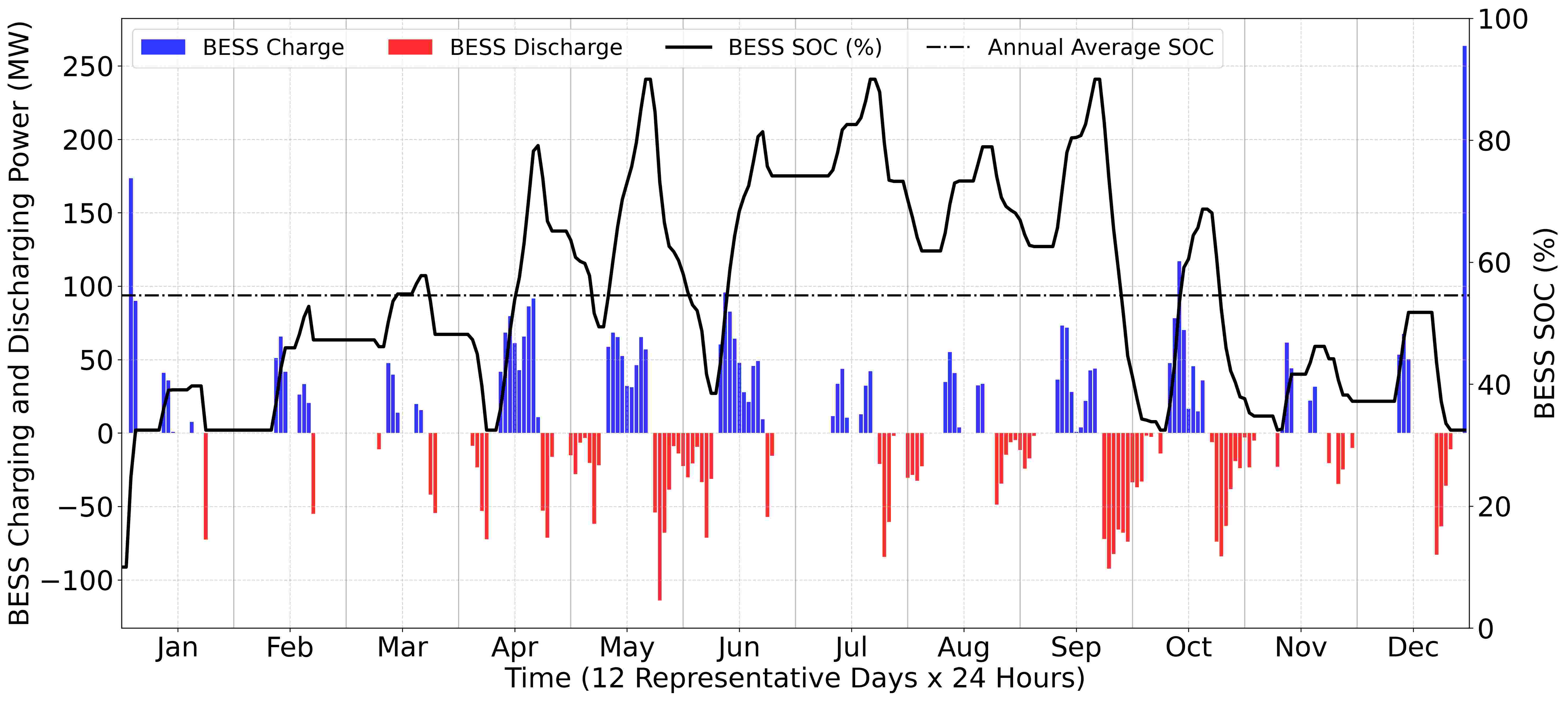}
\caption{Generation side BESS charging, discharging, and SOC.}
\label{Gen_BESS_SOC}
\end{figure}

\subsubsection{Optimal BESS Scheduling in Joint Energy-, Reserve-, Reliability-Constrained Optimization} Figure (\ref{Gen_BESS_SOC}) presents the optimal charging/discharging schedule and SOC profile of the generation-side BESS. The results clearly demonstrate that the BESS simultaneously supports energy dispatch and reserve provision. Periods of charging increase the SOC, whereas discharging periods reduce the stored energy to support system operation. The minimum SOC remains near 38\%, which corresponds to the energy reserved for contingency support, bridging requirements, and black-start capability. This reserved energy ensures that sufficient stored energy remains available to maintain system resiliency during generator outages and other abnormal operating conditions. The observed charging and discharging behavior is therefore driven not only by economic dispatch decisions but also by the reliability and resiliency requirements imposed within the optimization framework. Another BESS scheduling plot will be presented in reliability results section.

\begin{figure}[]
\center
\footnotesize
\captionsetup{font={footnotesize}}
\begin{tabular}{cc}
\subfloat[Hourly CFE]{\includegraphics[scale=0.112]{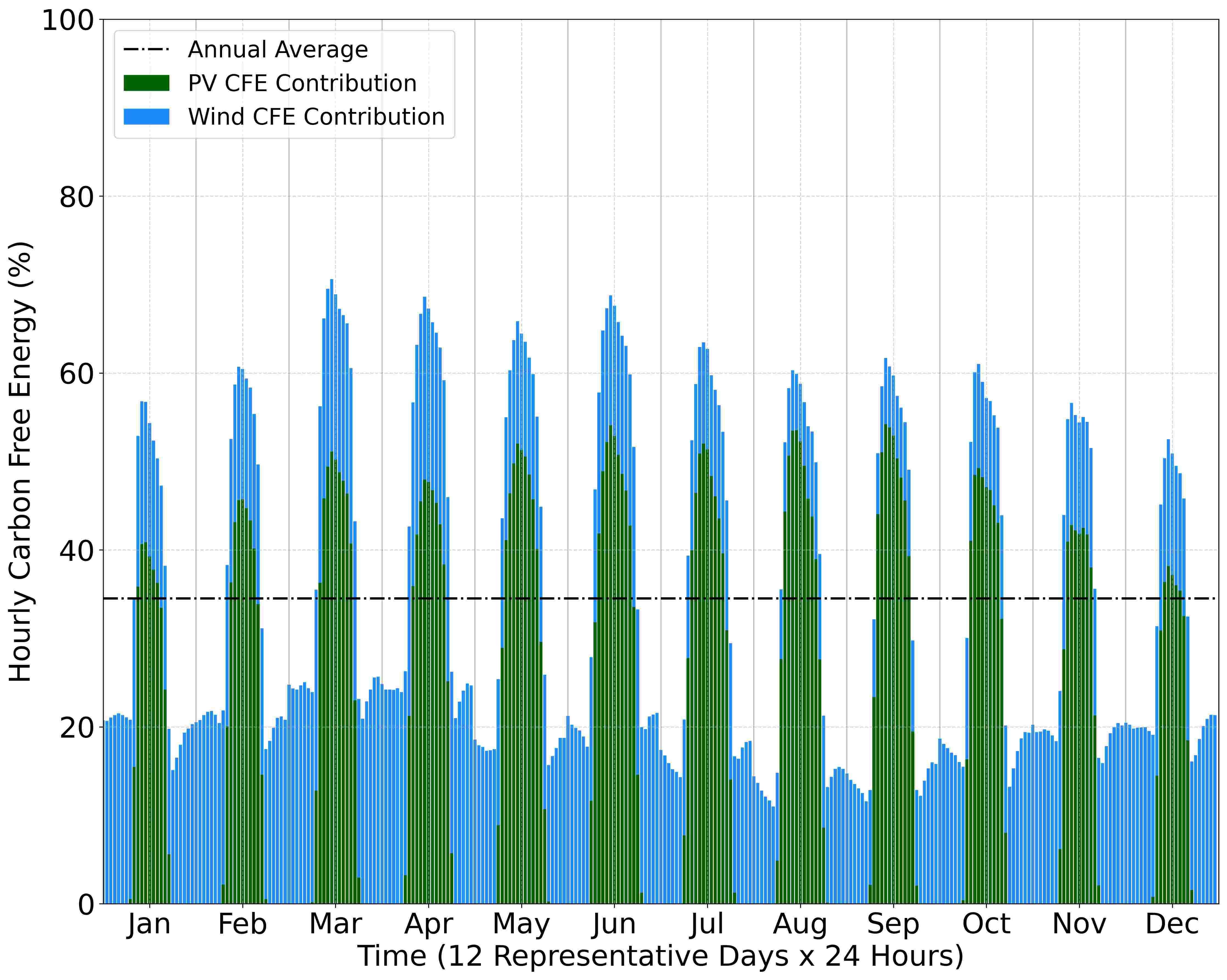}\label{CFE_hourlyy}}  
\subfloat[Monthly average CFE] {\includegraphics[scale=0.1175]{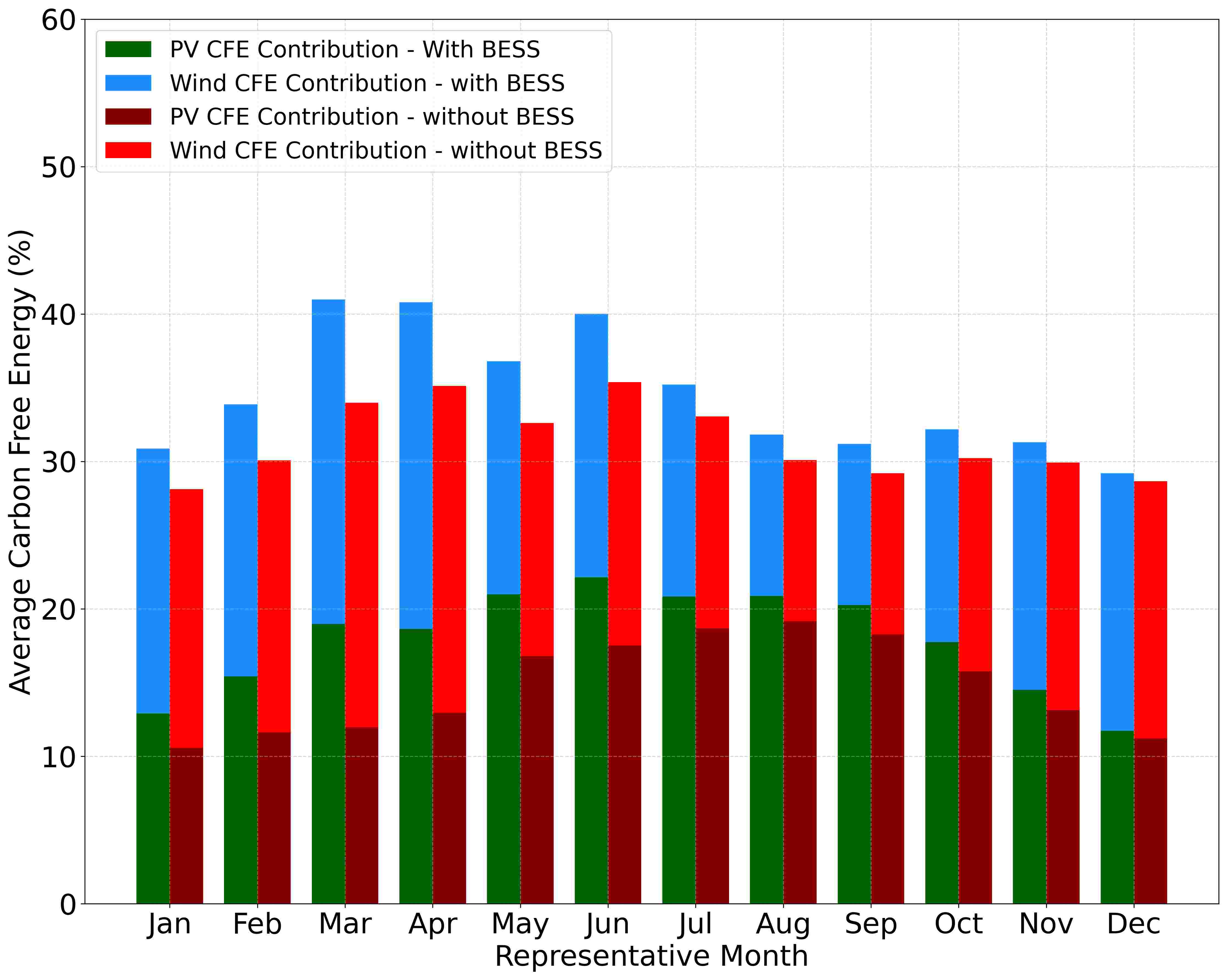}\label{monthly_average_CFE}}
\end{tabular}
\caption{Annual monthly and hourly CFE.}
\label{CFE_plots}
\end{figure}

\subsubsection{Carbon Free Energy Results} Figure (\ref{CFE_plots}) presents hourly CFE profile (with BESS) and monthly average CFE (with and without BESS) obtained from the system-level dispatch of BTM data center microgrid. Figure (\ref{CFE_hourlyy}) illustrates the hourly CFE profile and highlights the temporal variability of renewable resource availability throughout the month. The maximum hourly CFE reaches 70.60\%, which also occurs during March, reflecting periods when wind and solar PV generation are simultaneously available. During daytime, the combined dispatch of wind and solar resources frequently increases the CFE above 40\%. In contrast, during nighttime periods, the CFE contribution is primarily supplied by wind generation due to the absence of solar production. Overall, the average hourly CFE is 36.5\%, demonstrating the significant role of complementary wind and solar generation in increasing carbon-free energy penetration and reducing reliance on conventional generation sources.

In addition, Figure (\ref{monthly_average_CFE}) shows that the integration of BESS consistently improves carbon-free energy utilization across all representative months. The highest monthly CFE is observed in March, where the BESS-enabled case achieves 41.08\% CFE compared to 33.98\% without BESS, representing an absolute improvement of 7.1\%. Similar improvements are observed throughout the year, confirming the ability of BESS to increase renewable energy penetration and reduce dependence on conventional generation resources. The primary reason for this improvement is that, in the absence of BESS, thermal dispatchable units must remain online to satisfy energy, reserve, and reliability requirements. When these units are committed, their minimum operating limits force a portion of generation to be supplied by conventional resources even during periods of high renewable availability. As a result, renewable generation that cannot be accommodated by the system is curtailed, reducing the achievable CFE. With BESS installed, excess solar and wind energy can be captured instead of curtailed and later utilized for energy dispatch, reserve provision, or reliability support. This additional operational flexibility reduces renewable curtailment, decreases the need for conventional generation at minimum output levels, and increases the overall fraction of data center demand supplied by carbon-free resources.

\begin{figure}[]
\centering
\footnotesize
\captionsetup{justification=raggedright, singlelinecheck=false, font={footnotesize}}
\includegraphics[width=3.4in]{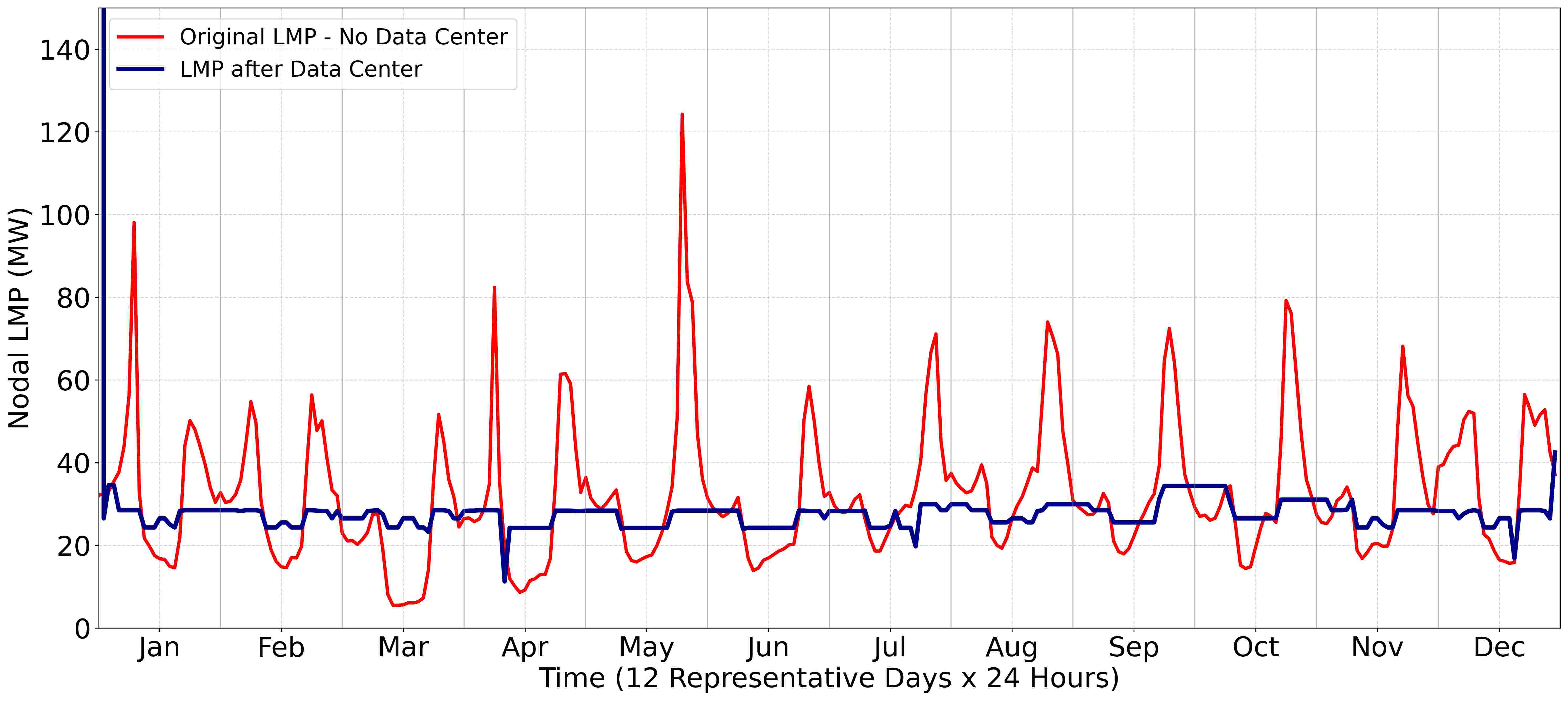}
\caption{Nodal system-level LMP before and after BTM data center with all generation assets.}
\label{LMP_node_BTMDC}
\end{figure}

\subsubsection{Nodal LMP Impact Analysis} 

Figure (\ref{LMP_node_BTMDC}) compares the original nodal LMP with the system-level LMP obtained from the optimized dispatch of the BTM data center microgrid. These two curves represent different quantities. The original nodal LMP reflects the market price that would be paid when importing electricity from the grid (when grid connection happens), whereas the system-level LMP represents the marginal cost of serving the data center load using the optimized portfolio of BTM resources. The figure shows that the system-level LMP remains relatively stable and periods with higher market prices indicate opportunities for BTM generation resources to economically supply the data center load and reduce exposure to costly grid purchases. Conversely, periods with low grid prices represent potential opportunities for grid-connected operation, where low-cost electricity can be imported and stored in the BESS for later use. The stored energy can subsequently be discharged during high-price periods, reducing future grid purchases and mitigating the impact of market price spikes. These results highlight that BESS provides value beyond reliability and reserve support by enabling energy arbitrage and enhancing operational flexibility. By strategically charging during low-price periods and discharging during high-price intervals, the BTM data center microgrid can further reduce operating costs while improving utilization of available energy resources. Additionally, It should be noted that the exceptionally high system-level LMP observed during the first hour does not represent a normal operating condition. An additional initialization hour was introduced prior to the 288-hour optimization horizon to transition the BTM generation assets from an offline state to their required operating conditions while respecting generator ramp-rate limitations. During this startup interval, available generation is temporarily constrained, resulting in the system marginal cost being set by the value of lost load (VOLL). Therefore, this initial price spike is an artifact of the model initialization process and is not representative of the steady-state economic operation of the BTM data center microgrid.

\begin{figure}[]
\centering
\footnotesize
\captionsetup{justification=raggedright, singlelinecheck=false, font={footnotesize}}
\includegraphics[width=3.4in]{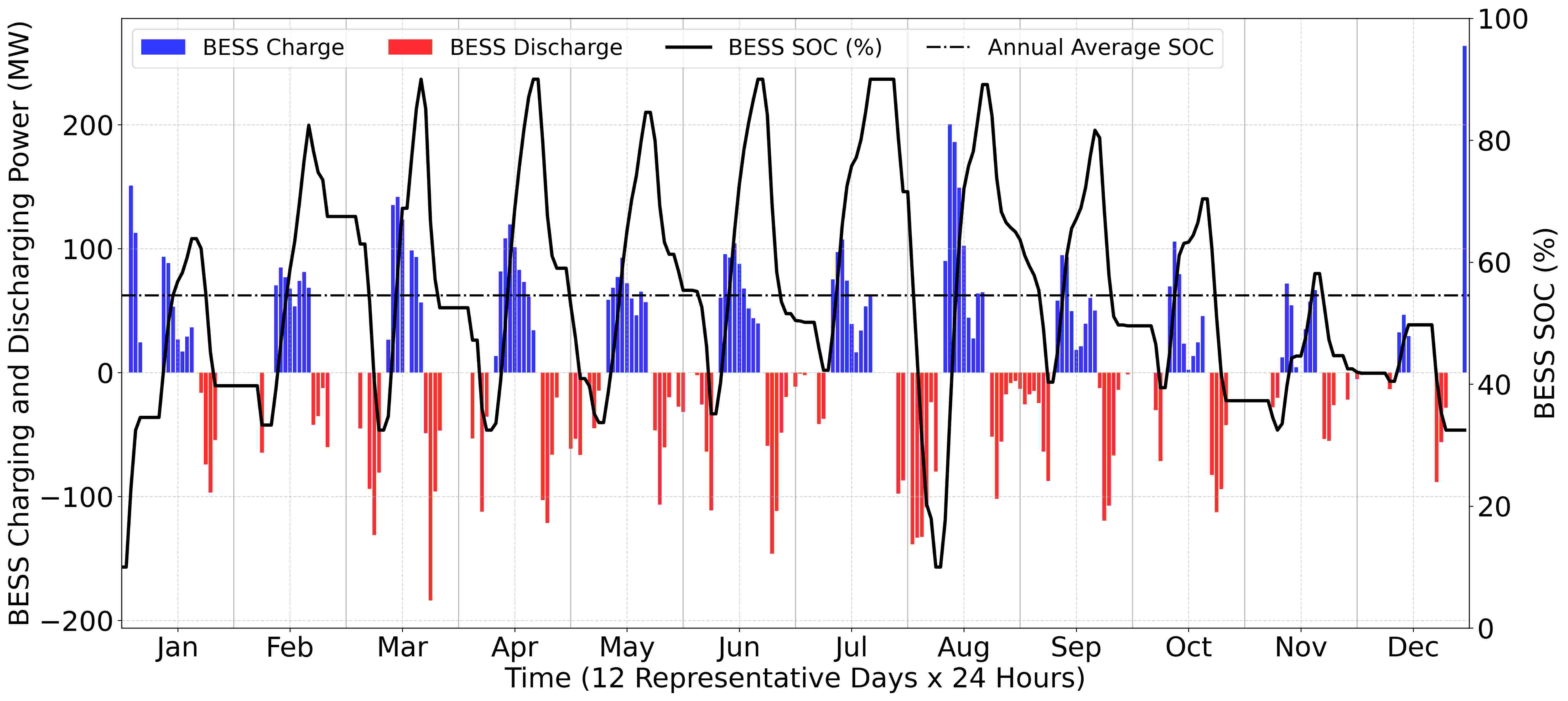}
\caption{Generation side BESS charging, discharging, and SOC with largest generator (GT3) outage for 10 hours starting at hour with peak load.}
\label{Gen_BESS_SOC_resilience}
\end{figure}

\subsubsection{Reliability and Resilience Analysis} 
Figure (\ref{Gen_BESS_SOC_resilience}) presents the optimal BESS dispatch schedule and corresponding SOC profile under the worst-case N-1 contingency, where the largest generating unit (GT3) experiences a forced outage for a 10-hour period beginning at the system peak-load hour (hour 166 - between July and August on plot). Prior to the outage event, the minimum SOC remains around 38\%, consisting of the base minimum SOC requirement of 10\% and an additional energy reserve maintained to satisfy the contingency-support requirement associated with the loss of the largest generating unit. Once the outage occurs, the SOC rapidly decreases from approximately 90\% to nearly 12\%, indicating that the BESS is immediately deployed to provide bridging energy and compensate for the lost generation capacity. During this period, the BESS maintains power balance and prevents load curtailment while supporting the remaining thermal units as they increase their output within their operational and ramp-rate constraints. After the outage is cleared and the generator returns to service, the BESS gradually restores its contingency reserve and the SOC returns to its normal operating range, increasing from the contingency-adjusted minimum level to above 80\%. These results demonstrate that the optimally sized BESS provides sufficient reserve and bridging capability to withstand the loss of the largest generating unit, while maintaining reliable system operation and supporting generator ramping requirements during contingency conditions.

To comprehensively evaluate the reliability and resilience performance of the proposed behind-the-meter data center microgrid, a deterministic N-1 contingency assessment was performed for all GT units as well as the FC modules. The analysis quantifies the ability of the microgrid to maintain continuous service during major generation outages while satisfying operating reserve requirements and preserving sufficient BESS flexibility. A set of reliability and resilience metrics commonly used in the power systems industry was calculated for each contingency scenario. These metrics collectively quantify service continuity, reserve adequacy, operational survivability, and post-disturbance recovery capability.

The energy not served (ENS) represents the total curtailed energy during a contingency and is calculated as

\begin{equation}
ENS=\sum_{t \in \mathcal{T}^{out}} P^{shed}_{[t]}\Delta t
\end{equation}

The loss-of-load hours (LOLH) metric quantifies the duration of service interruptions and is defined as

\begin{equation}
LOLH=\sum_{t \in \mathcal{T}^{out}}
\mathbb{I}\left(P^{shed}_{t}>0\right)\Delta t
\end{equation}

The percentage of load served during a contingency is calculated as

\begin{equation}
LS=\left(1-\frac{ENS}{\sum_{t \in \mathcal{T}^{out}}P^{load}_{t}\Delta t}\right)\times100
\end{equation}

Reserve adequacy is evaluated through the reserve shortage energy metric

\begin{equation}
RSE=\sum_{t \in \mathcal{T}^{out}}R^{Slack}_{t}\Delta t
\end{equation}

A resilience-based outage survival ratio is evaluated as

\begin{equation}
OSR=1-\frac{LOLH}{T^{out}}
\end{equation}

The proposed methodology was applied to all N-1 outage cases by sequentially forcing the outage of each gas turbine unit and the fuel cell system while re-optimizing the microgrid operation.

For every investigated N-1 outage condition, the optimization produced zero energy not served ($ENS=0$ MWh), zero loss-of-load hours ($LOLH=0$ h), and zero reserve shortage energy ($RSE=0$ MWh). Consequently, all facility demand was successfully supplied during the entire contingency period, resulting in a load-served percentage of 100\%. Furthermore, no load shedding events were observed, yielding an outage survival ratio of unity ($OSR=1$), which indicates complete ride-through capability under all simulated contingencies.

The results further demonstrate that the coordinated operation of the GTs, FC system, operating reserves, and BESS provided sufficient redundancy to satisfy both energy and reserve requirements during individual equipment outages. Therefore, the proposed microgrid configuration satisfies the analyzed deterministic N-1 reliability criterion while maintaining uninterrupted operation of the data center load and preserving adequate resilience against major generation contingencies.

\section{Conclusion}

This paper presented a comprehensive multi-phase framework for the planning, resilient design of BESS, and optimal operation of BTM hyperscale data center microgrids. The proposed methodology combines high-fidelity AI-driven data center load model, load-side BESS deployment for load smoothing and SSTI mitigation, generation-side BESS planning, and microgrid optimization within a unified framework. Unlike conventional approaches that rely on simplified load representations, the developed digital twin captures detailed IT and non-IT load behavior, second-level variability, transient burst events, cooling-system dynamics, and seasonal operating characteristics.

Simulation results demonstrated that future AI-oriented hyperscale facilities exhibit substantial short-term variability in addition to large seasonal demand fluctuations. The developed load model captured annual facility loading between 530.8 MW and 1048.7 MW, while identifying a maximum second-level load spike of 335.8 MW/s. To mitigate these rapid transients, the proposed load-side BESS sizing methodology identified a 383 MW, two-hour BESS capable of reliably absorbing the most severe load variations after accounting for efficiency losses, temperature derating, availability, end-of-life degradation, PCS losses, and design margins. The resulting smoothed load profile significantly reduced the exposure of generation resources to high-frequency load excursions which could result in SSTI challenges, if not controlled. For the generation side, the proposed resilience- and reserve-constrained optimization framework identified an optimal BESS size of 271 MW with a four-hour duration while coordinating FCs, GTs, solar PV, and wind resources. The optimized portfolio simultaneously satisfied economic, reserve, reliability, and resiliency requirements while improving renewable utilization and increasing carbon-free energy penetration through reduced renewable curtailment and enhanced operational flexibility. Finally, deterministic N-1 contingency analyses demonstrated the resilience of the proposed architecture. The coordinated operation of GTs, FCs, operating reserves, and BESS maintained uninterrupted service under all investigated contingency scenarios, resulting in zero energy not served, zero loss-of-load hours, zero reserve-shortage energy, and 100\% load served. These results confirm that the proposed framework provides a practical and effective pathway for addressing the emerging challenges of AI-driven load growth, renewable integration, reliability, resiliency, and operational flexibility in next-generation hyperscale data center microgrids. 

Future work will focus on incorporating uncertainty and risk quantification through distributionally robust chance-constrained optimization methods to address uncertainy of parameters, such as data center load and renewable energy generation. In addition, the optimization framework will be extended to integrate comprehensive reliability and availability assessments based on realistic maintenance schedules of BTM assets. Future research will also investigate the generation of large-scale scenarios for BTM data center microgrids and the application of machine learning and deep learning techniques to accurately predict optimization outcomes, enabling efficient real-time decision support and operational optimization.

\setstretch{1}

\bibliographystyle{IEEEtran}
\bibliography{mybib}

\vspace{-9cm}

\begin{IEEEbiography}[{\includegraphics[width=1in,height=1.25in,clip,keepaspectratio]{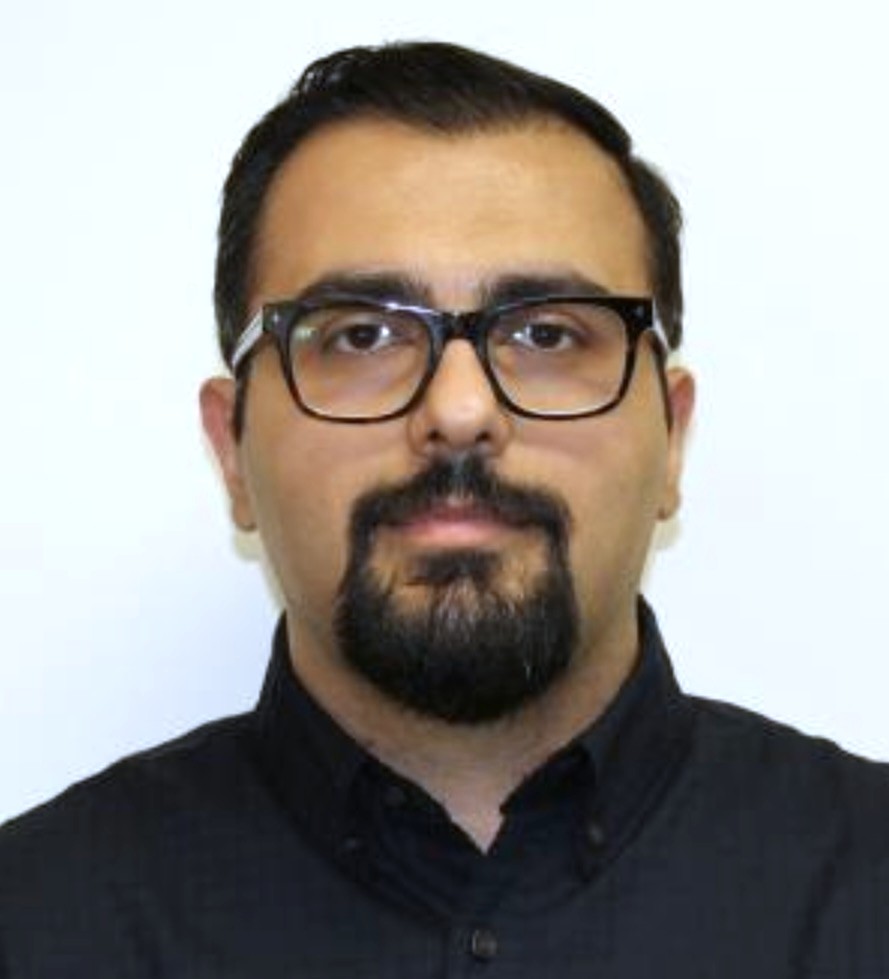}}]{Hamed Haggi (Member, IEEE)} received the Ph.D. degree in Electrical Engineering (Power Systems and Control) from the University of Central Florida (UCF), Orlando, FL, USA, in 2022. He is currently a Lead Optimization Engineer/Senior Technical Manager with WSP USA power and energy division. Prior to joining WSP in 2026, he was an Assistant Professor with the UCF. He has led and contributed to more than \$15 million in U.S. Department of Energy (DOE) and National Science Foundation (NSF) funded research and demonstration projects in collaboration with electric utilities, industry partners, and national laboratories. His research interests include power system operation, optimization, planning, renewable energy and energy storage systems, AI/ML, and the cyber-physical resilience and security of energy systems. He is an active reviewer for numerous IEEE journals and conferences, and an active member in IEEE PES, and IEEE IAS societies. He was the recipient of the Best Paper Award at the IEEE Power \& Energy Society General Meeting (2025), the Best Paper Award at the IEEE Power \& Energy Society Innovative Smart Grid Technologies (ISGT) Conference (2019), and the Best Paper Award at the International Power System Conference (PSC) (2017). He was also the recipient of multiple prestigious scholarships, fellowships, and research awards throughout his academic and professional career.
\end{IEEEbiography}
\vspace{-9cm}
\begin{IEEEbiography}[{\includegraphics[width=1in,height=1.25in,clip,keepaspectratio]{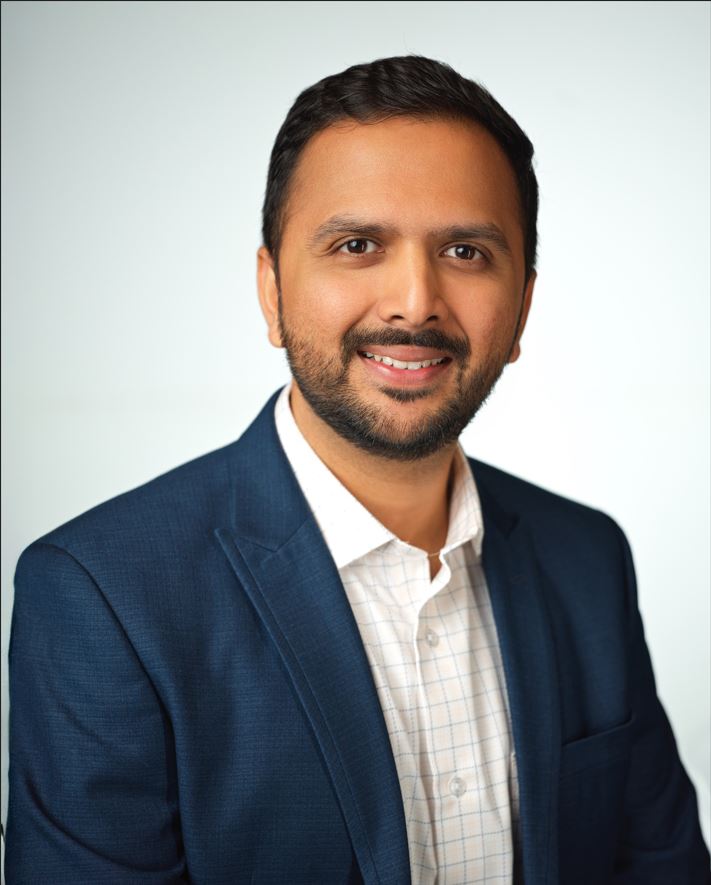}}]{Chinmay Morankar} (M’18, SM’23) became a Member (M) of IEEE in 2018 and Senior Member (SM) in 2023. He currently serves in Power 
and Energy group of WSP USA as VP of Systems Engineering. 
He is leading engineering designs for utility scale solar, wind, and 
energy storage projects. He is also playing a critical role in 
various microgrid and islanded data center projects in the United States. He is focused on providing Strategic Energy Infrastructure Planning services for the clients. He is a 
senior member of IEEE, and is part of technical committee for 
ACP R\&T. He has presented in various conferences. His 
professional preparation includes a Bachelor of Engineering in 
Mechanical Engineering from the University of Pune and an MS 
in Mechanical Engineering from the University of Colorado at 
Boulder.  
\end{IEEEbiography}

\end{document}